# Stabilizing Deep Tomographic Reconstruction

Weiwen Wu[1,5], Dianlin Hu[2], Wenxiang Cong[1], Hongming Shan[1,3], Shaoyu Wang[4], Chuang Niu[1], Pingkun Yan[1], Hengyong Yu[4,*], Varut Vardhanabhuti[5,*], Ge Wang[1,*]

[1]Biomedical Imaging Center, Center for Biotechnology and Interdisciplinary Studies, Department of Biomedical Engineering, Rensselaer Polytechnic Institute, Troy, NY, USA

[2]The Laboratory of Image Science and Technology, Southeast University, Nanjing, China

[3]Institute of Science and Technology for Brain-inspired Intelligence, Fudan University, Shanghai, China

[4]Department of Electrical & Computer Engineering, University of Massachusetts Lowell, Lowell, MA, USA

[5]Department of Diagnostic Radiology, Li Ka Shing Faculty of Medicine, The University of Hong Kong, Hong Kong SAR, China

**Abstract: Tomographic image reconstruction with deep learning is an emerging field, but a recent landmark study reveals that several deep reconstruction networks are unstable for computed tomography (CT) and magnetic resonance imaging (MRI). Specifically, three kinds of instabilities were reported: (1) strong image artefacts from tiny perturbations, (2) small features missing in a deeply reconstructed image, and (3) decreased imaging performance with increased input data. On the other hand, compressed sensing (CS) inspired reconstruction methods do not suffer from these instabilities because of their built-in kernel awareness. For deep reconstruction to realize its full potential and become a mainstream approach for tomographic imaging, it is thus critically important to meet this challenge by stabilizing deep reconstruction networks. Here we propose an Analytic Compressed Iterative Deep (ACID) framework to address this challenge. ACID synergizes a deep reconstruction network trained on big data, kernel awareness from CS-inspired processing, and iterative refinement to minimize the data residual relative to real measurement. Our study demonstrates that the deep reconstruction using ACID is accurate and stable, and sheds light on the converging mechanism of the ACID iteration under a Bounded Relative Error Norm (BREN) condition. In particular, the study shows that ACID-based reconstruction is resilient against adversarial attacks, superior to classic sparsity-regularized reconstruction alone, and eliminates the three kinds of instabilities. We anticipate that this integrative data-driven approach will help promote development and translation of deep tomographic image reconstruction networks into clinical applications.**

# Introduction

Medical imaging plays an integral role in modern medicine and grows rapidly over the past decades. In the USA, there are over 80 million computed tomography (CT) scans and 40 million magnetic resonance imaging (MRI) scans performed yearly [1, 2]. In a survey on medical innovations, it was reported that "*the most important innovation by a considerable margin is magnetic resonance imaging (MRI) and computed tomography (CT)*" [3]. Over the past several years, deep learning has attracted major attention. Since 2016, deep learning has been gradually adopted for tomographic imaging, known as deep tomographic imaging [4-6]. Traditionally, tomographic reconstruction algorithms are either analytic (*i.e.*, closed-form formulation) or iterative (*i.e.*, based on statistical and/or sparsity models). More recently with deep tomographic imaging, reconstruction algorithms employ deep neural networks, which are data-driven [7, 8]. This new type of reconstruction algorithms has generated tremendous excitement and promising results in many studies. Some examples are included in a recent review article [6].

While many researchers are devoted to catching this new wave of tomographic imaging research, there are concerns on deep tomographic reconstruction as well, with the landmark paper by Antun *et al.* [9] as the primary example. Specifically, Antun *et al*. [9] performed a systematic study to reveal the instabilities of a number of representative deep tomographic reconstruction networks, including AUTOMAP [10]. Their study demonstrates three kinds of network-based reconstruction vulnerabilities: (1) tiny perturbations on the input generating strong image artefacts (potentially, false positivity); (2) small structural features going undetected (false negativity); and (3) increased input data leading to decreased imaging performance. These critical findings are warnings and at the same time present opportunities of deep tomographic imaging research. Importantly, the study by Antun *et al.* [9] found that small structural changes (*e.g*., a small tumor) may not be always captured in the images reconstructed by the deep neural networks, but standard sparsity-regularized methods can capture these pathologies. It is worth noting that the issue of missing pathologies was one of the main concerns raised by radiologists in the Fast MRI challenge in 2019 [11].

Historically, a debate, challenge, or crisis typically inspired theoretical and methodological development. In the context of tomographic imaging, there are several such examples. In the earliest days of CT reconstruction, analytic reconstruction received a critique that given a finite number of projections, tomographic reconstruction is not uniquely determined, meaning that ghost structures can be reconstructed, which do not exist in reality but are consistent with the measured data [12]. Then, this problem was solved by regularization, such as enforcing the band-limitedness of the underlying signals [13]. Iterative reconstruction algorithms were initially criticized that image reconstruction was strongly influenced by the penalty term; in other words, what you reconstruct could be what you want to see. After selecting regularization terms and fine-tuning hyper-parameters, these shortcomings were addressed. Hence, such algorithms have been made into clinical applications [14, 15]. As far as compressed sensing (CS) is concerned, the validity of this theory is based on restricted isometry or robust null space properties. The correct sparse solution may not always be obtained. However, in practice it works successfully "*with an overwhelming probability*" [16]. Nevertheless, the sparsity constraint could be either too strong smearing features, or too weak resulting in artifacts. For example, a tumor-like structure could be introduced, and pathological vessels might be filtered out if the total variation is overly minimized, as demonstrated in purposely-designed numerical examples [17]. Despite the limitations, multiple sparsity-promoting reconstruction algorithms are used on commercial scanners, with excellent overall performance.

Now, the emerging deep tomographic imaging methods encounter challenges, as reported in [9]. In addition to extensive experimental data showing the instabilities of several deep reconstruction networks, the authors pointed out that these instabilities are fundamentally associated with the lack of "*kernel awareness*" [18] and are "*nontrivial to overcome*" [9]. On the other hand, their experiments show that CS-inspired reconstruction algorithms worked stably while their selected deep reconstruction network failed under the same conditions [9], since CS-based algorithms use sparse regularization that has "*at its heart a notion of kernel awareness*" [18].

This article focuses on the feasibility and principle of accurate and stable deep tomographic reconstruction, demonstrating that deep reconstruction networks can be stabilized in a hybrid model with a CS module embedded, and superior to CS-based reconstruction alone. Specifically, to overcome the instabilities of the deep reconstruction networks, here we propose an Analytic Compressed Iterative Deep (ACID) framework illustrated in Fig. 1(a). Given a deep reconstruction network $\Phi$ and measurement data $\boldsymbol{p}^{(\mathbf{0})}$, an image can first be reconstructed but it may miss fine details and introduce artifacts. Second, a CS-inspired module $\Theta$ enforces sparsity in the image domain [19], with a loss function covering both data fidelity and sparsity (such as total variation [20], low-rank [21], dictionary learning [22], *etc*.). Third, the forward imaging model projects the current image to synthesize tomographic data, which is generally different from the original data $\boldsymbol{p}^{(\mathbf{0})}$. The discrepancy is called a data residual that cannot be explained by the current image. From this data residual, an incremental image is reconstructed with the deep reconstruction network $\Phi$ and used to modify the current image aided by the sparsity-promoting CS module $\Theta$. This process can be repeated to prevent losing or falsifying features. As a meta-iterative scheme, the ACID reconstruction process cycles through these modules repeatedly. As a result, ACID finds a desirable solution in the intersection of the space of data-driven solutions, the space of sparse solutions, and the space of solutions subject to data constraints, as shown in Fig. 1(b). Because this integrative reconstruction scheme is uniquely empowered with both data-driven and sparsity priors, ACID would give a better solution than the classic sparsity-regularized reconstruction alone; for details, see the **Methods** section.

An important question is if the ACID iteration will converge to a desirable solution in the above-described intersection of the three spaces (Fig. 1(b)). The answer to this question is far from trivial. First, a deep learning network represents a non-convex optimization problem, which remains a huge open problem; see more details in the latest review [23]. The non-convex optimization problem in a general setting is of non-deterministic polynomial-time hardness (NP-hardness). To solve this problem with guaranteed convergence, practical assumptions must be made in almost all cases. These assumptions include changing a nonconvex formulation into a convex one under certain conditions, leveraging a problem-specific structure, and only seeking a local optimal solution. Specifically, the Lipschitz continuity is a common condition used to facilitate performing non-convex optimization tasks [24, 25].

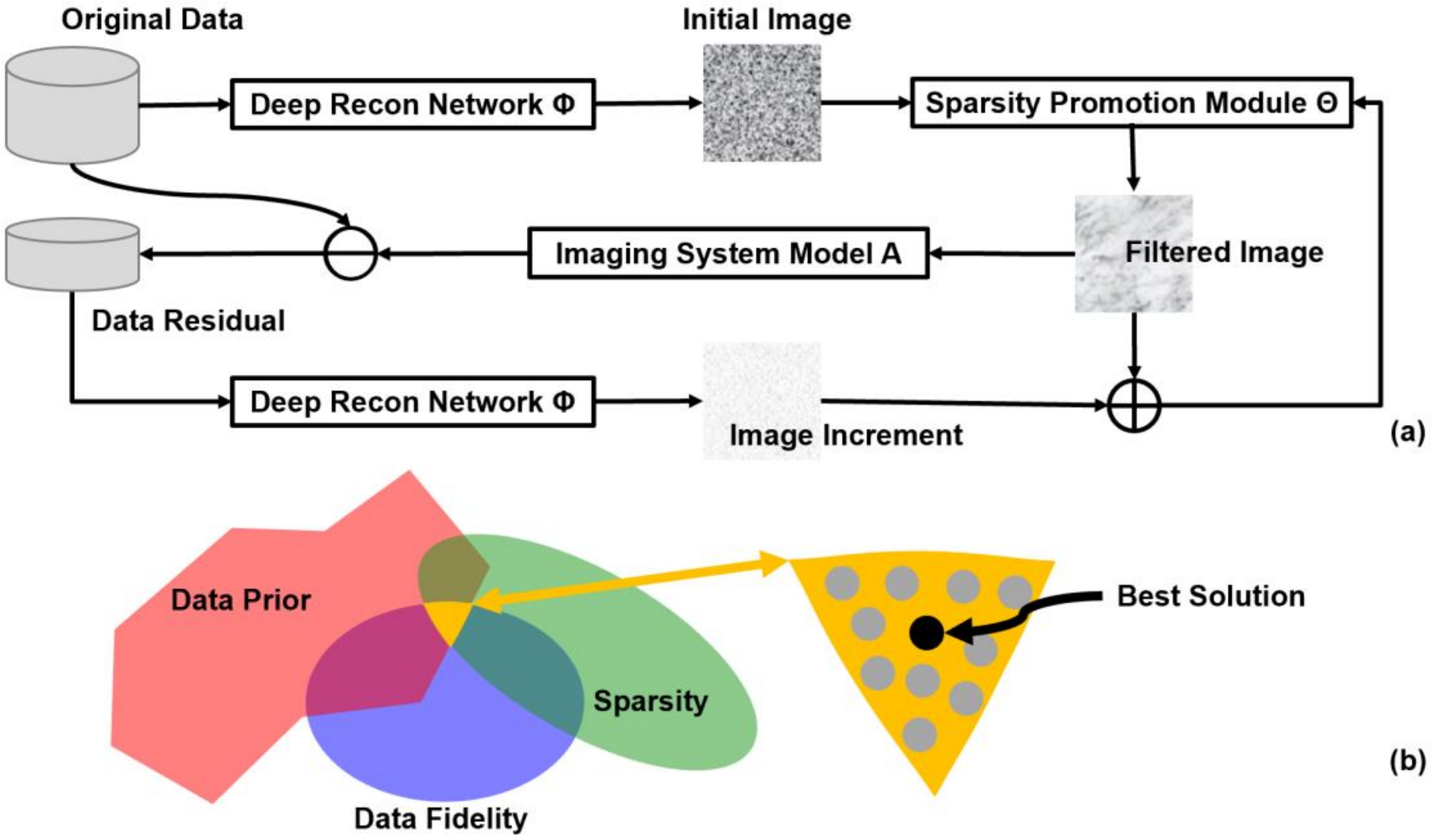

**Figure 1. ACID architecture for stabilizing deep tomographic image reconstruction**. Initially, the measurement data is reconstructed by the reconstruction network Φ. The current image is sparsified by the CS-inspired sparsity-promoting module Θ (briefly, the CS module). Tomographic data are then synthesized based on the sparsified image according to the system model $\boldsymbol{A}$, and compared to the measurement data to find a data residual. The residual data is processed by the modules Φ and Θ to update the current image. This process is repeated until a satisfactory image is obtained.

Given the theoretically immature status of the non-convex optimization, to understand our heuristically designed ACID system in terms of its convergence, we assume that a well-designed and well-trained deep reconstruction network satisfies our proposed Bounded Relative Error Norm (BREN) property, which is a special case of the Lipschitz continuity as detailed in the Supplementary Information (**SI**), Subsection **IV.G**. Based on the BREN property, the converging mechanism of the ACID iteration is revealed in our two independent analyses; see Section **IV** in the **SI** for details.

Here we would like to outline the key insight into the convergence of the ACID workflow. In reference to Fig. 1, we assume the BREN property of a deep reconstruction network Φ, as characterized by the ratio being less than 1 between the norm of the reconstruction error and the norm of the corresponding ground truth (assuming a nonzero norm without loss of generality); that is, the error component of the initial image reconstructed by the deep network Φ is less than the ground truth image in the $L_2$ norm. This error consists of both sparse and non-sparse components. The non-sparse component is effectively suppressed by the CS module Θ. The sparse errors are either observable or unobservable. The unobservable error is in the null space of the system matrix $\boldsymbol{A}$ and should be small relative to the ground truth image given the BREN property (the deep reconstruction network will effectively recover the null space component if it is properly designed and well trained). ACID can eliminate the observable error iteratively, owing to the BREN property. Specifically, the output of the module Θ is re-projected by the system matrix $\boldsymbol{A}$, and then the synthesized data are compared with the measured data. The difference is called the data residual due to the observable error component. To suppress this error component, we use the network Φ to reconstruct an incremental image and add it to the current image, and then refine the updated image with the CS module Θ. In this correction step, the desirable incremental image is the new ground truth image, and the BREN property remains valid that this step is a contraction mapping. In other words, the associated new observable error is less than the previous observable error, by the BREN property of the deep reconstruction network Φ. Repeating this process leads to the observable error diminishing exponentially fast (the BREN ratio less than 1). In doing so, the ACID solution will simultaneously incorporate data-driven knowledge, image sparsity preference, and measurement data consistency.

Note that, in a recent paper [26], a two-step deep learning strategy was analyzed for tomographic imaging, where a classical method was followed by a deep-network-based refinement to "*close the gap between practice and theory*" for that particular reconstruction workflow. The key idea is to use the null space network for data-driven regularization, achieving convergence based on the Lipschitz smoothness. It is underlined that our analysis on the convergence of ACID is in a similar spirit; for more details, please see **SI IV**.

# Results

Given the importance of the recent study on instabilities of some representative deep reconstruction networks [9], the main motivation of our work is to stabilize deep tomographic reconstruction. Hence, our experimental setup systematically mirrored what was described in [9], including datasets and their naming conventions, selected reconstruction networks, CS-based minimization benchmarks, and image quality metrics. As a result, the Ell-50 and DAGAN networks were chosen for CT and MRI reconstructions, respectively (details in the **Methods** section and **SI**). Both those CT and MRI networks were subjected to the instabilities reported in [9]. In addition to the system-level comparison, we performed an ablation study on the ACID workflow and investigated its own stability against adversarial attacks. For details, **SI III** and [9]. The full descriptions of the original cases of C1-C7, M1-M12 and A1-A4 are in **SI I**.

### (1) Stability with Small Structural Changes

We first demonstrated the performance of the ACID network with small structural changes. The Ell-50 network was employed, as a special FBPConvNet [27]. Fig. 2 shows representative results in two CT cases: C1 and C2 (the details in **SI I**). To examine the degrees of small image structure recovery allowed by all reconstruction methods, some text, the contour of a bird, and their mixture were used to simulate structural changes in CT images. It is observed in Fig. 2a-h that the proposed ACID network provided a superior performance owing to the synergistic fusion of deep learning, CS-based sparsification, and iterative refinement. In this case, Ell-50 served as the deep network in the ACID workflow.

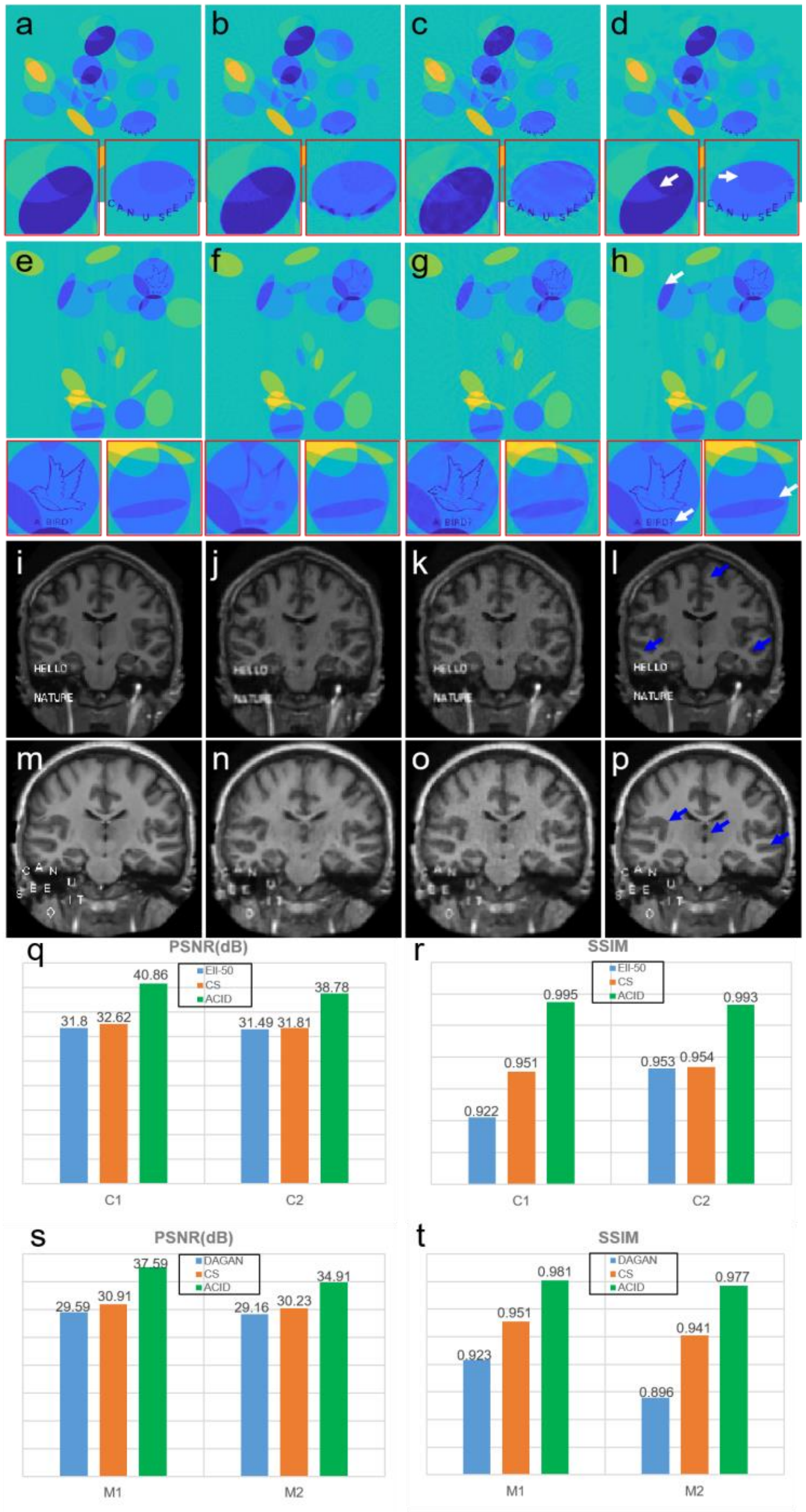

**Figure 2. Performance of ACID with small structural changes in the CT and MRI cases respectively.** Four phantoms with structural changes are reconstructed by ACID and competing techniques. **a**, **e**, **i** and **m** (first column) are the original images. **b** and **f**, **j** and **n** are the results with Ell-50 and DAGAN for CT and MRI respectively. **c**, **g**, **k** and **o** are the corresponding CS-inspired reconstruction results. **d**, **h**, **l** and **p** are the ACID counterparts. Each CT dataset contains 50 projections. The sub-sampling rate of MRI is 10%. The quantitative results are in **q**–**r** and **s-t** for CT and MRI respectively. The windows for C1 and C2 are [-150 150]HU and [-200 200]HU respectively. M1 and M2 results are normalized to [0 1].

It can be seen in Fig. 2 a-h that CS-inspired reconstructions produced better results than the Ell-50 network. This is consistent with the results reported in [9]. The CS-based reconstruction approach kept the structural changes. The text and bird were still identifiable in the CS-inspired reconstruction but became unclear in the Ell-50 results. In contrast, the text “CAN U SEE IT” and bird were well recovered using our ACID network. While the contour of the bird was compromised in the CS reconstruction, ACID produced better image quality than the CS method [28]. In terms of edge preservation, the Ell-50 reconstruction gave better sharpness overall than the corresponding CS reconstruction. Furthermore, ACID corrected the structural distortions seen in the Ell-50 and CS results.

A similar study was performed on MRI with small structural changes, as shown in Fig. 2 i-p. Since DAGAN [29] was used as a representative network in [9], we implemented it for this experiment. The text was added to brain MRI slices (M1 and M2 cases; more details in **SI I**). Fig. 2 i-l show the M1 results reconstructed from data sub-sampled at a rate of 10%. It is difficult to recognize the phrase “HELLO NATURE” in the DAGAN reconstruction. The structures were recovered by the CS method but with evident artifacts due to the low sub-sampling rate. In addition, the edges of “HELLO NATURE” were severely blurred, so was the text. On the other hand, our ACID network produced excellent results with the clearly visible words. To further show the power of ACID with small structural changes, another example (M2) in [9] was reproduced as Fig. 2 m-p. The text “CAN U SEE IT” was corrupted by both DAGAN and CS, rendering the insert hard to be read. Again, the text is easily seen in the ACID reconstruction. Indeed, compared with the DAGAN and CS results, the ACID reconstruction kept sharp edges and subtle features. The reconstructed results of M2 (similar to the DAGAN results in [9] but with different sub-sampling rate and pattern) also support the superior performance of ACID.

In brief, ACID exhibited superior stability with structural changes over the competitors, as quantified by the peak image signal-to-noise ratio (PSNR) and structural similarity (SSIM) (Fig. 2 q-t). In all these cases, ACID consistently obtained the highest PSNR and SSIM scores.

**(2) Stability against Adversarial Attacks**

A tiny perturbation could fool a deep neural network to make a highly undesirable prediction [9], which is known as an adversarial attack [30, 31]. To show the capability of the ACID approach against adversarial perturbations, the CT (cases C3 and C4) and MRI (cases M3 and M4) reconstructions under such perturbations are given in Fig. 3. Fig. 3 a-d show that Ell-50 network led to distorted edges as indicated by the arrows. Although the CS reconstruction had a stable performance against tiny perturbations, these distortions could not be fully corrected, with remaining sub-sampling artefacts. In contrast, this defect was well corrected by ACID. It is observed in Fig. 3 f-i that the artifacts marked by the arrows induced by perturbation distorted the image edges in the Ell-50 reconstruction. This could potentially result in a clinical misinterpretation. Although these artefacts were effectively eliminated in the CS reconstruction, CS-related new artifacts were introduced. Encouragingly, the corresponding edges and shapes were faithfully reproduced by ACID without significant artefacts. In addition, the text “CAN YOU SEE IT” was lost in the Ell-50 reconstruction. In contrast, our ACID results preserved the edges and letters. The worst MRI reconstruction results from tiny perturbations were obtained by DAGAN, as shown in Fig. 3 l and q. Compared with DAGAN, the CS-based reconstruction provided higher accuracy but still failed to preserve critical details such as edges, as shown in Fig. 3 m and r. On the other hand, our ACID network overcame these weaknesses. Fig. 3 u-v and w-x show the quantitative evaluation results, where ACID achieved the best performance.

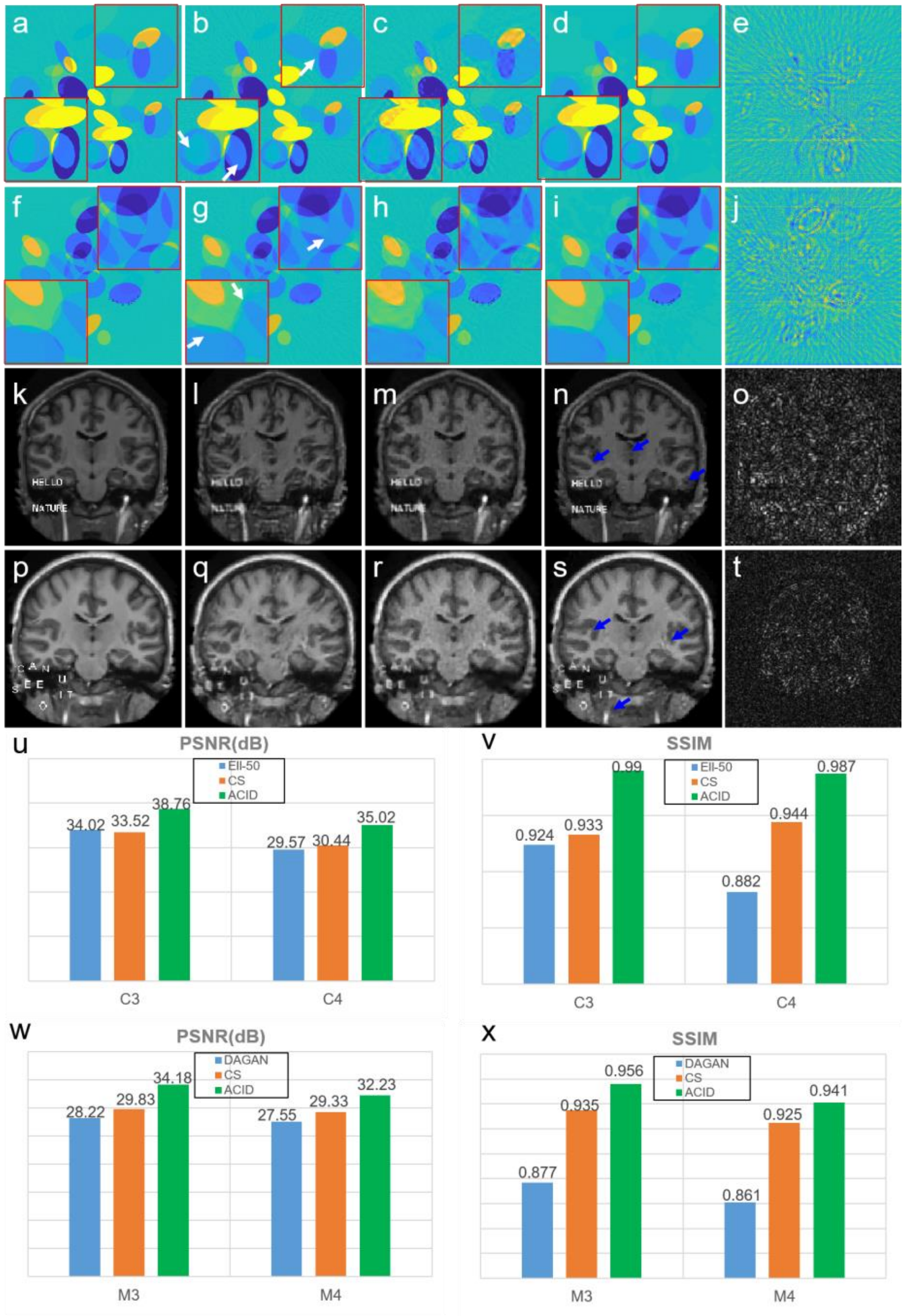

**Figure 3. Performance of ACID against adversarial attacks coupled with structural changes in the CT and MRI cases.** **a**, **f**, **k**, and **p** are the perturbed original images. **a**, **b**, **c** and **d** are the ground truth image, Ell-50, CS and ACID reconstructions in the C3 case (window [-80 80]HU). **f**, **g**, **h** and **i** are the counterparts in the C4 case (window [-150 150]HU). **k**, **l**, **m** and **n** are the ground truth image, DAGAN, CS and ACID reconstructions in the M3 case (normalized to [0, 1]. **p**, **q**, **r** and **s** are the counterparts in the M4 case (also normalized to [0 1]). **e**, **j**, **o** and **t** show the adversarial samples. The quantitative CT and MRI results are in **u**–**v** and **w-x** respectively. The display window for **e** and **j** is [-5 5]HU, and that for **o** and **t** is [-0.05 0.05].

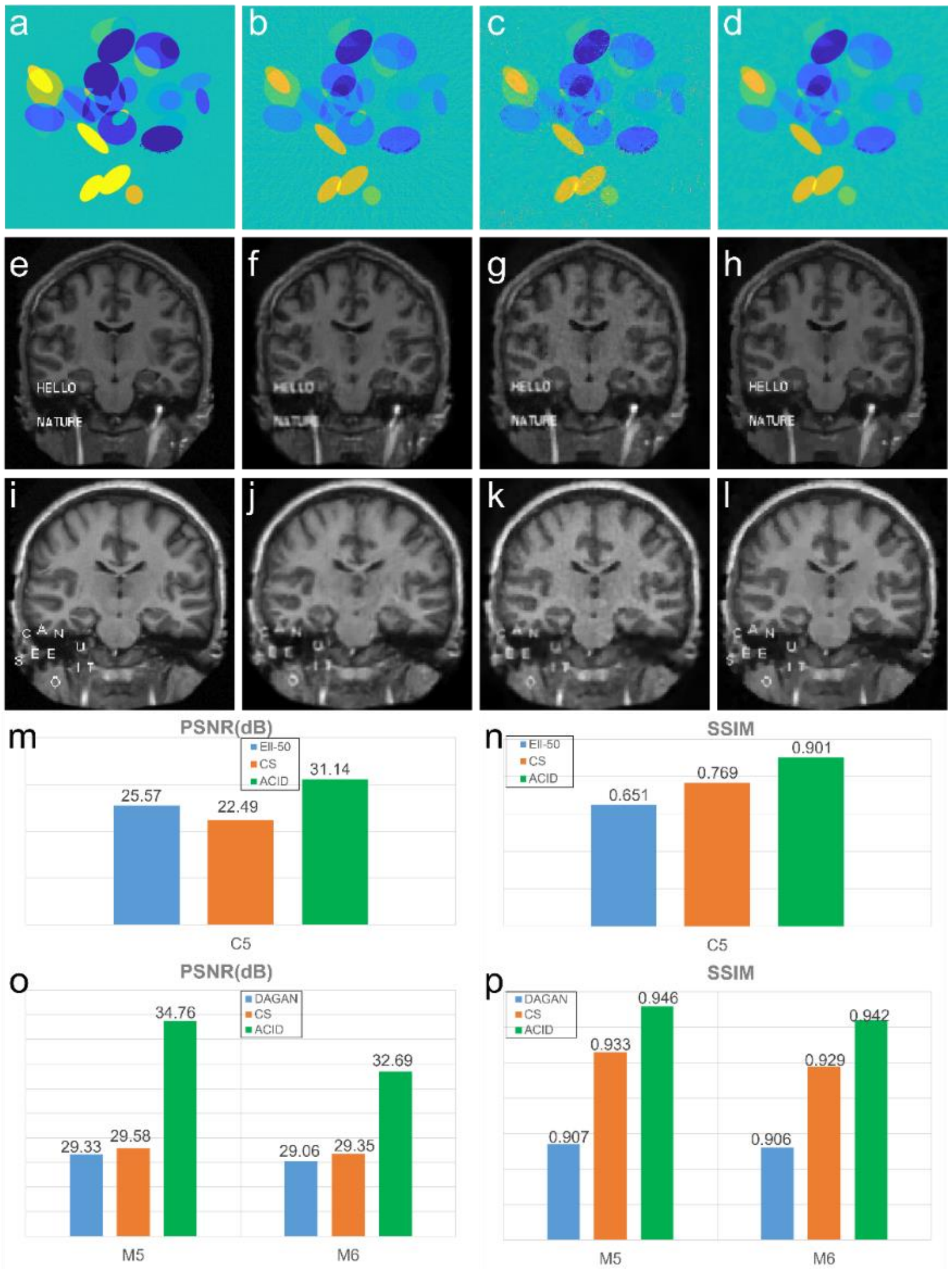

**Figure 4. Reconstruction results in the C5, M5 and M6 cases. a-d** present the ground truth, Ell-50, CS and ACID results on C5. **e-h** and **i-l** show the ground truth, DAGAN, CS and ACID results on M5 and M6 respectively. **m** and **n** list the PSNR and SSIM values on C5. **o** and **p** give the PSNR and SSIM results on M5 and M6 respectively.

To demonstrate the ACID performance in a practical setting, more experiments were performed in these CT and MRI cases with noisy data. The CT reconstruction results were obtained in the case C5, generated by adding Gaussian noise to the C1 data. Also, the reconstruction results were obtained in the experiments on M5 and M6, generated by adding Gaussian noise to the M1 and M2 datasets respectively. With the original networks (including Ell-50 and DAGAN) and CS methods, the image edges and other features were notably blurred. On the other hand, all the features including the embedded words were well recovered by ACID as shown in Fig. 4. It is observed that ACID gave better quantitative results than the competitors. Specifically, ACID suppressed image noise more effectively than the CS-based reconstruction method, even though the network was not trained for denoising. **SI III.D** reports detailed statistical results of ACID against noise using the strategy described in [32].

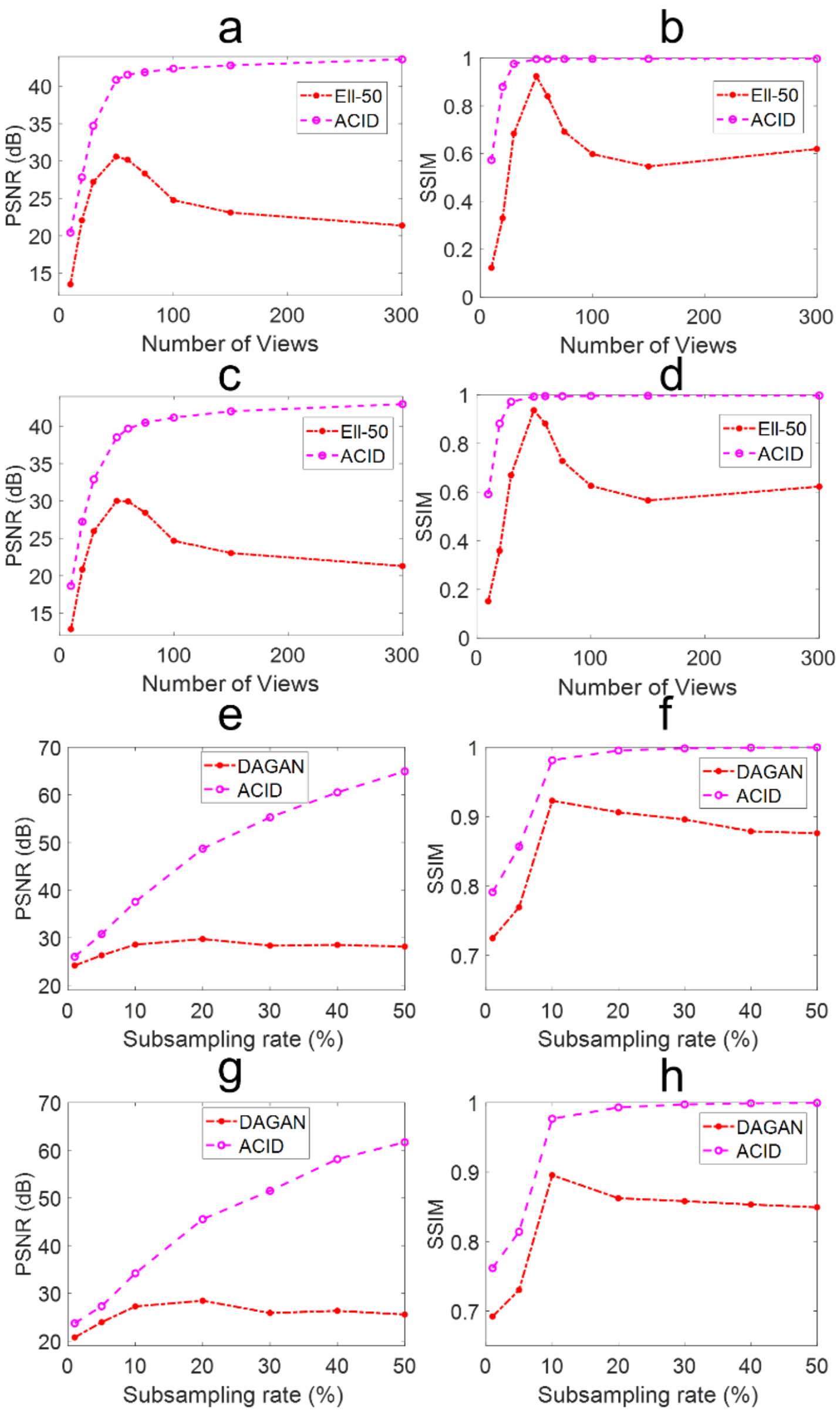

**Figure 5. Performance of ACID with more input data. a-b** and **c-d** contain the PSNR and SSIM curves with respect to the number of views in cases C1 and C2 respectively. **e-f** and **g-h** are the same type of curves with respect to different sub-sampling rates in cases M1 and M2 respectively.

## (3) Stability with More Input Data

Intuitively, a well-designed reconstruction scheme is expected to increase its performance monotonically as more input data become available. It was pointed out in [9] that the performance of some deep reconstruction networks, such as Ell-50 and DAGAN, degraded with more input data, which is certainly undesirable. To evaluate the performance of ACID with more input data, cases C1, C2, M1 and M2 were analyzed. The numbers of views in the CT cases were set to 10, 20, 30, 50, 60, 75, 100, 150 and 300, and in the MRI cases the sub-sampling rates were set to 1%, 5%, 10%, 20%, 30%, 40% and 50%, respectively. Fig.

5 shows that the performance of Ell-50 decreased with more projections than what were used for network training, being consistent with the observation in [9]. In contrast, ACID performed better with more views in terms of PSNR and SSIM. Similarly, the performance of DAGAN decreased when more data were collected at sampling rates higher than that used for DAGAN training, which agrees with the conclusion on DAGAN in [9]. Yet, ACID produced better reconstruction quality in terms of PSNR and SSIM. The performance of ACID substantially improved with more input data, indicating that our ACID generalizes well to more input data, similar to the CS methods.

**(4) Ablation Study on ACID**

ACID involves deep reconstruction, CS-inspired sparsification, analytic mapping, and iterative refinement. To understand the roles of these algorithmic ingredients, we evaluated their relative contributions to reconstruction quality. Specifically, we reconstructed images using the three simplified versions of ACID by removing/replacing individual key components. The three versions include (a) improving the initial deep reconstruction with CS-inspired sparsification without iteration, (b) replacing deep reconstruction with a conventional reconstruction method, and (c) abandoning the compressed sensing constraint. Fig. 6 shows that each of the simplified ACID variants compromised the ACID performance significantly.

**(5) Comparison with Classic Iteration-based Unrolled Networks**

The use of deep learning as a post-processor or an image-domain data-driven regularizer in a classic iterative reconstruction algorithm, as we suggested in [4], is inferior to ACID. It is mainly because the classic iterative reconstruction cannot take full advantage of data-driven prior, even if a deep learning image denoiser is used, such as in ADMM-net [33]. Different from existing iteration-based unrolled reconstruction networks that only use deep learning to refine an intermediate image already reconstructed using a classic iterative algorithm, ACID reconstructs an intermediate image with a deep network trained on big data and through iterative refinement. To highlight the merits of ACID, the classic ADMM-net was chosen for comparison [33]. The ADMM-net was trained on 20% sub-sampled data, with a radial sampling mask while the other settings being the same as that in [33]. Fig. 7 a-c show that ACID achieved the best reconstructed image quality, followed by ADMM-net and DAGAN sequentially. The phrase “HELLO NATURE” was blurred in the DAGAN reconstruction but became clearer in the ADMM-net reconstruction. However, the artifacts due to sub-sampling remain evident in the ADMM-net reconstruction. In Fig. 7 d-f, there are strong artifacts in the reconstructed image by DAGAN. On the other hand, the image quality from ADMM-net is better than that of DAGAN, such as in terms of edge sharpness. The image edges and features in the ACID images are overall the best, as shown in Fig. 7. To quantify the performance of these techniques, the PSNR and SSIM measures were computed, as shown in Fig. 7 g-h.

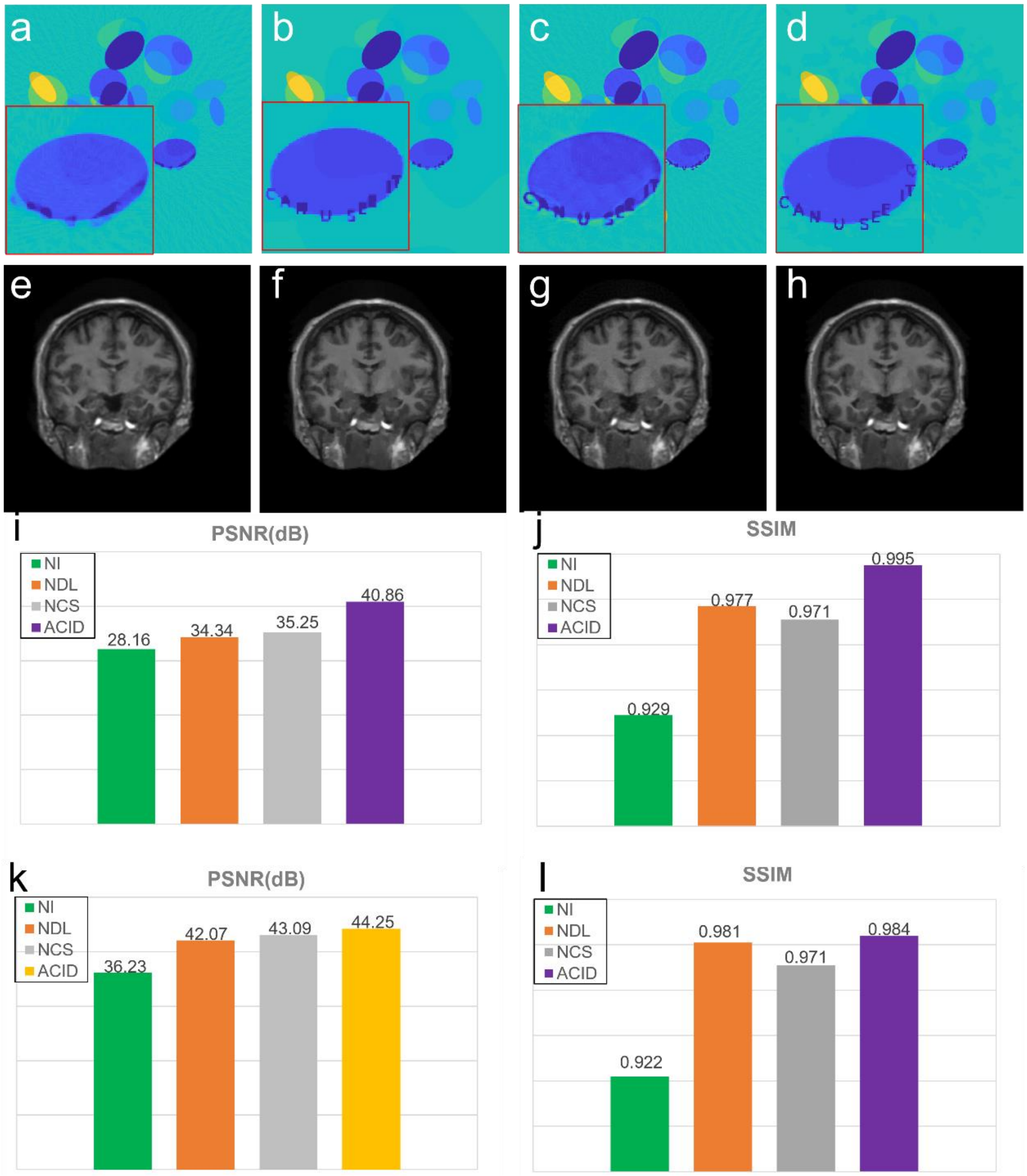

**Figure 6. ACID ablation study in terms of visual inspection and quantitative metrics in the cases C1 and M7.** NI denotes the reconstructed results by ACID without iterations (K=1). NDL and NCS denote ACID without deeply learned prior and CS-based sparsification, respectively. **a-d** represent the reconstructed results by NI, NDL, NCS and ACID in the C1 case. **e-h** are the reconstructed results by NI, NDL, NCS and ACID in the M7 case. **i-l** are the PSNR and SSIM results in these cases, respectively.

The ACID flowchart can be unfolded into the feedforward architecture. However, such an unfolded reconstruction network (similar to MRI-VN [34]) could be still subject to adversarial attacks, if kernel awareness is not somehow incorporated. Given the current GPU memory limit, it is often impractical to unfold the whole ACID workflow with a large-scale reconstruction network built-in up to 100 iterations or more.

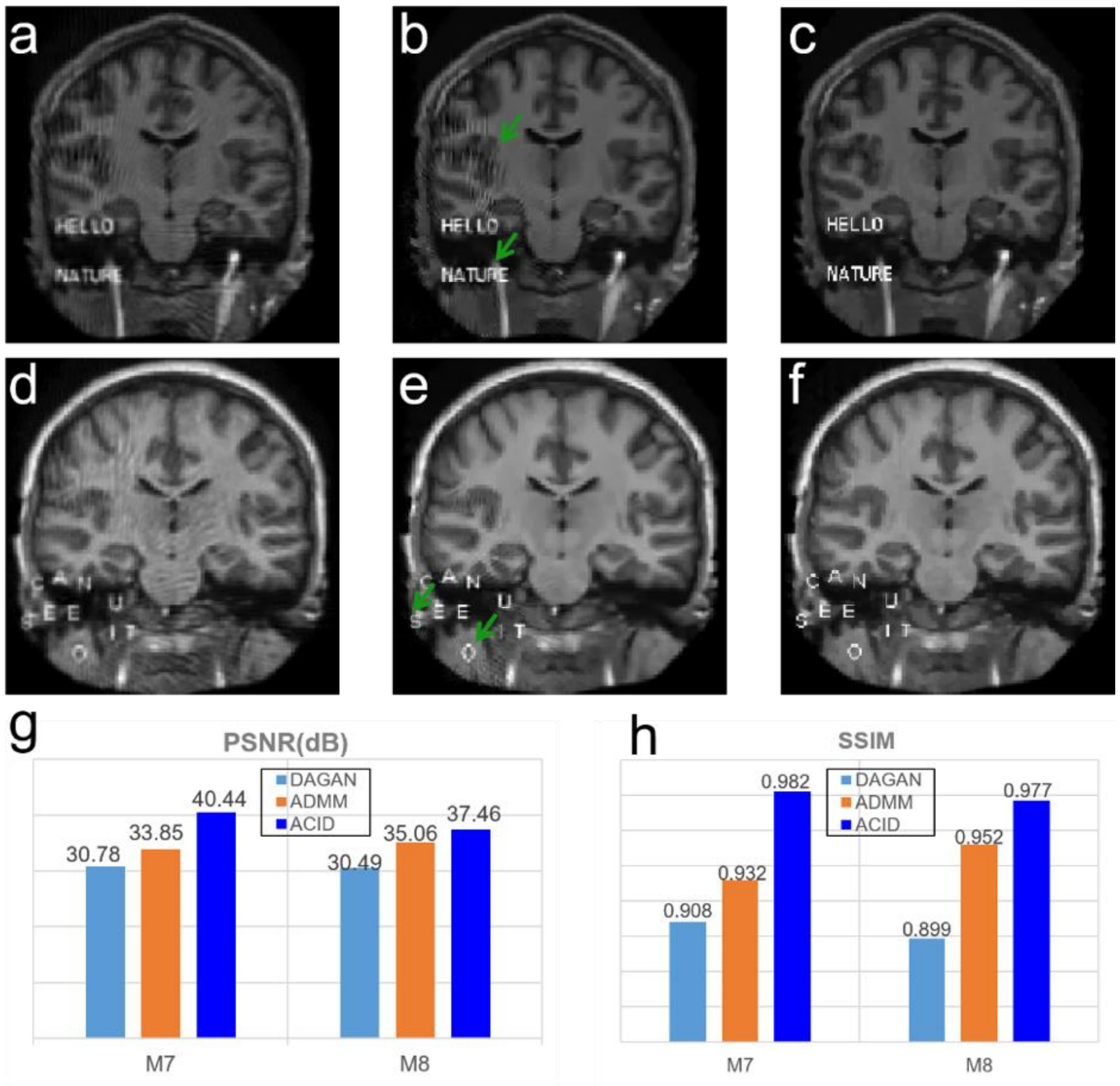

**Figure 7. Comparison of reconstruction performance relative to the ADMM-net. a**-**c** and **d-f** represent the results reconstructed by DAGAN, ADMM-net and ACID, respectively. **g-h** show the quantitative results in the M8 and M9 cases in terms of PSNR and SSIM respectively.

There are at least three differences between ACID and the unrolled reconstruction networks. First, large-scale trained networks, such as DAGAN [29] and Ell-50 [27], could be incorporated into the ACID framework as we have demonstrated in this study. However, if the ACID scheme is unrolled into a feedforward network, only small sub-networks could be integrated; *i.e*., an unrolled ACID network could only use relatively light networks such as multiple layer-CNNs [35, 36]. These sentences are not contradictory, because the ACID scheme is not an unrolled network. In fact, an unrolled network has a number of stages, each of which consumes a substantial amount of memory. Hence, the total size of the required memory is proportional to the number of stages. In contrast, ACID is computationally iterative, and thus the same memory space allocated for an iteration is reused for the next iteration. Therefore, a large-scale network can work with ACID but, when the ACID scheme is unrolled, only a light-weight network can be used for ACID reconstruction. Second, the size of images reconstructed by an unrolled networks is typically small. For example, the input image size is 256x256 pixels for ADMM-net [33], LEARN [35], and AirNet [37], limited by the memory size of GPU. The reconstructed low-resolution results could not satisfy the requirement of many clinical applications, especially for CT imaging tasks. Also, the unrolled networks were commonly designed for 2D imaging, difficult to be employed in 3D imaging geometry, given the memory increment proportional to the number of stages. Third, it has not been intended by others to incorporate the theoretically grounded sparsity regularization module in such an unrolled architecture. This could be due to the fact that some needed operations (for example, the image gradient $L_0$-norm [38]) could not be effectively implemented with compact feedforward networks which demanded big data and could not be easily trained. Nevertheless, ACID can stabilize these unrolled networks. For example, Fig. 8 demonstrates the results using ACID with a built-in model-based unrolled deep network (MoDL) [36]. MoDL performed well with structural changes but suffered from adversarial attacks [9]. Synergistically, ACID with MoDL built-in produced excellent image quality.

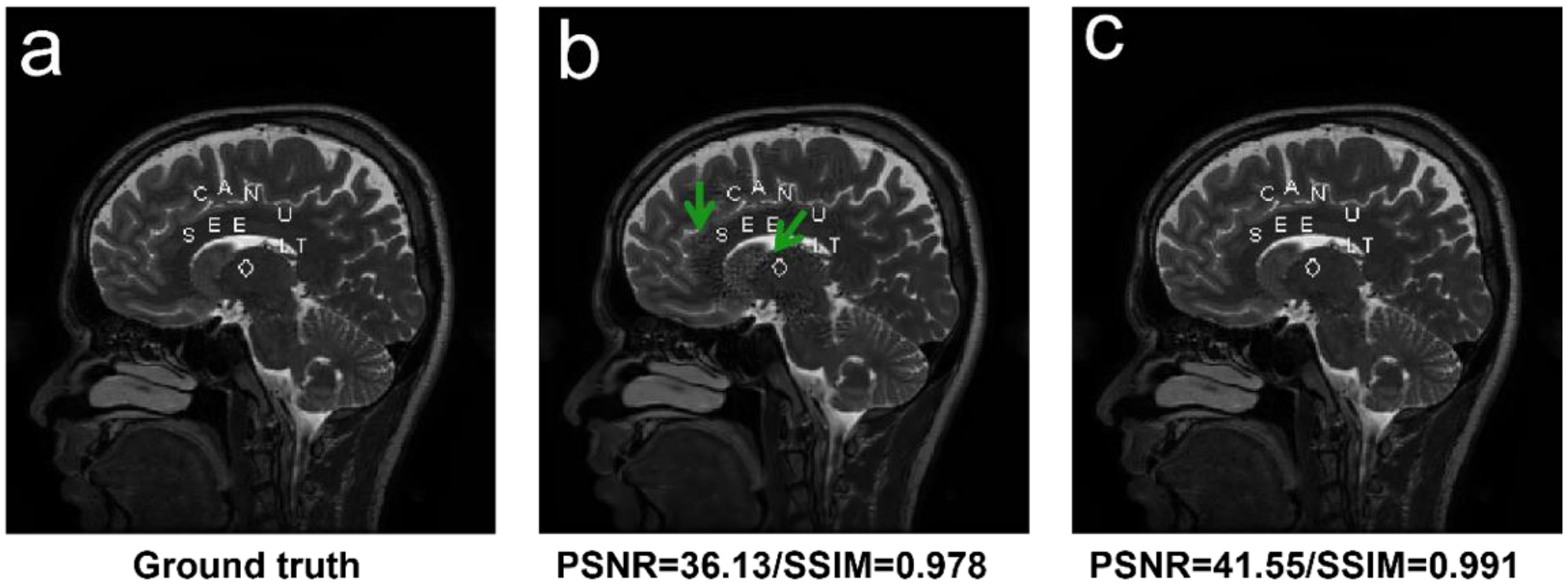

**Figure 8. Stabilization of MoDL using the ACID strategy. a-c** represent a representative reference, corresponding results reconstructed by MoDL and ACID (with MoDL built-in) respectively, where adversarial attacks were applied to the MoDL network, which was then successfully defended using the ACID scheme with MoDL built-in. The quantitative indices were computed in terms of PSNR and SSIM.

### (6) Adversarial Attacks to the ACID System

As demonstrated above, ACID can successfully stabilize an unstable network. Then, a natural question is if the whole ACID workflow itself is stable or not. To evaluate the stability of ACID in its entirety, we generated adversarial samples to attack the entire ACID system, with representative results in Fig. 9. Because ACID involves both deep reconstruction and sparsity minimization in the iterative framework, the adversarial attack mechanism is more complicated for ACID than that for a feedforward neural network. Please see **SI III** for details on the adversarial attacking method that we used to attack ACID. Using this adversarial method, C6, C7, M10-M12 images were perturbed to various degrees, being even greater in terms of the $L_2$-norm than what were used to attack individual deep reconstruction networks. Our ACID reconstruction results show that the structural features and inserted words were still clearly reproduced even after these adversarial attacks. Consistently, the PSNR and SSIM results of ACID were not significantly compromised by the adversarial attacks.

### (7) Stabilization of AUTOMAP

AUTOMAP, an important milestone in medical imaging, was used as another classic example in [9] to demonstrate the instabilities of deep tomographic reconstruction. To further test the stability of ACID, cases A1 and A2 with structural changes and A3 and A4 cases with adversarial attacks were employed, as shown in Fig. 10 (details on cases A1-A4 are in **SI I**). It is observed that AUTOMAP showed a good ability against structural changes but it suffered from adversarial attacks [9]. ACID produced significantly better image quality than AUTOMAP. Beyond the visual inspection, ACID achieved better PSNR and SSIM values than AUTOMAP. Please see our **SI** I for more details.

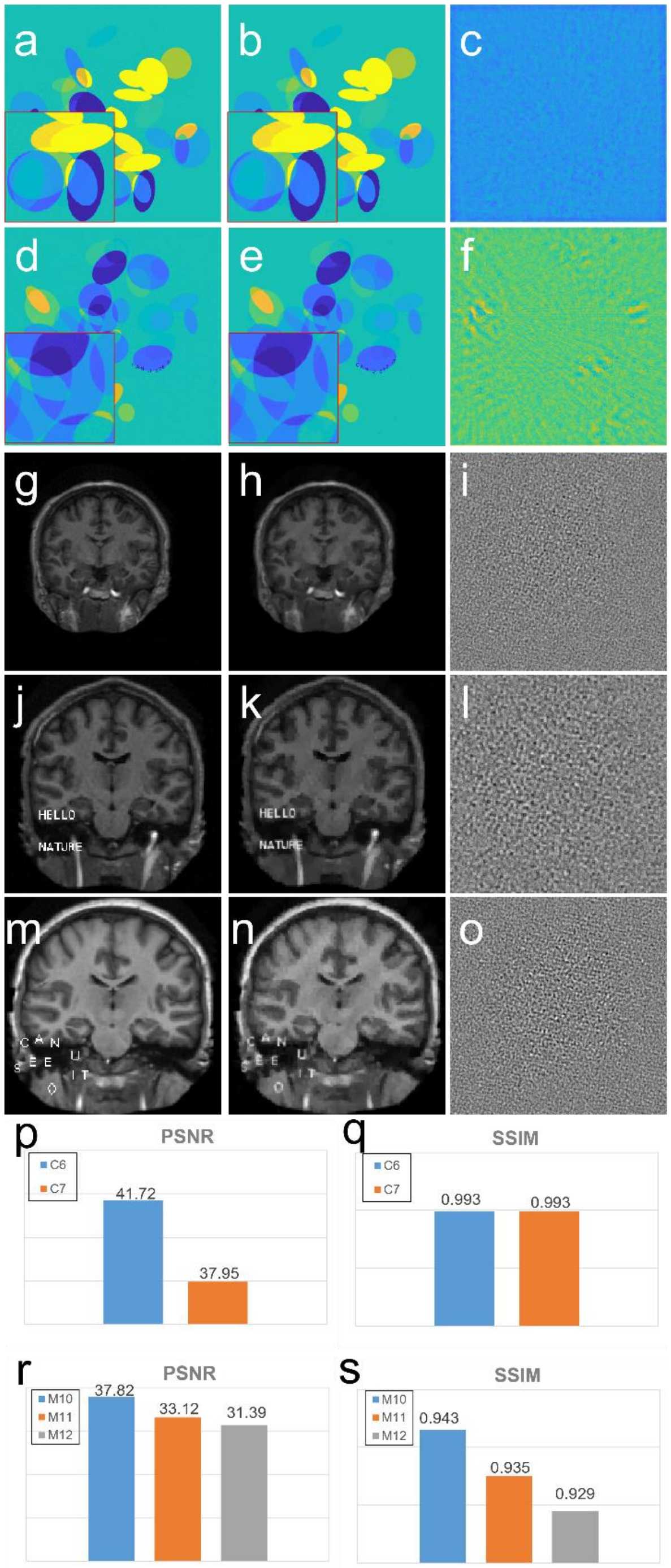

**Figure 9. ACID being resilient against adversarial attacks. a-c**, **d-f**, **g-i**, **j-l** and **m-o** are ACID results in C6, C7, and M10-12 cases respectively. The 1st -3rd columns represent the ground truth plus tiny perturbation, reconstructed images, and corresponding perturbations. **p-q** and **r-s** show the PSNR and SSIM results in the C6, C7, and M10-12 cases respectively.

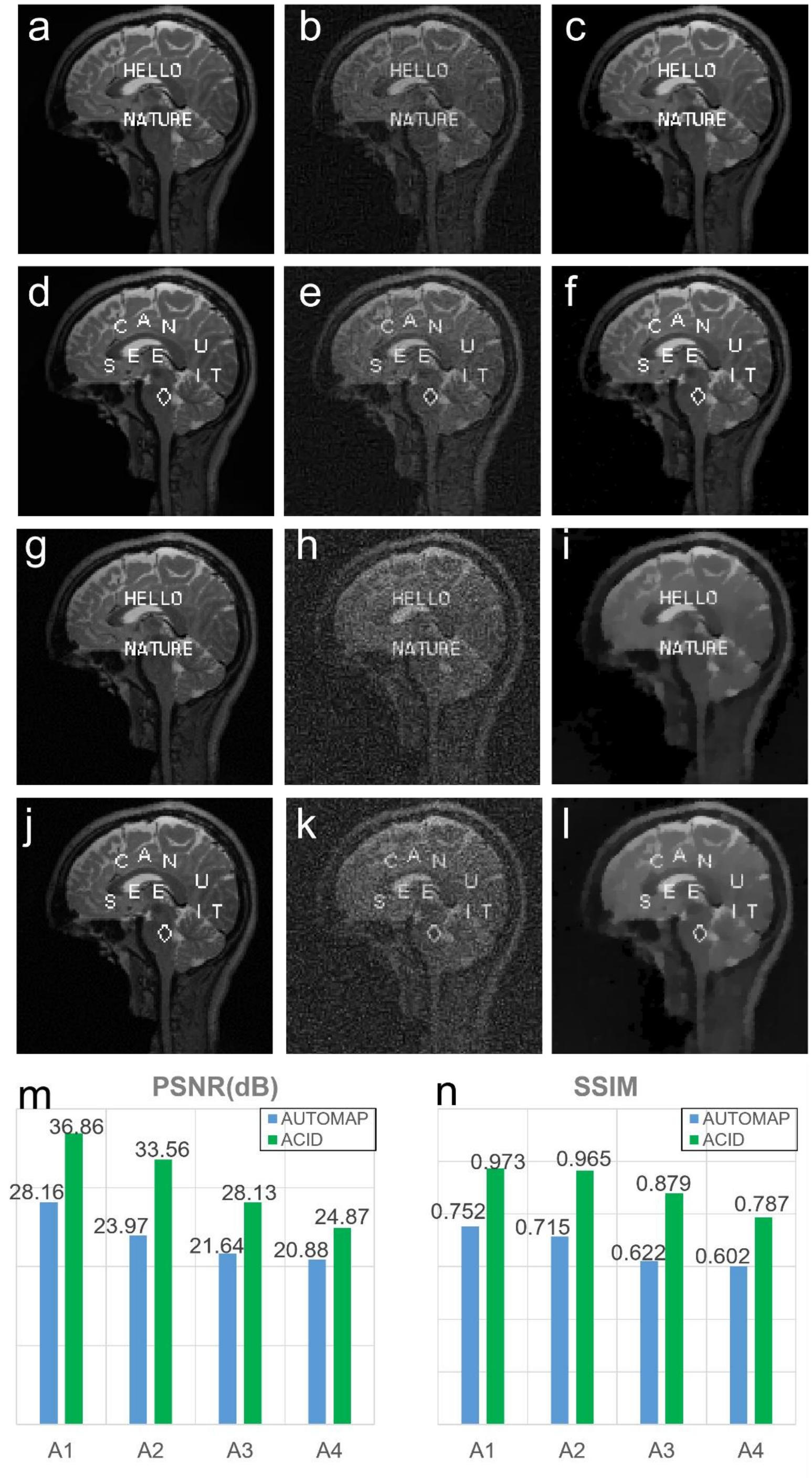

**Figure 10. Stabilization of AUTOMAP using ACID. a-c** and **d-f** represent the reconstruction results from structural changes, where **a** and **d**, **b** and **e**, **c** and **f** represent the reference, AUTOMAP and ACID (with AUTOMAP built-in) results respectively. **g-i** and **j-k** are the reconstruction results under adversarial attacks, where **g** and **j**, **h** and **k**, **i** and **l** denote the reference plus perturbation, AUTOMAP and ACID (with AUTOMAP built-in) results respectively. The PSNR and SSIM values are in **m** and **n**.

## Discussions and Conclusion

As clearly reviewed in **SI IV.A**, the "*kernel awareness*" [18] is important to avoid the so-called "*cardinal sin*". When input vectors are very close to the null space of the associated measurement matrix, if the input is slightly perturbed a large variation may be induced in the reconstructed image. If an algorithm lacks the kernel awareness, it will be intrinsically vulnerable, suffering from false-positive and false-negative errors; for mathematical rigor, please see Theorem 3.1 in [18]. For this reason, the deep tomographic networks were successfully attacked in [9]. On the other hand, sparsity-promoting algorithms were designed with the kernel awareness, leading to an accurate and stable recovery of underlying images, as also shown in [9]. As demonstrated by our results here, the kernel awareness has been embedded in the ACID scheme through both the CS-based sparsity constraint and the iterative refinement mechanism. Hence, ACID has a robust performance against noise, under adversarial attacks, and when the amount of input data is increased relative to what was used for network training.

It is important to understand how a CS-based image recovery algorithm implements the kernel awareness. The sparsity-constrained solution is iteratively obtained so that the search for the solution is within a low-dimensional manifold. That is, prior knowledge known as sparsity helps narrow down the solution space. Indeed, natural and medical images allow low-dimensional manifold representations [39]. It is critical to emphasize that a deep neural network is data-driven, and the resultant data-driven prior is rather powerful to constrain the solution space greatly for tomographic imaging. While sparsity prior is just one or a few mathematical expressions, deep prior is in a deep network topology with a large amount of parameters extracted from big data. These two kinds of priors are combined in our ACID workflow to have the merits from both sides. Because the combination of deep prior and sparsity prior is more informative than sparsity prior alone, ACID or similar networks would outperform classic algorithms including CS-inspired sparsity-promoting methods. Indeed, with a deep reconstruction capability, ACID outperforms the representative CS-based methods for image reconstruction, including dictionary learning reconstruction methods (details in **SI II.B**).

Unlike the establishment of the instabilities, it is mathematically insufficient to prove the general applicability of ACID using only a finite number of positive experimental results. Hence, a theoretical analysis is desirable on the convergence of the ACID iteration. Although a thorough characterization is rather challenging (since the field of non-convex optimization is still in its infancy), we have assumed an experimentally-motivated BREN property of the reconstruction network, which is a special case of the Lipschitz continuity that is widely used for non-convex optimization to establish various converging properties [24, 25]. The BREN property means that the relative error of a deep network-based reconstruction is under control in a $L_2$ norm. Based on BREN, we have made an initial effort to understand the converging mechanism of the ACID iteration in **SI IV**. Specifically, in **SI IV** we have provided not only (1) a heuristic analysis based on the simplification that the CS module allows a perfect sparsification but also (2) a mathematically denser analysis of the convergence under two approximations (the first approximation to invert an under-determined system matrix $\boldsymbol{A}$, and the other is to minimize

TV with a non-unitary transform $\boldsymbol{H}$. Why both the approximations are reasonable have been explained from the perspective of iterative reconstruction; see **SI IV** for details).

In conclusion, our proposed ACID workflow has synergized deep network-based reconstruction, CS-inspired sparsity regularization, analytic forward mapping, and iterative data residual correction to systematically overcome the instabilities of the deep reconstruction networks selected in [9] and achieved better results than the CS algorithms used in [9]. Furthermore, the converging behavior and solution characteristics of ACID have been described in **SI IV**. It is underlined that the ACID scheme is only an exemplary embodiment, and other hybrid reconstruction schemes of this type can be also investigated in a similar spirit [40, 41]. We anticipate that this integrative data-driven approach will help promote development and translation of deep tomographic image reconstruction networks into clinical applications.

# Methods

### (1) Heuristic ACID Scheme

In the imaging field, we often assume that the measurement $\boldsymbol{p}^{(0)} = \boldsymbol{A}\boldsymbol{f}^* + \boldsymbol{e}$, where $\boldsymbol{A} \in \mathbb{R}^{m\times N}$ is a measurement matrix (for example, $\boldsymbol{A}$ is the Radon transform for CT [42] and the Fourier transform for MRI [43]), $\boldsymbol{p}^{(0)} \in \mathbb{R}^m$ is an original dataset, $\boldsymbol{f}^* \in \mathbb{R}^m$ is the ground truth image, $\boldsymbol{e} \in \mathbb{R}^m$ is data noise, and most relevantly $m < N$, meaning that the inverse problem is underdetermined. In the under-deterministic case, additional prior knowledge must be introduced to recover the original image uniquely and stably. Typically, we assume that $\boldsymbol{H} \in \mathbb{R}^{N\times N}$ is an invertible transform, $\boldsymbol{A}$ satisfies the restricted isometry property (RIP) of order $s$ [44] (note that ACID works even without RIP but in that case the solution may or may not be unique; see **SI IV**), and $\boldsymbol{H}\boldsymbol{f}^*$ is s-sparse. We further assume that the function $\Phi(\cdot)$ models a properly designed and well-trained neural network with the BREN property that continuously maps measurement data to an image. To solve the problem of reconstructing $\boldsymbol{f}$ from measurement $\boldsymbol{p}^{(0)}$, the ACID scheme is heuristically derived from the following iterative solution (see **SI IV.E**):

$$\begin{cases} \boldsymbol{p}^{(k+1)} = \frac{\lambda\left(\mathbf{p}^{(0)} - \mathbf{A}\boldsymbol{f}^{(\boldsymbol{k})}\right)}{1+\lambda} \\ \boldsymbol{f}^{(\boldsymbol{k+1})} = \boldsymbol{H}^{-\mathbf{1}} S_\varepsilon\left(\boldsymbol{H}\left(\boldsymbol{f}^{(\boldsymbol{k})} + \frac{1}{\lambda}\Phi\left(\boldsymbol{p}^{(\boldsymbol{k+1})}\right)\right)\right) \end{cases} \tag{1}$$

where $k$ is the index for iteration, $k = 0,1,2,\cdots$, $\lambda > 0$ and $\varepsilon > 0$ are hyperparameters, $\boldsymbol{H}^{-\mathbf{1}}$ is the inverse transform of $\boldsymbol{H}$, and $S_\varepsilon(\cdot)$ is the soft-thresholding filtering kernel function defined as

$$S_\varepsilon(x) = \begin{cases} 0, & |x| < \varepsilon \\ x - sgn(x)\varepsilon & otherwise \end{cases}. \tag{2}$$

In our experiments, $\boldsymbol{H}\boldsymbol{f}$ is specialized as a discrete gradient transform, and $\boldsymbol{H}^{-\mathbf{1}}$ is interpreted as a pseudo-inverse [45] (see **SI IV.F**). Under the same conditions described in [45], the ACID iteration would converge to a feasible solution subject to an uncertainty range proportional to the noise level (under the conditions and approximations discussed in **SI IV.E** and **F**).

**(2) Selected Unstable Networks Stabilized in the ACID Framework**

The Ell-50 and DAGAN networks are two examples of unstable deep reconstruction networks chosen to validate the effectiveness of ACID, both of which were used in [9] and suffered from the three kinds of instabilities. In addition, the results from stabilizing AUTOMAP, a milestone deep tomographic network, were also included.

The projection data for Ell-50-based CT reconstruction were generated using the *radon* function in MATLAB R2017b, where 50 indicates the number of projections. For fair comparison, we only used the trained networks in [27], which were the same as those used in [9]. The test image for case C1 was provided by the authors of [9], which can be downloaded from the related website [9]. Case C2 with the bird icon and text "A BIRD?" was provided by the authors of [18] and downloaded from the specified website in [9]. The test images are of 512×512 pixels containing structural features without any perturbation. To generate adversarial attacks, the proposed method in [9] was employed. Then, we obtained C3 and C4 images by adding perturbations to C1 and C2 images respectively. Furthermore, Gaussian noise with zero mean and deviation 15 HU over the pixel value range was superimposed in case C1 to obtain case C5, and adversarial attacking was performed on the whole ACID workflow by perturbing C1 and C2 images to generate C6 and C7 images respectively. More details on the datasets and implementation details are in **SI I-III**.

To evaluate ACID in the MRI case, the DAGAN network was employed [29], which was proposed for single-coil MRI reconstruction. In this study, we set the sub-sampling rate to 10% and sub-sampled the resultant images with the 2D Gaussian sampling pattern. Also, the DAGAN network was re-trained, with the same hyperparameters and training datasets as those used in [29]. The test images were a series of brain images, each of which consists of 256×256 pixels. Case M1 was randomly chosen from the test dataset [29], then the phrase "HELLO NATURE" was placed in the image as structural changes. Case M2 was obtained in the same way as in [9], where the sentence "CAN U SEE IT" and "◊" were added to the original image. Also, we applied the same attacking technique used in [9] to generate adversarial samples. These perturbations were added to M1 and M2 images to obtain M3 and M4 images respectively [9]. Furthermore, the Gaussian noise with zero mean and deviation of 15 over the pixel value range [0, 255] was superimposed to cases M1 and M2 to obtain M5 and M6 images. M7 was randomly chosen from the DAGAN test dataset, which can be freely downloaded [29]. Additionally, cases M8 and M9 were generated by putting a radial mask of a 20% sub-sampling rate on the M1 and M2 images, which were used to compare ACID with ADMM-net [33]. Cases M10 and M12 were generated by directly perturbing the entire ACID system. The comparative results are given in **SI III**.

ACID was then compared with AUTOMAP for MRI reconstruction. The AUTOMAP network was tested on sub-sampled single-coil data. The trained AUTOMAP weights used in our experiments were provided by the authors of [10]. The AUTOMAP network took a vectorized sub-sampled measurement data as its input. First, the complex *k*-space data were computed using the discrete Fourier transform of an MRI image. Then, sub-sampled *k*-space data were generated with a sub-sampling mask. Lastly, the measurement data were reshaped into a vector and fed into the AUTOMAP network. In this study, the images of 128×128 pixels at a sub-sampling rate of 60% were used for testing. The original image was provided in [9], "HELLO NATURE", "CAN U SEE IT" and "◊" were added to the test image to generate A1 and A2 with the structural changes. The perturbations were added to images A1 and A2 to obtain images A3 and A4 [9]. For representative results, please see **SI III**.

### (3) Image Quality Assessment

To quantitatively compare the results obtained with different reconstruction methods, the peak signal-to-noise ratio (PSNR) was employed to measure the difference between a reconstructed image and the corresponding ground truth image. Also, the structural similarity (SSIM) was used to assess the similarity between images. For qualitative analysis, the reconstructed results were visually inspected for structural changes (*i.e.*, the inserted text, bird, and patterns) and artifacts induced by perturbation. In this context, we mainly focused on details such as edges and integrity such as overall appearance.

To highlight the merits and stability of the ACID scheme, the representative CS-based methods served as the baseline. For CT, the sparsity-regularized method combining X-lets (shearlets) and TV was used [46], which is consistent with the selection in [9]. For MRI, the total generalized variation (TGV) method was chosen [47]. All the parameters including the number of iterations for these CS methods were optimized for fair comparison, as further detailed in **SI I-III**.

### (4) Numerical Verification of Convergence

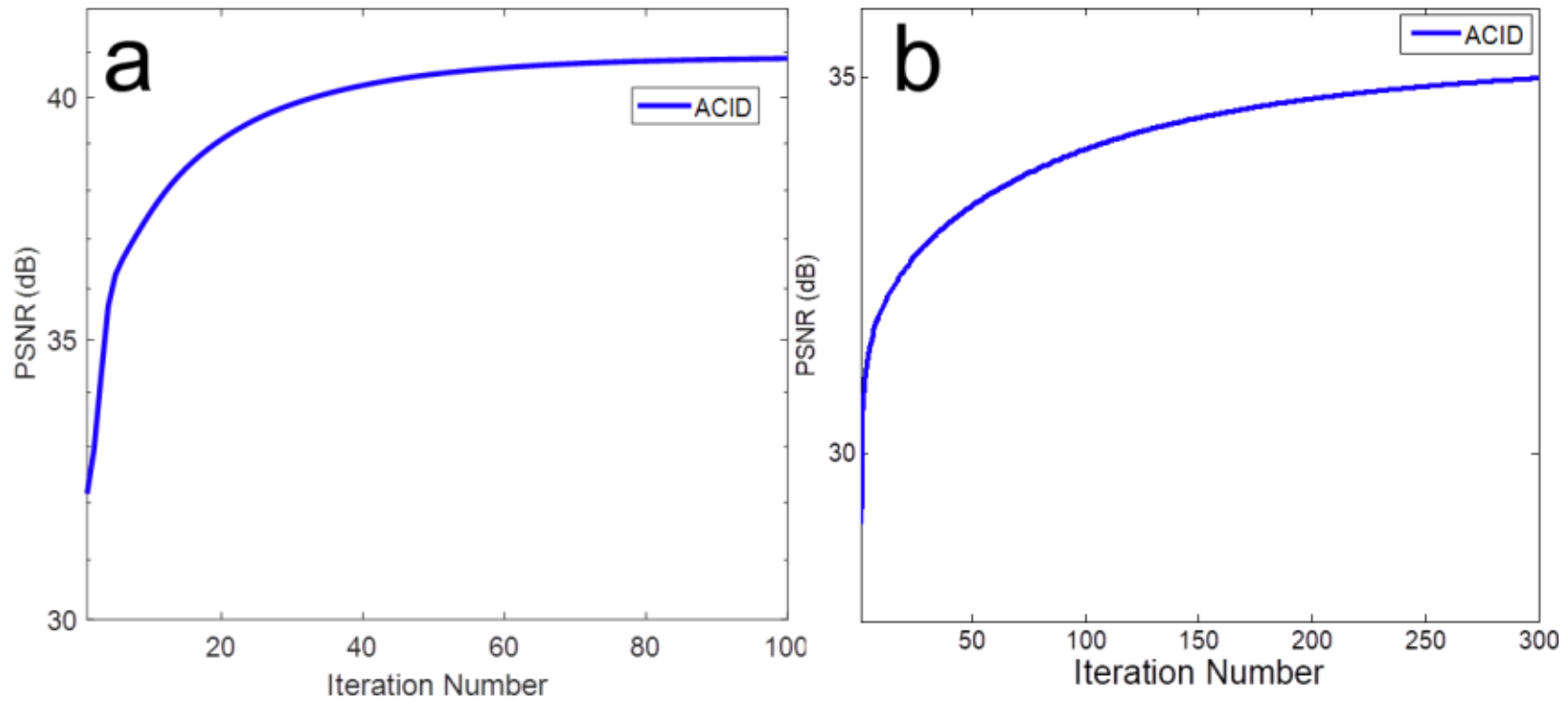

**Figure 11. Convergence of the ACID iteration in terms of PSNR**. **a** and **b** show the convergence curves in the C1 and M2 cases, respectively.

To verify the convergence of the ACID iteration, we numerically investigated the convergence rate and computational cost. We used PSNR as the metric to reflect the convergence of ACID (Fig. 11). It is seen that the ACID iteration converged after about 30 iterations for CT and became stable after 250 iterations for MRI. In this study, we set the number of iterations to 100 and 300 for CT and MRI, respectively. Additionally, we also empirically showed the convergence of ACID in terms of the Lipschitz constant; see **SI III** for details.

### (5) ACID Parameterization

The ACID method mainly involves two parameters: $\lambda$ and $\varepsilon$ as defined in **SI IV.E**. These parameters were optimized based on our quantitative and qualitative analyses, as summarized in Table 1.

**Table 1. Parameters optimized in the experiments.**

| Variables | C1 | C2 | C3 | C4 | C5 | C6 | C7 | M1 |
|---|---|---|---|---|---|---|---|---|
| $\varepsilon(10^{-3})$ | 0.700 | 0.700 | 0.500 | 1.100 | 0.35 | 0.100 | 0.700 | 0.333 |
| λ | 0.76 | 0.76 | 14 | 2.4 | 60 | 3.0 | 0.76 | 0.1 |
| Variables | M2 | M3 | M4 | M5 | M6 | M7 | M8 | M9 |
| $\varepsilon(10^{-3})$ | 0.333 | 0.500 | 0.500 | 0.667 | 0.667 | 0.333 | 0.500 | 0.500 |
| λ | 0.1 | 0.01 | 0.01 | 0.01 | 0.01 | 0.1 | 0.01 | 0.01 |
| Variables | M10 | M11 | M12 | A1 | A2 | A3 | A4 | |
| $\varepsilon(10^{-3})$ | 0.100 | 0.100 | 0.100 | 0.33 | 0.33 | 2.0 | 2.0 | |
| λ | 0.01 | 0.01 | 0.01 | 3.10 | 3.10 | 6.10 | 6.10 | |

**Acknowledgement:** W. Wu was partially supported by the Li Ka Shing Medical Foundation. V. Vardhanabhuti is partially supported by the Li Ka Shing Medical Foundation. W. Wu, H. Shan, W. Cong, C. Niu, H. Yu and G. Wang are partially supported by NIH grants U01EB017140, R01EB026646, R01CA233888, R01CA237267, and R01HL151561.

**Corresponding Authors:** Hengyong Yu, Varut Vardhanabhuti, and Ge Wang.

**Author Contributions:**
G. Wang initiated the project, and supervised the team in collaboration with H. Yu and V. Vardhanabhuti;
W. Wu, H. Yu, G. Wang designed the ACID network;
W. Wu and D. Hu conducted the experiments;
H. Yu, W. Cong, G. Wang established the mathematical model and performed the theoretical analysis;
W. Wu, H. Yu and G. Wang drafted the manuscript;
W. Wu, D. Hu and H. Shan worked on user-friendly codes/data sharing; and all co-authors participated in discussions, contributed technical points, and revised the manuscript iteratively.

**Data and code availability:** The codes, trained networks, test datasets and reconstruction results are publicly available on Zenode (https://zenodo.org/record/5497811).

# Supplementary Information for

# "Stabilizing Deep Tomographic Reconstruction"

Weiwen Wu[1,5], Dianlin Hu[2], Wenxiang Cong[1], Hongming Shan[1,3], Shaoyu Wang[4], Chuang Niu[1], Pingkun Yan[1], Hengyong Yu[4,*], Varut Vardhanabhuti[5,*], Ge Wang[1,*]

[1]Biomedical Imaging Center, Center for Biotechnology and Interdisciplinary Studies, Department of Biomedical Engineering, Rensselaer Polytechnic Institute, Troy, NY, USA

[2]The Laboratory of Image Science and Technology, Southeast University, Nanjing, China

[3]Institute of Science and Technology for Brain-inspired Intelligence, Fudan University, Shanghai, China

[4]Department of Electrical & Computer Engineering, University of Massachusetts Lowell, Lowell, MA, USA

[5]Department of Diagnostic Radiology, Li Ka Shing Faculty of Medicine, The University of Hong Kong, Hong Kong SAR, China

**Corresponding Authors:** Hengyong_yu@uml.edu, varv@hku.hk, and wangg6@rpi.edu

# Table of Contents

# Summary

This supplemental information (**SI**) consists of four sections **I-IV**. The purpose of **SI** is to provide sufficient details for reproducible research and initial theoretical insights into the stability and convergence of the ACID scheme.

In **SI I**, we describe the selected reconstruction networks including Ell-50, DAGAN, AUTOMAP, and ADMM-net, along with the associated datasets. For each of these networks, we focus on the architecture, training and testing details.

In **SI II**, we review compressed sensing (CS) methods for image reconstruction. In addition to the popular total variation minimization approach, we present an advanced dictionary learning algorithm for comparison with the ACID scheme, along with relevant results in the CT and MRI cases.

In **SI III**, we share the pseudo-codes of the ACID scheme. Then, we present adversarial attack algorithms to perturb the selected reconstruction networks and the whole ACID workflow respectively. Furthermore, we show the numerical convergence of the ACID iteration in terms of the Lipschitz constant, and the experimental results on local stability of ACID against noise.

In **IV**, we review the definition of kernel awareness and explain why deep learning can be unstable. Also, we comment on non-convex optimization and deep reconstruction. Then, we introduce the Bounded Relative Error Norm (BREN, previously called BEN) property, and report our theoretical work consisting of two independent parts:

(1) a heuristic analysis on the convergence of ACID (with the simplification that the CS module achieves a perfect sparsification), and

(2) a mathematically denser analysis (with the two approximations: (a) $\boldsymbol{A}^{T}$ is viewed as an inverse $\boldsymbol{A}^{-1}$ and (b) a non-unitary $\boldsymbol{H}$, which are well motivated and heuristically reasonable in the perspective of iterative reconstruction, to be explained in this section, but these approximations have compromised mathematical rigor).

Finally, we clarify the consistency between the Lipschitz constant and BREN.

.

# I. Deep Networks & Datasets

## I.A. Ell-50 Network

**I.A.1. Narrative.** Ell-50 is a special form of FBPConvNet, which is a classic neural network for CT imaging proposed in [1]. The FBPConvNet with multiple-solution decomposition and residual learning [2] was proposed to remove sparse-data artifacts and preserve image features and structures. The reconstruction performance of the FBPConvNet was validated, outperforming the total variation-regularized iterative reconstruction using the realistic phantoms. Besides, it was very fast to reconstruct an image on GPUs. In this study, the training dataset mainly contains ellipses with different intensities, sizes and locations. The network is named Ell-50, indicating that the measurements were collected from 50 different views. This network was trained by the authors of [1], which can be freely downloaded (https://github.com/panakino/FBPConvNet).

**I.A.2. Network Architecture.** The Ell-50 network was trained to reconstruct $\boldsymbol{f}$ from measurements $\boldsymbol{p} = \boldsymbol{A}\boldsymbol{f}$, where $\boldsymbol{A}$ represents a subsampling system matrix, with which only 50 uniformly spaced radial lines are collected. Because the FBPConvNet is an image post-processing network, it is trained from filtered backprojection (FBP) reconstruction images rather than directly learning a mapping from $\boldsymbol{p}$ to $\boldsymbol{f}$. The network first employs FBP to convert $\boldsymbol{p}$ to $\hat{\boldsymbol{f}} = \boldsymbol{A}^{+}\boldsymbol{p}$ where $\mathbf{A}^{+}$ represents the FBP and is considered as the first layer of the neural network.

The FBPConvNet is a useful model based on U-Net [3], which is considered as an encoder-decoder pair. The main features of U-Net based FBPConvNet are summarized as the following three features: multilevel decomposition, multichannel filtering, and skip connections (including concatenation operator and residual learning). The network input is an image with 512 ×512 pixels, where it is first down-sampled 5 times for encoding, and then the resultant low-dimensional image features are up-sampled to 512 ×512 pixels. Besides, the skip concatenation operator is employed in this network. The Ell-50 network consists of convolutional and deconvolutional layers, and each convolutional and deconvolutional layer is followed by batch normalization (BN) and ReLU layers. The sizes of filters and stride in the Ell-50 network were set to 3×3 and 1×1, respectively. Moreover, the Ell-50 network details are shown in Fig. S1.

**I.A.3. Network Training.** The few-view and full-view FBP images are treated as the input and ground-truth of the Ell-50, respectively. In this study, the network was implemented using the MatConvNet [4] toolbox with a slight modification to train and evaluate the performance. To prevent the divergence of the cost function, the MatConvNet [4] toolbox was slightly modified by clipping the computed gradients to a fixed range [5]. In this study, we only used the pre-trained network weights of [1] that were publicly available at GitHub (https://github.com/panakino/FBPConvNet). Such a configuration is consistent with the literature [6]. The loss function plays an important role in controlling the image quality, and the mean square error (MSE) between the network output and the ground truth is considered in Ell-50. Since the employed network was performed on a TITAN Black GPU graphic processor (NVIDIA Corporation), the total training time took about 15 hours with 101 epochs [6]. Regarding the learning rate, it was decreased logarithmically from 0.01 to 0.001. Besides, the batch size, momentum, and clipping value were set as 1, 0.99 and $10^{-2}$, respectively. For the Ell-50 network, it was implemented in MATLAB with the MatConvNet platform based on Window 10 system with one NVIDIA TITAN XP graphics processing units (GPUs) installed on a PC (16 CPUs @3.70GHz, 32.0GB RAM and 8.0GB VRAM).

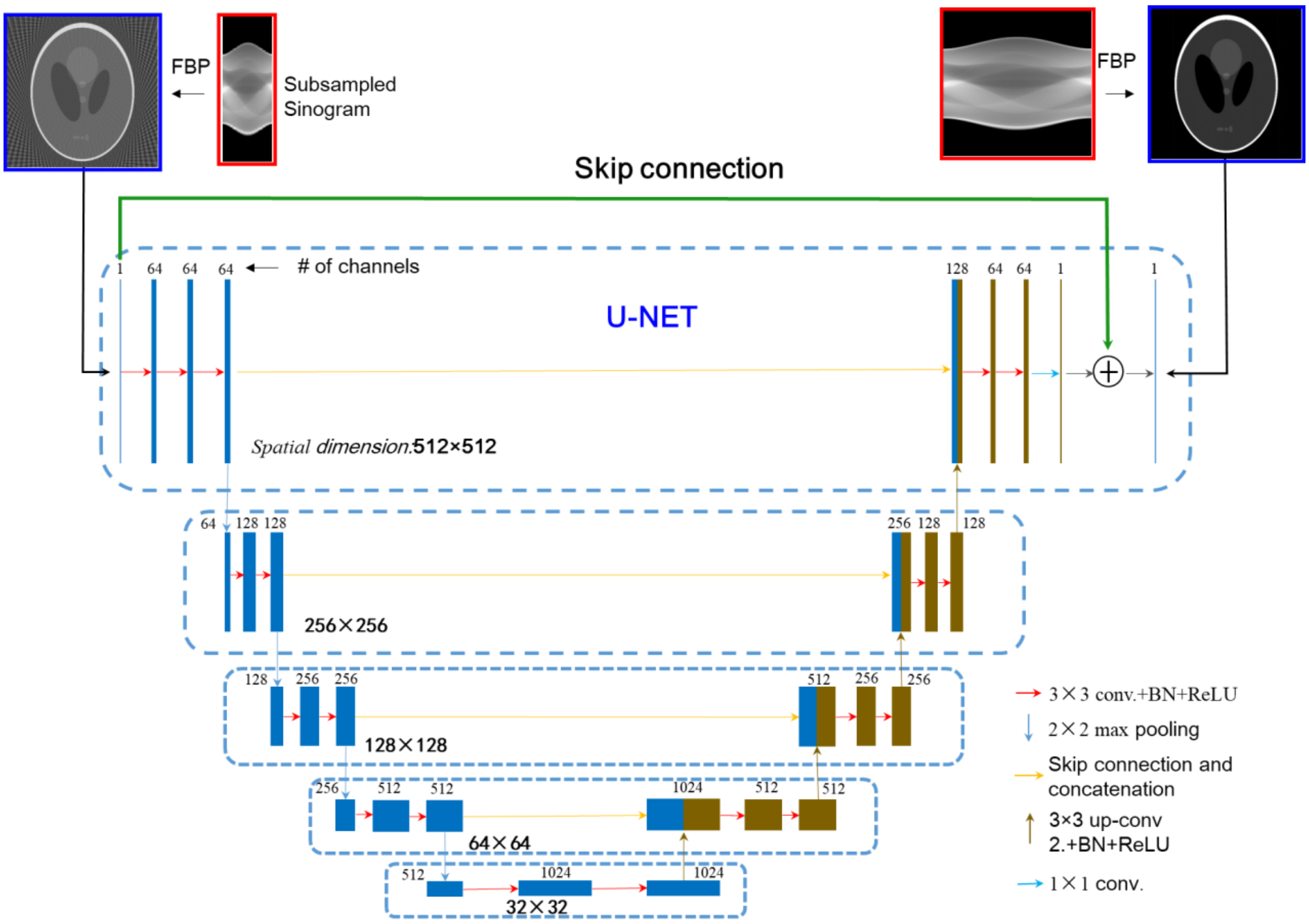

Fig. S1. Architecture for Ell-50.

**I.A.4. Training Data**. Regarding the training dataset, the number of training images is 475. The training images were reconstructed via FBP with sparse-view measurement. The dynamic range of the reconstructed images was controlled to the range of [−500, 500] HU. Since only ellipses with different sizes, locations and intensities were simulated, the projections were accurately and analytically computed [7]. The scanning geometry was set to produce parallel beams [8]. The number of full projections and the number of detector units were set as 1,000 and 729, respectively. Especially, the functions of *radon* and *iradon* in MATLAB were employed to realize the projection and backprojection operations. For sparse-data reconstruction, only 50 views were extracted from full projections, and then FBP reconstructed images were input to the selected networks in this paper. This case is one of typical sparse views reconstruction [9-11]. The ground truths are FBP images from full projections (i.e., 1000 views).

**I.A.5. Test Data**. To demonstrate the instability of neural network (i.e., Ell-50), the additional symbol "♥" and the text "CAN YOU SEE IT" were first embedded in the original image, which was provided by the authors of [6]. These artificial features were to mimic the structure changes and further validate the instability of the neural network in this case (see Fig. S2). In this study, the image with the symbol "♥" and text "CAN YOU SEE IT" was also treated as case C1 to validate the instability of Ell-50 and the stability of our proposed ACID method. Besides, a slightly complicated phantom with the inserted logo of a bird and text "A BIRD?" was provided by the authors of [12] and downloaded from [6], which is defined as case C2. The test image consists of $512 \times 512$ pixels, and it contains structural features without tiny perturbation. To generate adversarial attacks, the proposed method in [6] was employed to induce tiny perturbations. For case C3, an original image was randomly selected from the test datasets of https://github.com/panakino/FBPConvNet, which contains no perturbation. Here, the tiny perturbation is added to the original image with the same technique used in [6], and then we obtained the case C3. Regarding case C4, the same technique used in [6] was employed to generate the tiny perturbation and then embedded into case C1 to obtain C4. Furthermore, a Gaussian noise image with zero mean and 15 deviations in HU over the pixel value range was superimposed to case 1 to obtain C5 image. To validate the ability of ACID against adversarial attacks, the adversarial samples (see Section III in this supplementary information for details) for the whole ACID were generated and added into the C3 and C1 images respectively to obtain the cases of C6 and C7. The searched adversarial attacks in the whole ACID flowchart are greater than those used in a single neural network (i.e., Ell-50) in terms of $L_2$-norm.

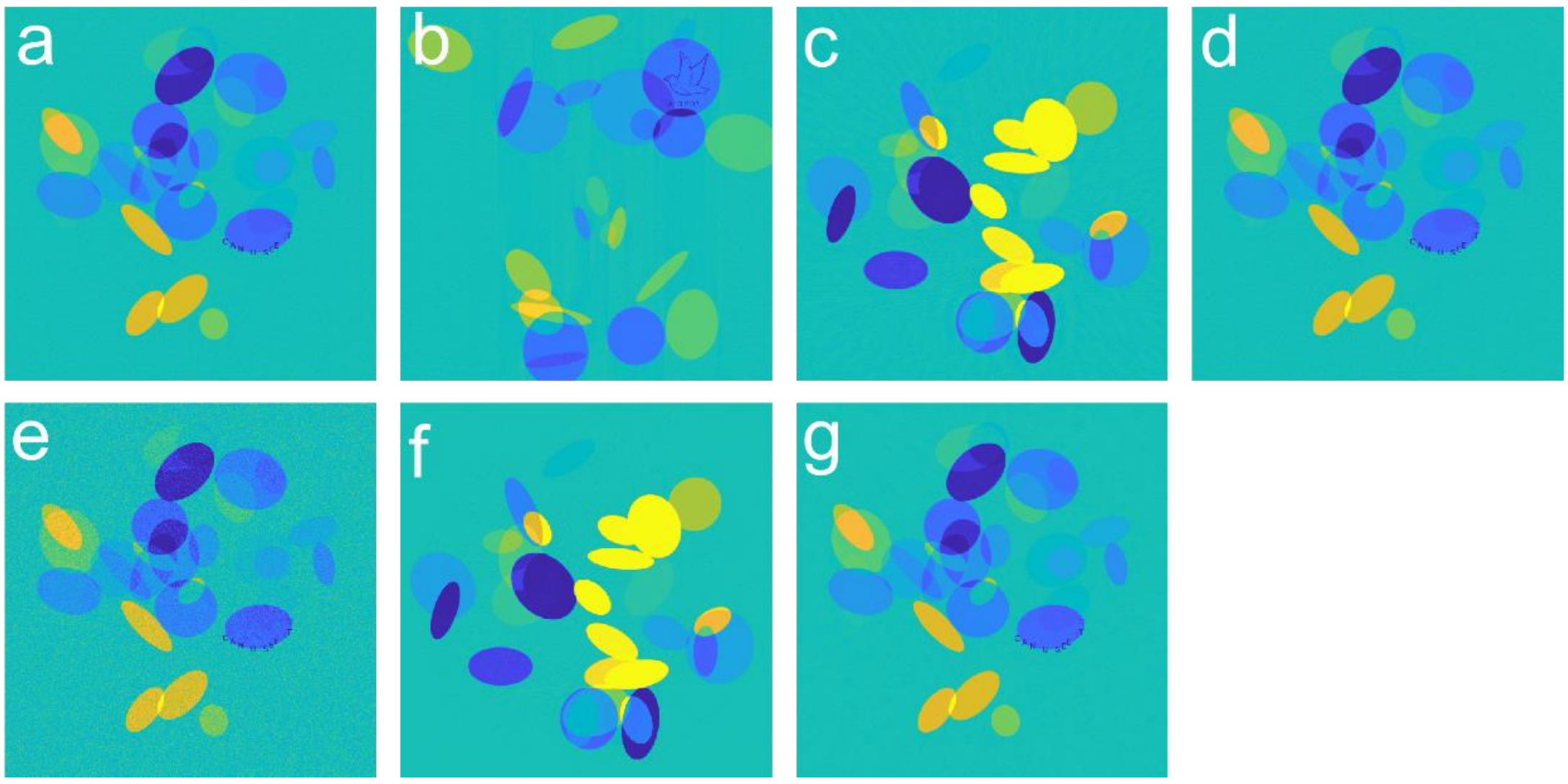

Fig. S2. Test images used to validate the effectiveness of ACID in stabilizing Ell-50 for CT study. (a)-(g) correspond to the C1-C7 cases, respectively.

## I.B. DAGAN Network

**I.B.1. Narrative.** The DAGAN network is to reduce aliasing artifacts with the U-Net [3] based generator [13]. To enhance the ability of the reconstruction method in preserving image texture and edges, DAGAN incorporates an innovative content loss and adversarial loss. Besides, it also introduces frequency-domain features to encourage coherence in image and frequency domains. Compared with the traditional CS-based and some other deep learning methods [14-16], the DAGAN method achieved superior performance in retaining image details. Besides, as one of the post-processing methods, the speed of DAGAN reconstruction is very fast. In this study, the DAGAN network was tested on a single-coil MRI with 10% and 20% subsampling rates. The trained weights are not available online, however, the authors of [13] provide the implementation details of DAGAN. With this help, we retrained the DAGAN with different subsampling rates and masks. The architecture, training parameters, and test data are summarized in the following subsections.

**I.B.2. Network Architecture.** The DAGAN [13] network was proposed for fast MRI reconstruction from subsampled measurement data. In the case of DAGAN, the measurement data is $\boldsymbol{p} = \boldsymbol{Af}$, where $\boldsymbol{A}$ is the subsampled discrete Fourier transform. The aim of DAGAN is to recover $\boldsymbol{f}$ from degraded image $\hat{\boldsymbol{f}}$ that is reconstructed directly via inverse Fourier transform from the zero-filled measurement data.

To restore high-quality MR images from measurement, DAGAN adopted a conditional generative adversarial network (GAN) [17-19] model. It consists of two modules: generator and discriminator. The generator is to recover the image, and the discriminator is to distinguish the recovered image and the ground-truth. The goal is to make the discriminator fail, and hence improve the recovered image quality. The authors in [13] provided three variants of DAGAN, and we selected the full model version (Pixel-Frequency-Perceptual-GAN-Refinement) in our experiments. According to parameter settings in [13] and the codes provided by the authors of [6], we retrained the DAGAN.

The architecture of the generator is illustrated in Fig. S3. It adopted the basic U-Net type structure, which contains 8 convolutional layers and 8 deconvolutional layers. All of them are followed by batch normalization layers to accelerate training convergence and overcome overfitting. The leaky ReLU layers are adopted as the activation function with a slope equal to 0.2 when the input is less than 0. Additionally, skip connections are employed to concentrate on encoder and decoder features to gain reconstruction details and promote the information flow. The hyperbolic tangent function is applied as the activate function for the output of the last convolutional layer. Then a global skip connection, adding the input data and the output of the hyperbolic tangent function together, is then clipped by a ramp function to scale the output of the generator to the range [-1,1]. The global skip connection can accelerate the training convergence and improve the performance of the network. The DAGAN network architecture was shown in Fig. S3. For more detailed information on the DAGAN network, please refer to [13].

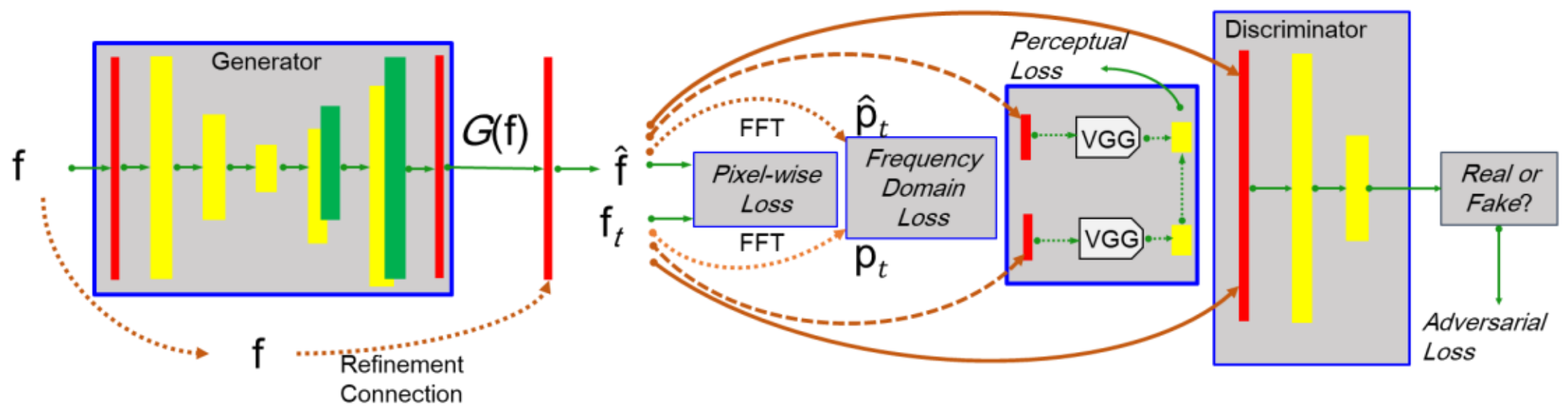

Fig. S3. The architecture of the DAGAN network.

**I.B.3. Training Parameters.** The loss function of DAGAN is formulated as

follows:

$$L_{DAGAN} = \sigma_1 L_{Img} + \sigma_2 L_{frq} + \sigma_3 L_{VGG} + \sigma_4 L_D \;\; s.t.\, \sigma_1, \sigma_2, \sigma_3, \sigma_4 > 0 , \quad \text{(S.1.1)}$$

where $L_{Img}$ computes the Euclidean distance in the image domain between the generated image and ground truth, and $L_{frq}$ accounts for the counterpart in the k-space. To constrain the similarity loss $L_{VGG}$ in the feature space, the trained VGG-16 was used to optimize the $L_2$-distance between feature maps of the generated image and ground truth, which is the same as [20]. In particular, the feature maps generated of the conv4 layer in VGG-16 were used to calculate $L_{VGG}$. Last, $L_D$ is the adversarial loss using a cross entropy to make the generated image more realistic. $\sigma_1, \sigma_2, \sigma_3, \sigma_4$ are the hyper-parameters to balance different constraint terms. According to [13], they were set to 15, 0.1, 0.0025 and 1, respectively. The generator and the discriminator were optimized using the Adam algorithm [21] with $\beta_1 = 0.5$ and $\beta_2 = 0.999$. Specifically, the learning rate was initially set to 0.0001, which was decreased half every 5 epochs, and the batch size is 25. To prevent overfitting, an early stopping strategy was adopted via measuring the loss $L_{frq}$ on the validation set, and the stopping number was set to 10.

**I.B.4. Training Data.** The datasets for training the DAGAN network were provided by the MICCAI 2013 Grand Challenge and are publicly available in https://my.vanderbilt.edu/masi/workshops/. More details about the training datasets are given in https://github.com/tensorlayer/DAGAN. Specifically, to exclude the negative influence on the DAGAN network, all the images that have more *T*% background pixels were dropped. In our experiments, the threshold *T* was set to 90. After data preprocessing, there are 15,912 images for training and 4,977 images for testing. All the images are T1-weighted brain MR images. Again, the data augmentation methods were applied to eliminate overfitting, including image flipping, rotation, shifting, and so on [13].

In the experiments, the DAGAN is to recover images from 10% subsampling rate using a 2D Gaussian mask and the radial mask of a 20% subsampling rate, respectively. Two models of the DAGAN network were trained for these two subsampling masks. All the codes were implemented with TensorLayer and Tensorflow frameworks [13].

**I.B.5. Testing Data.** To test the robustness of DAGAN in terms of small structural changes, adversarial attacks and noise, the symbols “HELLO NATURE” and “CAN YOU SEE IT” were embedded in two different original images, which are denoted as Cases M1 and M2, respectively. Specifically, the image with the symbol “CAN YOU SEE IT” was provided by the authors of [6] (download in https://github.com/vegarant/Invfool). The original image with the symbol “HELLO NATURE” was produced (downloaded from https://github.com/tensorlayer/DAGAN). In cases M1 and M2, there are two test images used to demonstrate the instability of the DAGAN network with respect to small structural changes. Next, to explore the performance of the DAGAN network in terms of adversarial attacks and small structural changes, the tiny perturbations derived from [6] were added into cases M1 and M2 to generate cases M3 and M4. Last, to test the DAGAN network in terms of anti-noising, the noise was superimposed to cases M1 and M2 to obtain cases M5 and M6. In our ablation study of ACID, we randomly selected one original image as M7 from the DAGAN test dataset (https://github.com/tensorlayer/DAGAN).

Furthermore, cases M8 and M9 were generated by applying the radial mask of a 20% subsampling rate on the M1 and M2 images, which were used to compare the performances of ACID and the classic Alternating Direction Method of Multipliers (ADMM)-net [22]. Regarding the stability of ACID, the tiny perturbations from ACID were added into M7, M1 and M2, and then the images with tiny perturbations were marked as M10, M11 and M12. The tiny perturbations from M11 and M12 are greater than the perturbations within M3 and M4 in terms of the $L_2$-norm. Except for M8 and M9, all the rest of the images were recovered from the k-space data collected at a 10% subsampling rate using the Gaussian mask. All the images from the references of M1-M12 are shown in Fig. S4.

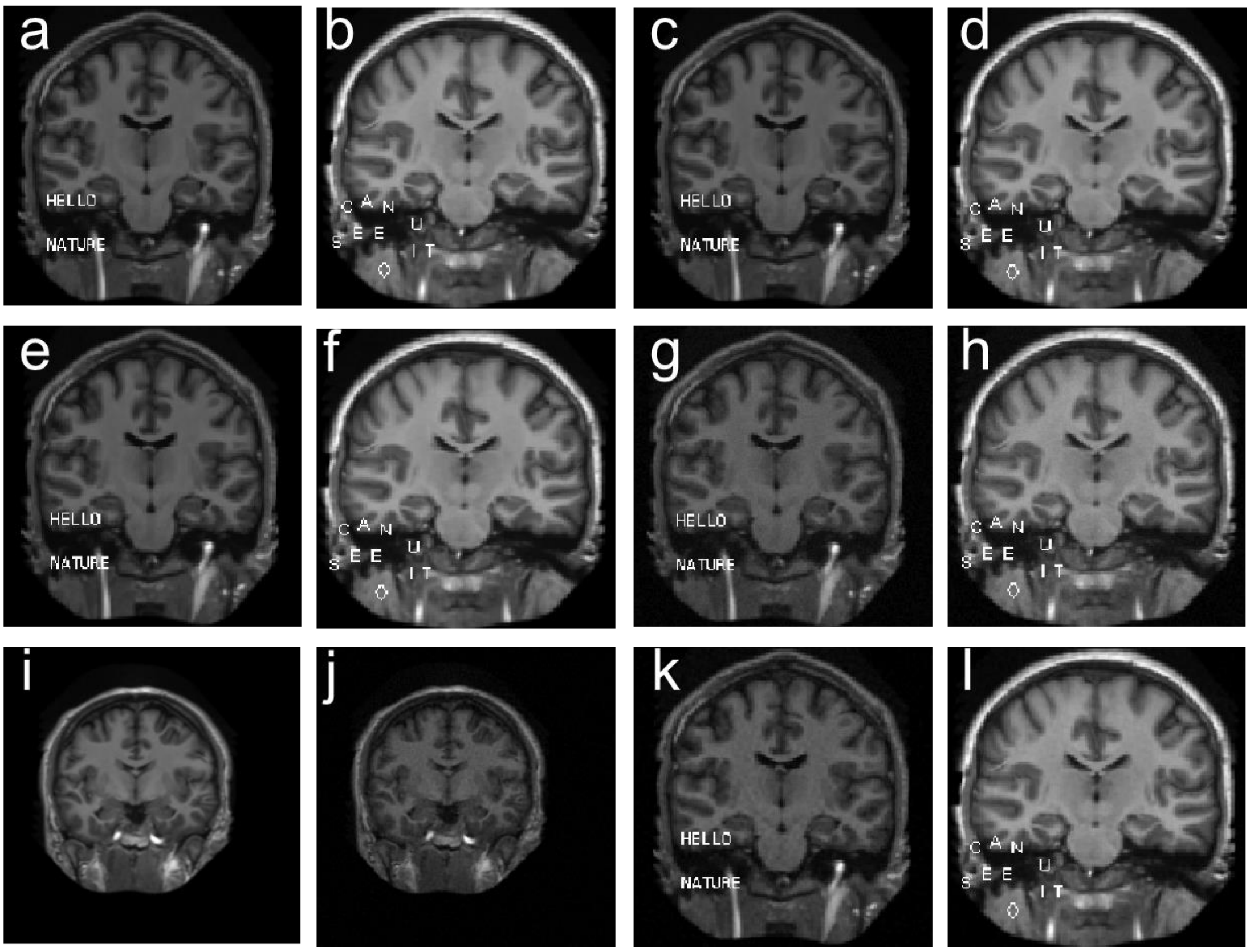

Fig. S4. Test images for showing the instability of neural networks. (**a**)-(**l**) correspond to the M1-M12 cases, respectively.

## I.C. AUTOMAP Network

**I.C.1. Narrative.** Proposed as a framework for image reconstruction, the automated transform by manifold approximation (AUTOMAP) transfers sensor data to a high-quality image with a mapping function between the sensor and image domains [23]. The AUTOMAP demonstrated its advantages in various magnetic resonance imaging acquisition modes using the same architecture and hyperparameters. In this study, the AUTOMAP neural network was tested on the single-coil MRI with subsampled data. The trained ATUOMAP used in our experiments is provided by [6]. The architecture, training details, and test data of AUTOMAP are in the following sub-sections.

**I.C.2. Network Architecture.** The AUTOMAP [23] presents a framework for image reconstruction by translating sensor-domain signals into the image domain directly via domain-transform manifold learning. For MRI reconstruction, four subsampling strategies were applied to access the performance of the AUTOMAP, which are Radon projection, spiral non-Cartesian Fourier, under-sampled Cartesian Fourier, and misaligned Fourier.

The AUTOMAP network takes a vectorized measurement data as input which is sub-sampled from the full-sampled k-space data. First, we can obtain the complex k-space data using the discrete Fourier transform on the MR images. Then, the subsampled k-space data are generated via a subsampling mask. Next, these measurement data are reshaped into vectors. Last, the vectorized measurement data are fed into the AUTOMAP network. In this paper, the images with the size of 128×128 and 60% subsampling rate are tested for MRI reconstruction. There are two fully connected layers in the AUTOMAP network, which have 25,000 and $128^2$ nodes, respectively. The activate function of the first fully connected layer is the hyperbolic tangent function, and the output of the second fully connected layer then subtracts the mean value of itself. Next, it is reshaped into a feature map with the same size as the reconstructed image. Furthermore, two convolutional layers are applied to extract essential features from their input data. Each of them contains 64 filters with a size of 5×5 and the stride of 1×1. The activation function of the first convolution layer is a hyperbolic tangent function and the other is rectified linear unit (ReLU). The last convolutional layer has one filter with the size of 7×7 and a stride of 1×1. The output of the network is the corresponding reconstruction image. The trained weights were provided by the authors of [6].

**I.C.3. Training Parameters.** The whole optimization problem of the AUTOMAP is defined as follow:

$$L_{AUTOMAP} = L_{rec} + \lambda_1 L_{fea} \ . \tag{S.1.2}$$

The loss function of AUTOMAP $L_{AUTOMAP}$ consists of two terms, i.e., $L_{rec}$ and $L_{fea}$ . $L_{rec}$ is employed to evaluate the Euclidean distance between the predicted image provided by the AUTOMAP network and the ground-truth image. $L_{fea}$ is $\ell_1$-norm to constrain the feature maps produced by the activation function of the second convolutional layer. $\lambda_1 > 0$ is to balance the two terms. The total loss function is optimized by the RMSProp algorithm with momentum 0 and decay 0.9 [23]. The learning rate is 0.00002 and the batch size is 100. The network was trained and stopped after 100 epochs.

**I.C.4. Training Data.** Selected in the MGH-USC HCP public dataset (http://www.humanconnectomeproject.org/data/), there are 50,000 images from 131 subjects in total. Specifically, the training images are 128×128 matrices, which were subsampled from the central part cropped from the original image. Meanwhile, all the training datasets were scaled to a given range. In the Fourier space, the subsampled measurement data were produced by a Poisson-disk mask of a 60% subsampling rate.

To improve the generalization ability of the AUTOMAP network, the data augmentation strategy was applied. 1.0% multiplicative noise was added to the input to promote manifold learning during the course of network training, and it is beneficial for the trained network learning robust representations from corrupted inputs. In fact, the specific additional noise distribution of the corruption process is not subject to the additive Gaussian noise during the process of evaluation. The corresponding training datasets with the size of 128×128 are cropped from original MR images by using four types of reflections. All the related codes were implemented in the TensorFlow framework [6].

**I.C.5. Testing Data.** To validate the instability of the AUTOMAP network, the symbol "♥" was first added to the original MR image, which was also provided by the author [6]. This simple symbol was used to simulate small structural changes in the patient and then test the instability of the AUTOMAP network reconstruction. All the test data were downloaded from [6]. In addition, the "HELLO NATURE", "CAN U SEE IT" and "◊" were added to the original test image to generate A1 and A2 with the structural changes. The resultant tiny perturbations were added to A1 and A2 images to obtain A3 and A4 images (see Fig. S5).

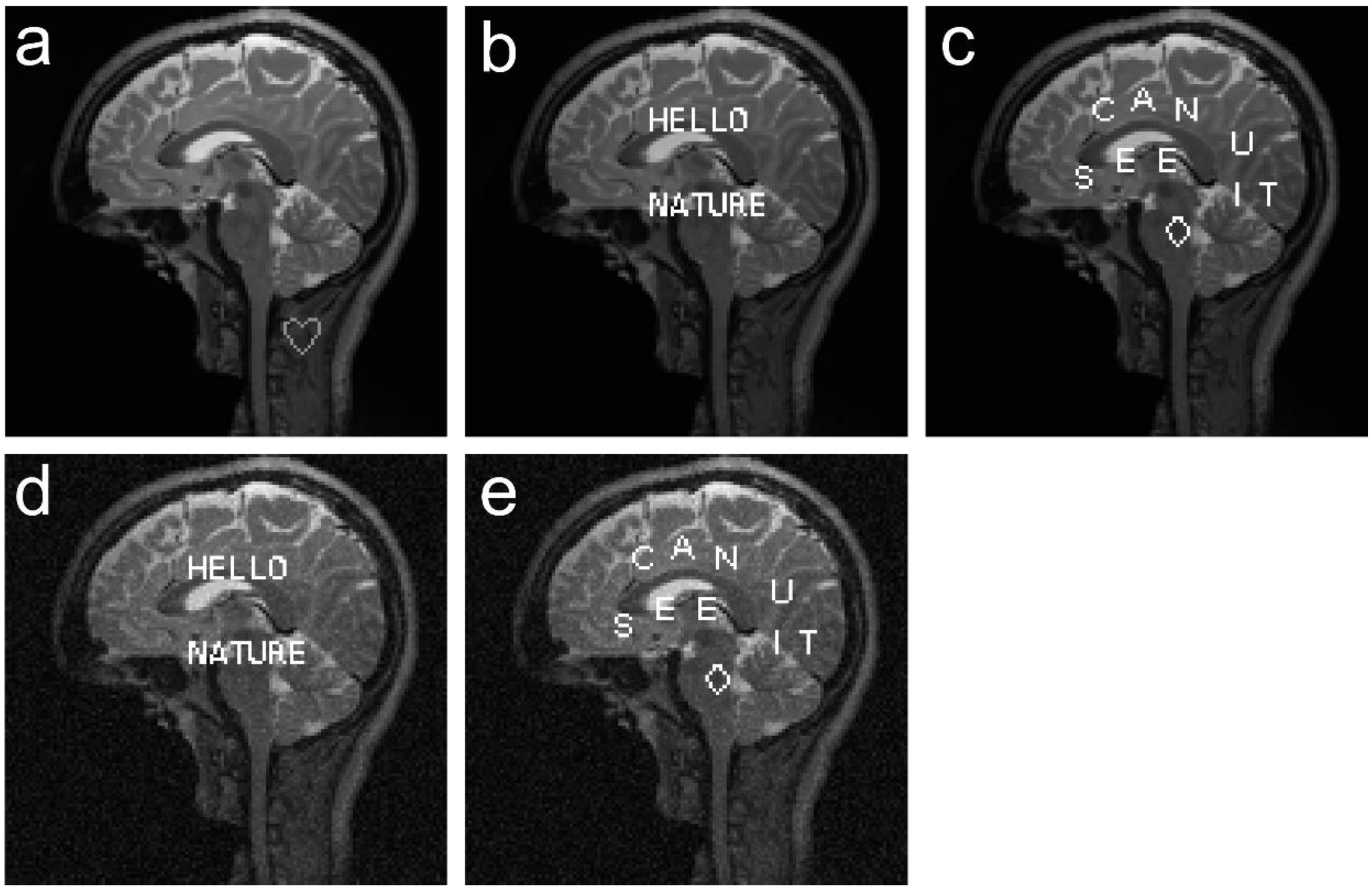

Fig. S5. Test images with a structural change and tiny perturbations for evaluation of the AUTOMAP stability. (a) was the test image provided in [6], and (b)-(e) represent the test images of A1-A4.

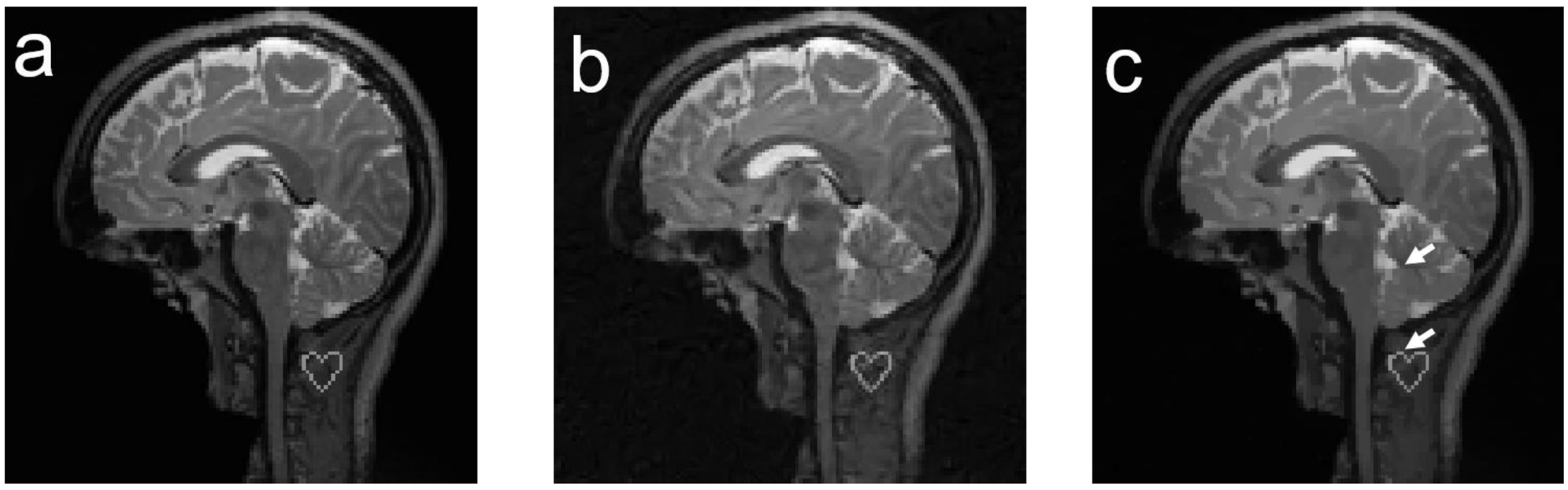

Fig. S6. ACID deep reconstruction with the embedded AUTOMAP network. (a) represents the original brain phantom, (b) and (c) represent the reconstructed results by AUTOMAP and ACID respectively.

**I.C.6. Reconstruction Results.** Here, to demonstrate the advantages of ACID, a typical reconstruction network, AUTOMAP, was selected as an example, and the reconstruction results of Fig. S5 (a) are in Fig. S6. As shown in Fig. S6, ACID produced significantly better image quality than AUTOMAP. The PSNR was improved by ACID to 36.0 dB, well above 27.8 dB of AUTOMAP. Also, the SSIM of ACID reached 0.971, while the counterpart of AUTOMAP was 0.730. It further demonstrates that ACID achieves better image quality than AUTOMAP. The related reconstruction results of A1-A4 are in the main body of the paper.

## I.D. ADMM-net

**I.D.1 Narrative.** Inspired by the traditional Alternating Direction Method of Multipliers (ADMM) iterative optimization algorithm for CS-based MRI [24], the ADMM-net defined over a data flow graph was first proposed in [22]. Regarding the training procedure, the network parameters (e.g. image shrinkage functions, transforms) are trained into an end-to-end architecture using the L-BFGS algorithm [25]. Regarding the testing step, it needs a similar computational overhead with the ADMM. However, there is only one parameter to be chosen initially in the ADMM-net since others are automatically learned during the training step. The superior experiments on MRI image reconstruction demonstrate the advantages over fast MRI imaging and higher image quality. In this study, the ADMM-net is tested on the single-coil MRI with 20% subsampling and the trained weights are provided by the authors of [22]. The architecture, training parameters and test data of ADMM-net are summarized in the following sections. The workflow of ADMM-net is given in Fig. S7.

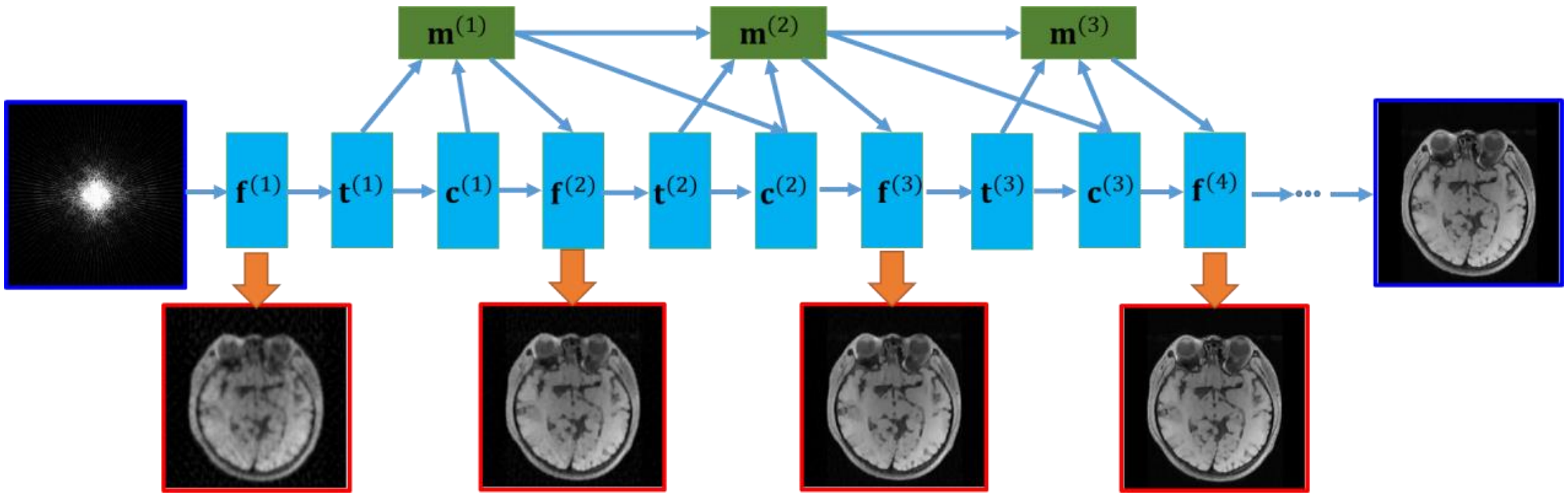

Fig. S7. The flowchart of the ADMM-net. $\boldsymbol{f}^{(k)}$, $\boldsymbol{c}^{(k)}$, $\boldsymbol{t}^{(k)}$ and $\boldsymbol{m}^{(k)}$ represent the construction layer, convolution layer, nonlinear transform layer, and multiplier update layer in the k-th stage.

**I.D.2. Network Architecture.** ADMM-net [22] is a classical unrolled iterative optimization algorithm for MRI reconstruction. Different from the traditional compressed sensing (CS) based methods [26] and data-driven based methods, ADMM-net can be trained end-to-end by incorporating a physic-guided model, and it achieves excellent performance in MR imaging with much less computational cost. The ADMM-net is derived from the ADMM algorithm via solving the sub-problem with deep learning networks. The CS-MRI model can be described as:

$$\underset{\boldsymbol{f}}{\operatorname{argmin}} \frac{1}{2}\left\|\boldsymbol{A}\boldsymbol{f}-\boldsymbol{p}^{(0)}\right\|_F^2+\sum_{l=1}^{L}\lambda_l g(D_l(\boldsymbol{f})) \quad \text{(S.1.3)}$$

where $\boldsymbol{f} \in C^N$ is the MR image to be reconstructed, $\boldsymbol{p}^{(0)} \in C^H$ denotes the under-sampled measurement data, $\boldsymbol{A}$ is the Fourier translation based system matrix with an under-sampled mask, $D_l$ represents the transform operation, $g$ is the regularization function, and $\lambda_l$ is the regularization parameter. By introducing $t_l = D_l(\boldsymbol{f})$ , $l = 1, \dots, L$ , (S.1.3) is converted into the following constraint optimization problem:

$$\underset{\boldsymbol{f},\{\mathrm{t}_l\}_{l=1}^{L}}{\operatorname{argmin}} \frac{1}{2}\left\|\boldsymbol{A}\boldsymbol{f} - \boldsymbol{p}^{(0)}\right\|_F^2 + \sum_{l=1}^{L}\lambda_l g(\mathrm{t}_l), \quad \mathrm{t}_l = D_l(\boldsymbol{f}), l = 1, \dots L. \qquad \text{(S.1.4)}$$

(S.1.4) is a constraint programming procedure and it can be further converted into the following unconstraint problem

$$\underset{\mathbf{f},\{\mathrm{t}_l\}_{l=1}^{L},\{\alpha_l\}_{l=1}^{L}}{\operatorname{argmin}} \ \frac{1}{2}\left\|\boldsymbol{A}\boldsymbol{f} - \boldsymbol{p}^{(0)}\right\|_F^2 + \sum_{l=1}^{L}\lambda_l g(\mathrm{t}_l) - \sum_{l=1}^{L}\langle \mathrm{t}_l - D_l(\boldsymbol{f}), \alpha_l\rangle + \frac{1}{2}\sum_{l=1}^{L}\gamma_l \|\mathrm{t}_l - D_l(\boldsymbol{f})\|_F^2, \qquad \text{(S.1.5)}$$

where $\alpha_l$ $(l = 1, \dots, L)$ are Lagrange multipliers and $\gamma_l$ $(l = 1, \dots, L)$ are the corresponding penalty parameters. (S.1.5) can be solved using the ADMM algorithm [27] as the following three sub-problems:

$$\boldsymbol{f}^{(k+1)} = \underset{\boldsymbol{f}}{\operatorname{argmin}} \frac{1}{2}\left\|\mathbf{Af} - \boldsymbol{p}^{(0)}\right\|_F^2 - \sum_{l=1}^{L}\langle \mathrm{t}_l^{(k)} - D_l(\mathbf{f}), \alpha_l^{(k)}\rangle + \frac{1}{2}\sum_{l=1}^{L}\gamma_l \left\|\mathrm{t}_l^{(k)} - D_l(\boldsymbol{f})\right\|_F^2 \quad , \qquad \text{(S.1.6)}$$

$$\mathrm{t}_l^{(\mathrm{k}+1)} = \underset{\{\mathrm{t}_l\}_{l=1}^{L}}{\operatorname{argmin}} \lambda_l g(\mathrm{t}_l) - \langle \mathrm{t}_l - D_l(\boldsymbol{f}^{(\mathrm{k}+1)}), \alpha_l\rangle + \frac{1}{2}\gamma_l \left\|\mathrm{t}_l - D_l(\boldsymbol{f}^{(\mathrm{k}+1)})\right\|_F^2, \ l = 1, \dots, L, \qquad \text{(S.1.7)}$$

$$\alpha_l^{(\mathrm{k}+1)} = \alpha_l^{(\mathrm{k})} + \mathrm{t}_l^{(\mathrm{k}+1)} - D_l(\boldsymbol{f}^{(k+1)}), l = 1, \dots, L. \qquad \text{(S.1.8)}$$

Finally, these three sub-problems can be updated iteratively using deep neural blocks. Regarding the ADMM-net, the above optimization with one separate variable update can be generalized as four type layers: reconstruction layer ($\boldsymbol{f}^{(k+1)}$), convolutional layer ($\{D_l(\boldsymbol{f}^{(k+1)})\}_{l=1}^{L}$), non-linear layer ($\{t_l^{(\mathrm{k}+1)}\}_{l=1}^{L}$), and multiplier update layer ($\{\alpha_l^{(\mathrm{k}+1)}\}_{l=1}^{L}$). More details related to the construction and organization of the layers in ADMM-net can be referred to [22]. The ADMM-net takes the sub-sampled k-space data as the input and finally generates the reconstructed image through the iterative process.

**I.D.3. Training Parameters.** The ADMM-net adopts the normalized mean square error (NMSE) as the loss function to optimize the neural network. The image reconstructed from fully-sampled k-space data was used as the reference image and the corresponding under-sampling data in the k-space was used as the input. The loss function is defined as

$$L_{NMSE} = \frac{1}{N_1}\sum_{n_1=1}^{N_1} \frac{\sqrt{\|\hat{\boldsymbol{f}}_{n_1}(\Theta) - \boldsymbol{f}_{n_1}\|_F^2}}{\sqrt{\|\boldsymbol{f}_{n_1}\|_F^2}} \quad , \qquad \text{(S.1.9)}$$

where $\hat{\boldsymbol{f}}_{n_1}$ and $\boldsymbol{f}_{n_1}$ are the generated image from ADMM-net and the reference image (as the label), respectively. $N_1$ is the number of training samples. Θ denotes parameters needed to be optimized in ADMM-net. The L-BFGS algorithm was used to minimize the loss function $L_{NMSE}$.

**I.D.4. Training Data.** The ADMM-net is trained with brain and chest MR image datasets (https://my.vanderbilt.edu/masi/workshops/). For each dataset, 100 images were randomly selected for training and 50 images for testing. In our experiments, all the under-sampled k-space data were generated with the radial mask of a 20% subsampling rate, as shown in Fig. S8. All the codes are in MATLAB with Intel core i7-4790k CPU, and the training and testing datasets were downloaded from https://github.com/yangyan92/Deep-ADMM-Net.

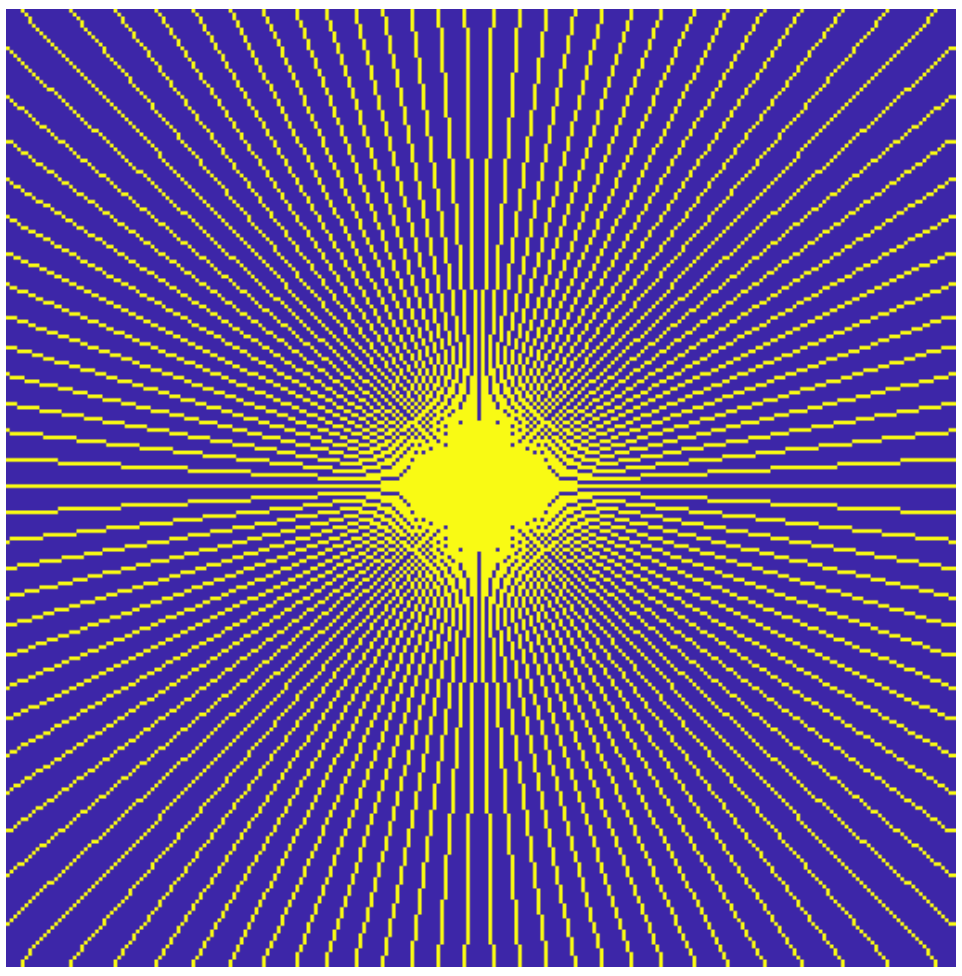

Fig. S8. Radial sampling mask of a 20% subsampling rate.

**I.D.5. Testing Data.** To validate the performance of ADMM-net about small structural changes, the same images as those in Fig. 2 in the main text of the paper with the symbols "CAN YOU SEE IT" and "HELLO NATURE" were employed with the radial mask of a 20% subsampling rate.

# II. CS Based Reconstruction Methods

## II.A. CS-inspired Reconstruction

**II.A.1. Narrative.** To demonstrate the advantages of our ACID in terms of stability against the benchmark compressed sensing (CS)-based methods [28-31], the related experiments are performed, and the reconstruction results are provided using the established methods [32-34]. Since the total variation minimization (individual or combination) is popular in the image reconstruction field with consideration of sparsity prior, it is respectively chosen for CT and MRI image reconstruction in this study. The specific details are given as follows.

**II.A.2. X-ray CT Reconstruction.** The re-weighting technique [35] combining both shearlets and TV in [36] and [37] was proposed to validate the stability in [6]. In this study, it served as a state-of-the-art CS-based comparison method for CT. The details can be found in [37] and [6]. Here we only provide a brief summary as follows.

The mathematical model for this method is formulated as

$$\underset{\boldsymbol{f}}{argmin}\frac{\omega}{2}\|\boldsymbol{Af}-\boldsymbol{p}\|_F^2+\sum_{j=1}^{J}\vartheta_j\left\|W_j\psi_j\boldsymbol{f}\right\|_1+TGV_{\varrho}(\boldsymbol{f}), \quad \text{(S.2.1)}$$

where $\vartheta_j$ represents the *j*-th balance factor, $W_j$ is a diagonal matrix, and $\psi_j$ represents the *j*-th subband from the corresponding shearlet transformation. The $\mathrm{TGV}_{\varrho}(\mathbf{f})$ stands for the total generalized variation with the parameter $\varrho$. $\mathrm{TGV}_{\varrho}(\mathbf{f})$ is combined with the components from the first and second orders of the total variation of the reconstructed image. Furthermore, the parameters $\varrho$ is introduced to balance these two terms. $\omega > 0$ is to balance data fidelity and regularization of sparsity prior.

To solve the optimization problem (S.2.1), $\boldsymbol{d} = \psi'\boldsymbol{f}$ is introduced to represent the matrix format of $\sum_{j=1}^{J} \psi_j \boldsymbol{f}$ and (S.2.1) is split into three sub-problems:

$$\{\boldsymbol{f}^{(k+1)}, \boldsymbol{d}^{(k+1)}\} = \underset{\boldsymbol{f},\boldsymbol{d}}{argmin} \frac{\omega}{2} \|\boldsymbol{Af} - \boldsymbol{p}\|_F^2 + \|W\boldsymbol{d}\|_1 + \frac{\omega_1}{2} \left\|\boldsymbol{d} - \psi'\boldsymbol{f} - \boldsymbol{b}^{(k)}\right\|_F^2 + TGV_\varrho(\boldsymbol{f}), \quad \text{(S.2.2)}$$

$$\boldsymbol{b}^{(k+1)} = \boldsymbol{b}^{(k)} + \psi'\boldsymbol{f}^{(k+1)} - \boldsymbol{d}^{(k+1)}, \quad \text{(S.2.3)}$$

$$\boldsymbol{p}^{(k+1)} = \boldsymbol{p}^{(k)} + \boldsymbol{p} - \boldsymbol{Af}^{(k+1)}. \quad \text{(S.2.4)}$$

where $W$ is the matrix format of $\vartheta_j W_j$. In (S.2.2)-(S.2.4), the four variables are updated iteratively. First, the minimization problem in (S.2.2) is optimized utilizing the multiple non-linear block Gauss-Seidel iterations [38]. Compared with the original re-weighting strategy in [35], the weights in $W$ are not only updated after convergence to the solution of (S.2.2), but also are put into the following split process. This unique weight updating strategy is further described in [6]. In this work, the same strategy and the configuration in [6] were used (including the parameters, the number of iterations, etc.). Note that the number of iterations and the regularization parameters can be further optimized.

**II.A.3. MRI Reconstruction.** By extending the iteratively regularized Gauss–Newton method (IRGN) with variational penalties [39, 40], the total generalized variation (TGV) based IRGN (IRGN-TGV) was proposed in [41], and better reconstruction quality was achieved by combining estimation of image and coil sensitivities with TGV regularization. Indeed, the IRGN-TGV had superior noise suppression because of the TGV regularization. In addition, the IRGN-TGV approach can remove sampling artifacts arising from pseudorandom and radial sampling patterns. In this study, it was employed as a state-of-the-art to perform CS-based MRI experiments. Here we also give a brief summary of this method.

Mathematically, MRI is a typical inverse problem with the sampling operator ***A*** and the correspondingly k-space data ***p*** from the receivers. Besides, the spin density is given as ***h***, and ***c*** represents the unknown set of coil sensitivities. For the current iteration index k with the given $\boldsymbol{f}^{(k)} := \left(\boldsymbol{h}^{(k)}, \boldsymbol{c}^{(k)}\right)$, the solution $\Delta\boldsymbol{f} := (\Delta\boldsymbol{h}, \Delta\boldsymbol{c})$ is sought to minimize the following objective function:

$$\underset{\Delta\boldsymbol{f}}{argmin} \frac{1}{2} \left\|\boldsymbol{A}'\left(\boldsymbol{f}^{(k)}\right)\Delta\boldsymbol{f} + \boldsymbol{A}\left(\boldsymbol{f}^{(k)}\right) - \boldsymbol{y}\right\|_F^2 + \frac{\alpha_k}{2} \left\|W_1\left(\boldsymbol{c}^{(k)} + \Delta\boldsymbol{c}\right)\right\|_1 + \rho_k R\left(\boldsymbol{h}^{(k)} + \Delta\boldsymbol{h}\right). \quad \text{(S.2.5)}$$

Given $\alpha_k > 0$ , $\rho_k > 0$, we have $\boldsymbol{f}^{(k+1)} := \boldsymbol{f}^{(k)} + \Delta\boldsymbol{f}$, $\alpha_{k+1} = q_a \alpha_k$ and $\rho_{k+1} = q_b \rho_k$ and $0 < q_a, q_b < 1$. $\boldsymbol{A}'\left(\boldsymbol{f}^{(k)}\right)$ represents the derivative of $\boldsymbol{A}\left(\boldsymbol{f}^{(k)}\right)$ with respect to $\boldsymbol{f}^{(k)}$. The term $W_1\left(\boldsymbol{c}^{(k)} + \Delta\boldsymbol{c}\right)$ represents the penalty on the Fourier coefficients, and $R$ is a regularization term. In the original IRGN method, the conventional $L^2$ was considered. Since the TV regularization can introduce stair-casing artifacts and reduce the image quality if the penalty parameter is too large, the authors in [41] considered the second-order TGV (total generalization variation, TGV), which is a generalized TV. Compared with the conventional TV, the TGV avoids stair-casing in regions of smooth signal changes and improves the image quality [30, 42]. Therefore, the authors of [41] employed TGV in IRGN and then generated IRGN-TGV for MRI. More details are in [30] and the corresponding code can be downloaded from https://www.tugraz.at/fileadmin/user_upload/Institute/IMT/files/misc/irgntv.zip. The parameters can be further tuned, depending on experimental designs [30].

## II.B. Dictionary Learning-Based Reconstruction

**II.B.1. Narrative.** As a successful example, dictionary learning-based methods were developed for tomographic reconstruction, including MRI [43-45], Optical Coherence Tomography [46-49] and CT [33, 50-52]. Dictionary learning based reconstruction methods explored the intrinsic properties using the trained dictionary with initial reconstruction results. The reconstruction process is usually divided into two steps: dictionary learning and image reconstruction. Without loss of generality, we compare the dictionary learning-based reconstruction method with our proposed ACID for CT and MRI.

**II.B.2. Dictionary Learning Model.** A number of image patches $\boldsymbol{f}_{i_d} \in \boldsymbol{\mathcal{R}}^{s\times s}$, $i_d = 1, \dots, I_d$, are extracted from the training datasets $\boldsymbol{f}$ , and $s$ represents the size of image patches. The set of $\boldsymbol{f}_{i_d}$, $i_d = 1, \dots, I_d$ is employed to train the global dictionary $\boldsymbol{D}_{ic} \in \boldsymbol{\mathcal{R}}^{S\times T_d}$, where $S = s \times s$ and $T_d$ is the number of atoms. The aim of dictionary learning is to search representation coefficients with sparse-level space constrained by $\boldsymbol{q} \in \boldsymbol{\mathcal{R}}^{T_d \times I_d}$ based on the dictionary $\boldsymbol{D}_{ic}$. It can be explained by solving the following optimization expression:

$$\{\boldsymbol{D}_{ic}^*, \boldsymbol{q}^*\} = argmin\ \frac{1}{2}\sum_{i_d=1}^{I_d}\left\| \boldsymbol{f}_{i_d} - \boldsymbol{D}_{ic}\boldsymbol{q}_{i_d}\right\|_2^2 \quad \text{s.t.}\ \left\|\boldsymbol{q}_{i_d}\right\|_0 \le L_{dl}, \qquad \text{(S.2.6)}$$

where $L_{dl}$ is the sparsity level of dictionary learning, $\|\cdot\|_0$ represents the quasi-$l_0$ norm, $\boldsymbol{q}_{i_d} \in \boldsymbol{\mathcal{R}}^{T_d\times 1}$ represents sparse representation coefficients for the $i_d$–th image patch. (S.2.6) is a constrained problem, and it is equivalent to the following unconstrained problem under a certain condition:

$$\{\boldsymbol{D}_{ic}^*, \boldsymbol{q}^*\} = argmin\ \left(\sum_{i_d=1}^{I_d}\left(\frac{1}{2}\left\|\boldsymbol{f}_{i_d} - \boldsymbol{D}_{ic}\boldsymbol{q}_{i_d}\right\|_2^2 + h_{i_d}\left\|\boldsymbol{q}_{i_d}\right\|_0\right)\right), \qquad \text{(S.2.7)}$$

where $h_{i_d}$ represents a Lagrange multiplier, which needs to be optimized. Furthermore, (S.2.7) can be solved by an alternating minimization scheme. First, we need to update $\boldsymbol{q}_{i_d}$ with a fixed dictionary $\boldsymbol{D}_{ic}$,

$$\boldsymbol{q}^* = argmin \sum_{i_d=1}^{I_d}\left(\frac{1}{2}\left\|\boldsymbol{f}_{i_d} - \boldsymbol{D}_{ic}\boldsymbol{q}_{i_d}\right\|_2^2 + h_{i_d}\left\|\boldsymbol{q}_{i_d}\right\|_0\right). \qquad \text{(S.2.8)}$$

(S.2.8) can be solved using the matching pursuit (MP) [53] or orthogonal matching pursuit (OMP) algorithm [54].Then, we can update the dictionary with a fixed sparse representation coefficients $\boldsymbol{q}$. Many methods can be employed to train the dictionary $\boldsymbol{D}_{ic}$, such as K-SVD [55], discriminate K-SVD [56], coupled dictionary training [57], online learning technique [58] and online robust learning [59]. In this study, the K-SVD was employed.

**II.B.3. Dictionary Learning-Based CT Reconstruction.** The conventional dictionary learning was first employed to MR reconstruction from under-sampled k-space data [43]. Then, the dictionary learning was utilized to low-dose CT imaging in our previous work [33], few-view CT reconstruction [28] and material decomposition [60]. In this study, we only consider the dictionary learning-based sparse data CT reconstruction. The mathematical model of dictionary learning-based CT reconstruction can be written as follows:

$$\underset{\boldsymbol{f},\boldsymbol{q}}{\operatorname{argmin}} \frac{1}{2}\left\|\boldsymbol{p}^{(\mathbf{0})} - \boldsymbol{A}\boldsymbol{f}\right\|_2^2 + \varsigma \sum_{i_d=1}^{I_d}\left(\frac{1}{2}\left\|\wp_{i_d}\boldsymbol{f} - \boldsymbol{D}_{ic}\boldsymbol{q}_{i_d}\right\|_2^2 + h_{i_d}\left\|\boldsymbol{q}_{i_d}\right\|_0\right), \quad \text{(S.2.9)}$$

where $\varsigma > 0$ represents the regularization penalty parameter. $\wp_{i_d}$ is an operator to extract $i_d$–th image patch from $\boldsymbol{f}$. Regarding the optimization of (S.2.9), there are many strategies to reach such a goal. Here, the split-Bregman method is used to obtain its solution. First, we introduce a new variable $\boldsymbol{b}$ to replace $\boldsymbol{f}$ and (S.2.9) can be converted into the following constraint programming problem

$$\underset{\boldsymbol{f},\boldsymbol{q}}{\operatorname{argmin}} \frac{1}{2}\left\|\boldsymbol{p}^{(\mathbf{0})} - \boldsymbol{A}\boldsymbol{f}\right\|_2^2 + \varsigma \sum_{i_d=1}^{I_d}\left(\frac{1}{2}\left\|\wp_{i_d}\boldsymbol{b} - \boldsymbol{D}_{ic}\boldsymbol{q}_{i_d}\right\|_2^2 + h_{i_d}\left\|\boldsymbol{q}_{i_d}\right\|_0\right), s.t., \boldsymbol{f} = \boldsymbol{b}. \quad \text{(S.2.10)}$$

To optimize (S.2.10), it can be further converted into

$$\underset{\boldsymbol{f},\boldsymbol{b},\boldsymbol{\chi},\boldsymbol{q}}{\operatorname{argmin}} \frac{1}{2}\left\|\boldsymbol{p}^{(\mathbf{0})} - \boldsymbol{A}\boldsymbol{f}\right\|_2^2 + \varsigma \sum_{i_d=1}^{I_d}\left(\frac{1}{2}\left\|\wp_{i_d}\boldsymbol{b} - \boldsymbol{D}_{ic}\boldsymbol{q}_{i_d}\right\|_2^2 + h_{i_d}\left\|\boldsymbol{q}_{i_d}\right\|_0\right) + \frac{\varsigma_1}{2}\left\|\boldsymbol{f} - \boldsymbol{b} - \boldsymbol{\chi}\right\|_2^2, \quad \text{(S.2.11)}$$

where $\varsigma_1 > 0$ represents the coupling factor, and $\boldsymbol{\chi}$ is the error feedback. In (S.2.11), there are four variables $\boldsymbol{f}, \boldsymbol{b}, \boldsymbol{q}$ and $\boldsymbol{\chi}$. It can be split into the following three sub-problems:

$$\underset{\boldsymbol{f}}{\operatorname{argmin}} \frac{1}{2}\left\|\boldsymbol{p}^{(\mathbf{0})} - \boldsymbol{A}\boldsymbol{f}\right\|_2^2 + \frac{\varsigma_1}{2}\left\|\boldsymbol{f} - \boldsymbol{b}^{(k)} - \boldsymbol{\chi}^{(k)}\right\|_2^2, \quad \text{(S.2.12)}$$

$$\underset{\boldsymbol{b},\boldsymbol{q}}{\operatorname{argmin}} \sum_{i_d=1}^{I_d}\left(\frac{1}{2}\left\|\wp_{i_d}\boldsymbol{b} - \boldsymbol{D}_{ic}\boldsymbol{q}_{i_d}\right\|_2^2 + h_{i_d}\left\|\boldsymbol{q}_{i_d}\right\|_0\right) + \frac{\varsigma_1}{2}\left\|\boldsymbol{f}^{(k+1)} - \boldsymbol{b} - \boldsymbol{\chi}^{(k)}\right\|_2^2, \quad \text{(S.2.13)}$$

$$\boldsymbol{\chi}^{(k+1)} = \boldsymbol{\chi}^{(k)} - \tau_d\left(\boldsymbol{f}^{(k+1)} - \boldsymbol{b}^{(k+1)}\right), \quad \text{(S.2.14)}$$

where $\tau_d > 0$ represents the step length and it was set to 1 in this study. Regarding (S.2.12), it is solved by using the separable surrogate method [61]

$$\boldsymbol{f}_{j_1j_2}^{(k+1)} = \boldsymbol{f}_{j_1j_2}^{(k)} - \frac{\left[\boldsymbol{A}^T\left(\boldsymbol{A}\boldsymbol{f}^{(k)} - \boldsymbol{p}^{(\mathbf{0})}\right)\right]_{j_1j_2} + \varsigma_1\left[\boldsymbol{f}^{(k)} - \boldsymbol{b}^{(k)} - \boldsymbol{\chi}^{(k)}\right]_{j_1j_2}}{\left[\boldsymbol{A}^T\boldsymbol{A} + \varsigma_1\right]_{j_1j_2}}, \quad \text{(S.2.15)}$$

where$[.]_{j_1j_2}$ represents the $(j_1, j_2)^{th}$ pixel in the matrix. In practice, (S.2.15) was performed using two steps:

$$\boldsymbol{f}_{j_1j_2}^{\left(k+\frac{1}{2}\right)} = \boldsymbol{f}_{j_1j_2}^{(k)} - \frac{\left[\boldsymbol{A}^T\left(\boldsymbol{A}\boldsymbol{f}^{(k)} - \boldsymbol{p}^{(\mathbf{0})}\right)\right]_{j_1j_2}}{\left[\boldsymbol{A}^T\boldsymbol{A} + \varsigma_1\right]_{j_1j_2}}, \quad \text{(S.2.16)}$$

and

$$\boldsymbol{f}_{j_1j_2}^{(k+1)} = \boldsymbol{f}_{j_1j_2}^{\left(k+\frac{1}{2}\right)} - \frac{\varsigma_1\left[\boldsymbol{f}^{(k)} - \boldsymbol{b}^{(k)} - \boldsymbol{\chi}^{(k)}\right]_{j_1j_2}}{\left[\boldsymbol{A}^T\boldsymbol{A} + \varsigma_1\right]_{j_1j_2}}. \quad \text{(S.2.17)}$$

In fact, the number of iterations for $\boldsymbol{f}_{j_1j_2}^{\left(k+\frac{1}{2}\right)}$ in (S.2.16) needs to be set to a good number (it was set as 10 in this study), and then $\boldsymbol{f}_{j_1j_2}^{(k+1)}$ is updated. Since $\varsigma_1$ is specific to scanning geometry, it is normalized into a new parameter $\gamma_1$ so that $\varsigma_1 = \gamma_1\|\boldsymbol{A}^T\boldsymbol{A}\|$; that is, we only need to select a geometrically-invariant $\gamma_1$. Regarding the optimization of (S.2.13), it is a typical dictionary learning-based signal recovery problem, and there are a large number of algorithms to solve this problem [55, 57]. To control the image recovery via dictionary learning, the parameters of sparsity level $L_{dl}$ and the precision level $\zeta$ should be chosen; for more details, see our previous studies [28, 33, 62].

**II.B.4. Dictionary Learning-Based MRI Reconstruction.** The conventional dictionary learning methods are common for MR reconstruction [43, 63-65]. In this study, the dictionary learning-based MRI (DLMRI) in [43] was employed to highlight the advantages of the ACID with built-in DAGAN and TV. Regarding the reconstruction process of DLMRI, it is similar to the process for CT reconstruction. It is also divided into two steps: dictionary learning and image updating. Regarding the dictionary learning step, both MRI and CT are the same except that training images are different. Again, the dictionary used in CT reconstruction was trained from FBP results or updated results within the iteration process. In contrast, the dictionary utilized in MRI was trained from the inverse Fourier transform results. Regarding the image updating step, it is not necessary to update the image based on the fast Fourier transform. More details can be found in [43].

**II.B.5. Experimental Results.** To validate the outperformance of ACID in comparison with the dictionary learning-based CT reconstruction method (DLCT), we repeated the experiments design for the cases C1 and C2. Here, we adopt the FBP method to reconstruct images. Then, the FBP results were employed to train the dictionaries. In this study, only 1.0x$10^4$ image patches were extracted from FBP images to train the dictionary by the K-SVD algorithm. The size of extracted image patches was set to $6 \times 6$. The dictionary $\boldsymbol{D}_{ic}$ is overcomplete, and it can benefit the sparsity level enforcement. The number of atoms was set to 512. The sparsity level $L_{dl}$ in the dictionary training can be set empirically, and it was chosen as 6. The number of iterations for the training dictionary was set to 100.

Note that the total variation is still treated as the compressed sensing-based sparsity for the built-in component in the ACID. Here, the parameters of $\gamma_1$, $L_{dl}$ and $\zeta$ in DLCT were set to 0.22, 8 and 0.06, respectively. The number of outer iterations was set to 200. The implementation environment for training and reconstruction is the same as Ell-50. Specifically, the computational costs of dictionary training and reconstruction consume 139 and 561 seconds. However, the whole ACID with the built-in Ell-50 consumes about 70.5 seconds. In other words, the ACID is faster than the DLCT method.

The reconstruction results from DLCT and ACID with C1 and C2 are in Fig. S9. It is observed that DLCT provides higher image quality than that obtained by the CS method. However, it is still worse than those obtained by the ACID. Besides, the proposed ACID method obtains better image edges and avoids blurred artifacts compared with the DLCT method. Especially, the insert texts in DLCT results (i.e., “CAN U SEE IT” and “A BIRD?”) are very blurry, and they failed to be discriminated against. These texts are clearly observed in ACID results. Regarding small features (*i.e.*, the symbol “♥”), they are totally missing in the DLCT result. However, they were still recovered by the ACID. In terms of quantitative assessment, our proposed ACID obtained the best results remarkably. More details for codes and test data are at https://zenodo.org/record/5497811.

To show the advantages of ACID with the built-in DAGAN, the reconstruction results in the M2 case from ACID and DLMRI are given in Fig. S10. The DLMRI obtained higher image quality than that obtained by the conventional CS method in the main text. However, it is still worse than those obtained by the ACID. Besides, the proposed ACID method obtained better image edges and avoided blurry artifacts compared with the DLMRI method. Especially, the inserted texts in DLCT results (i.e., “CAN U SEE IT”) are very blurry, and they could not be discriminated. These texts are clearly observed in the ACID results. The small symbol was totally blurred in DLMRI result, which was still recovered by ACID. In terms of quantitative assessment, our proposed ACID obtained better results than those achieved by DLMRI method. The MATLAB code of the MRIDL method can be downloaded from http://www.ifp.illinois.edu/~yoram/DLMRI-Lab/DLMRI.html. The reconstruction parameters within DLMRI were optimized. Regarding the computational cost, under the same computing environment, DLMRI took 606.3 seconds, which is higher than that of CS reconstruction methods in the main text (i.e., 127.8 seconds). Our proposed ACID only took 9.2 seconds.

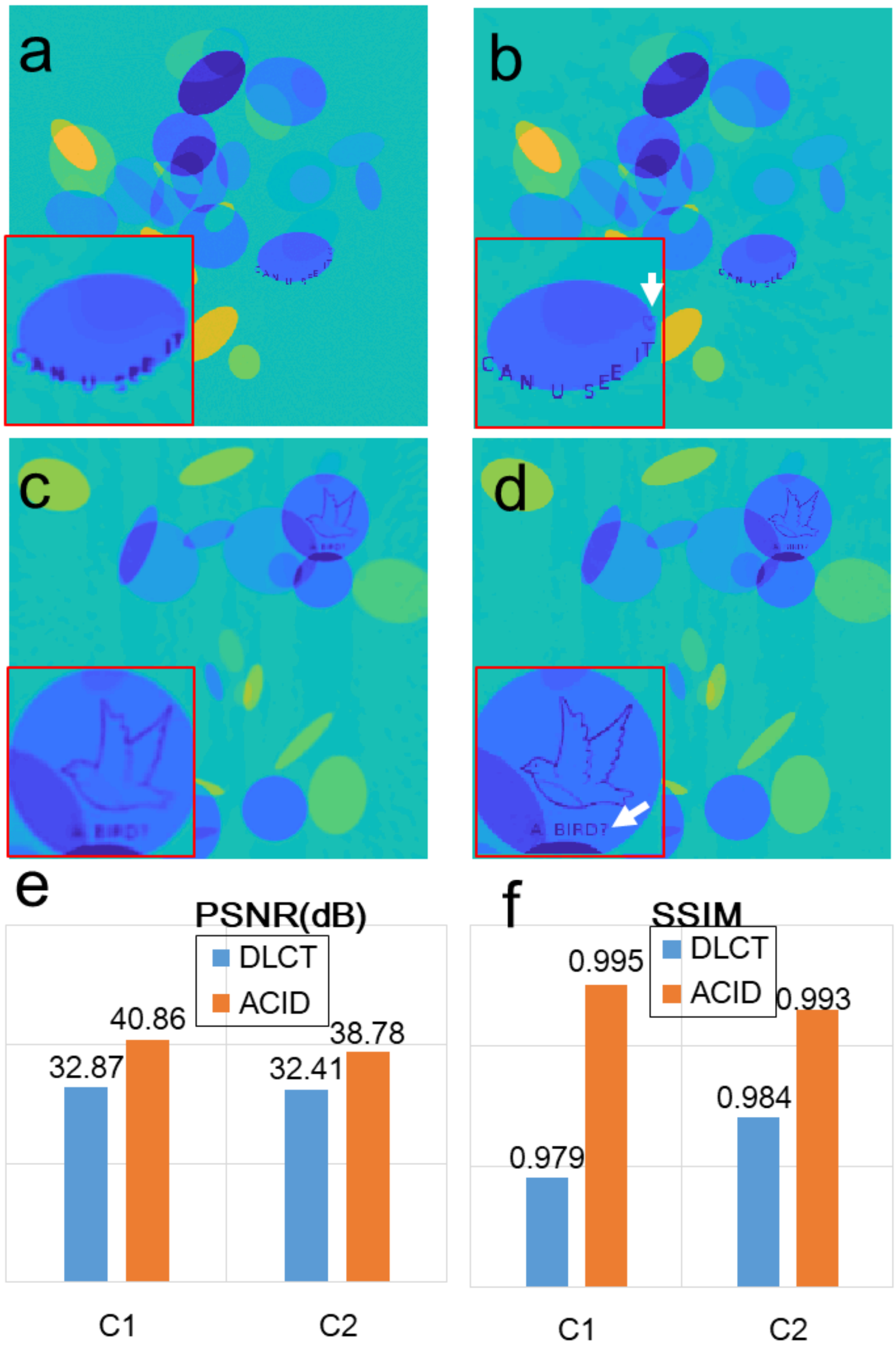

Fig. S9. Comparison study on the DLCT and ACID methods. **a** and **c**, **b** and **d** are reconstructed results from DLCT and ACID, respectively. **e** and **f** represent the quantitative results in terms of PSNR and SSIM.

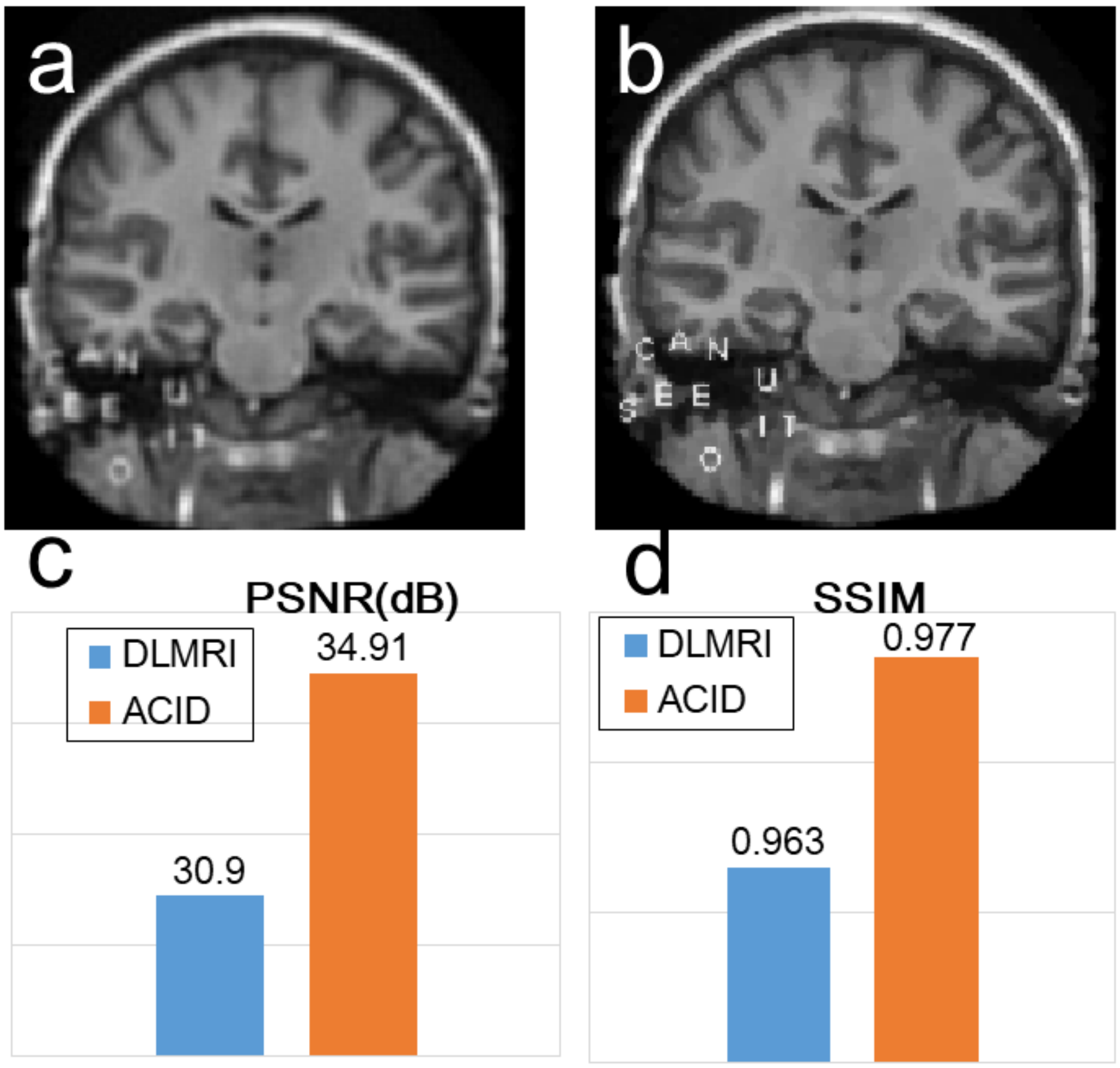

Fig. S10. Comparison study on the DLMRI and ACID methods. **a** and **b** are the reconstructed results from DLCT and ACID respectively. **c** and **d** represent the quantitative results in terms of PSNR and SSIM.

# III. ACID Architecture & Adversarial Attacks

## III.A. ACID Implementation

For an implementation of the whole ACID system, it is considered as an iterative framework and listed in Algorithm 1. In the whole ACID framework, we need to have input data $\boldsymbol{p}^{(0)}$, a neural network $\Phi$ and a system matrix $\boldsymbol{A}$. Then, we should specify the stopping condition; i.e., the maximum number of iterations *K*. Finally, the parameters $\lambda$ and $\varepsilon$ should be given to control the iterative process and the regularization strength, all of which can be empirically picked up. When k=1, we need to compute $\Phi(\boldsymbol{p}^{(0)})$ and then normalize $\Phi(\boldsymbol{p}^{(0)})$. The goal of the normalization operator is to facilitate the adjustment of the regularization parameters for different applications. Then, we obtain the updated $\boldsymbol{b}^{(1)}$ using the second formula in (1) in the main body of this paper. Next, $\boldsymbol{f}^{(1)}$ is updated by de-normalizing $\boldsymbol{b}^{(1)}$. When 1<k<K+1, we need to compute the residual data by $\boldsymbol{p}^{(k+1)} = \frac{\lambda(\boldsymbol{p}^{(0)} - \boldsymbol{A}\boldsymbol{f}^{(k)})}{1+\lambda}$. Since the residual data are not in the dynamic range of the original data, the residual data should be normalized into the original range to make sure the efficiency of the neural network (Line #9 in Algorithm 1). After the neural network predicts a residual image, the de-normalization operator should be applied on the prediction to ensure the consistency of the reconstruction results. Then, $\boldsymbol{f}^{(k)} + \frac{1}{\lambda}\Phi(\boldsymbol{p}^{(k+1)})$ is normalized and fed to the compressed sensing-based regularization module to encourage image sparsity. Finally, we obtain the updated image $\boldsymbol{f}^{(k+1)}$ after the de-normalization. More details on our codes and other materials are available at https://zenodo.org/record/5497811.

Algorithm 1. Pseudocode of the ACID workflow.

Input: Data $\boldsymbol{p}^{(0)}$, neural network $\Phi$, system matrix $\boldsymbol{A}$, maximum number of iterations $K$, auxiliary parameters $\lambda$, $\varepsilon$, and k=1;
If k<K+1 do
  if k=1 do
    Computing $\Phi(\boldsymbol{p}^{(0)})$;
    Normalizing $\Phi(\boldsymbol{p}^{(0)})$;
    Updating $\boldsymbol{b}^{(1)}$ where the normalized $\Phi(\boldsymbol{p}^{(0)})$ is treated as the input;
    Updating $\boldsymbol{f}^{(1)}$ by de-normalizing $\boldsymbol{b}^{(1)}$;
  else do
    Computing residual data using $\boldsymbol{p}^{(k+1)} = \frac{\lambda(\boldsymbol{p}^{(0)} - \boldsymbol{A}\boldsymbol{f}^{(k)})}{1+\lambda}$;
    Normalizing the residual data $\boldsymbol{p}^{(k+1)}$ into the input range of neural network to obtain $\overline{\boldsymbol{p}}^{(k+1)}$ ;
    Inputting $\overline{\boldsymbol{p}}^{(k+1)}$ into the neural network $\Phi$ and obtaining $\Phi(\overline{\boldsymbol{p}}^{(k+1)})$;
    De-normalizing $\Phi(\overline{\boldsymbol{p}}^{(k+1)})$ to obtain $\Phi(\boldsymbol{p}^{(k+1)})$;
    Normalizing $\boldsymbol{f}^{(k)} + \frac{1}{\lambda}\Phi(\boldsymbol{p}^{(k+1)})$
    Updating $\boldsymbol{b}^{(k+1)}$;
    Updating $\boldsymbol{f}^{(k+1)}$ by de-normalizing $\boldsymbol{b}^{(k+1)}$;
  end
end
return $\boldsymbol{f}^{(K)}$
Output: Reconstructed image $\boldsymbol{f}^{(K)}$

## III.B. Adversarial Attacks

**III.B.1. Attacking Model**. In the image reconstruction field, the continuous imaging system [66, 67] can be discretized into a linear model

$$\boldsymbol{p} = \boldsymbol{A}\boldsymbol{f}, \qquad \boldsymbol{A} \in R^{m\times N}, \tag{S.3.1}$$

where $\boldsymbol{p}$ represents collected data, $m$ and $N$ defines the size of the system matrix $\boldsymbol{A}$. The aim of image reconstruction is to reconstruct $\boldsymbol{f}$ from $\boldsymbol{p}$ for a given system matrix $\boldsymbol{A}$. $\boldsymbol{A}$ is the Fourier transform for MRI and the Radon transform for CT [68].

To assess the stability of image reconstruction, it is necessary to compute a tiny perturbation or adversarial attack [6, 69, 70]. In this context, the researchers [6] first computed a tiny perturbation for the following model (S.3.1):

$$\text{Reconstruct } \boldsymbol{f} \text{ from } \boldsymbol{p} = \boldsymbol{A}\boldsymbol{f}, \qquad \boldsymbol{A} \in \mathrm{R}^{m\times N}. \tag{S.3.2}$$

For a well-trained neural network $\Phi: \mathrm{R}^m \rightarrow \mathrm{R}^N$ to solve (S.3.2), similar to the adversarial attack in image classification [71], we compute the instabilities by formulating the following problem based on [6]

$$\hat{\boldsymbol{e}}(\boldsymbol{f}) \in \underset{\boldsymbol{e}}{\operatorname{argmin}} \|\boldsymbol{e}\|, \quad s.t. \|\Phi(\boldsymbol{p} + \boldsymbol{A}\boldsymbol{e}) - \Phi(\boldsymbol{p})\| \geq \epsilon. \tag{S.3.3}$$

With (S.3.3), $\boldsymbol{p} = \boldsymbol{A}\boldsymbol{f}$ cannot always be true when $\epsilon > 0$. One can consider the constrained Lasso variant of (S.3.3) as follows:

$$\hat{\boldsymbol{e}}(\boldsymbol{f}) \in \underset{\boldsymbol{e}}{\operatorname{argmax}} \|\Phi(\boldsymbol{p} + \boldsymbol{A}\boldsymbol{e}) - \Phi(\boldsymbol{p})\|, \quad s.t. \|\boldsymbol{e}\| \leq \sigma \tag{S.3.4}$$

There is no infeasibility issue for (S.3.4). An unconstrained Lasso inspired version of (S.3.4) is given by

$$\boldsymbol{e}^*(\boldsymbol{p}) \in \underset{\mathbf{e}}{\operatorname{argmax}} \frac{1}{2} \|\Phi(\boldsymbol{p} + \boldsymbol{A}\boldsymbol{e}) - \Phi(\boldsymbol{p})\|_2^2 - \frac{\gamma}{2} \|\boldsymbol{e}\|_2^2. \tag{S.3.5}$$

With $\Phi(\boldsymbol{p}) = l(\boldsymbol{f})$, (S.3.5) is further converted to

$$\boldsymbol{e}^{*}(\boldsymbol{p}) \in \underset{\mathbf{e}}{\operatorname{argmax}} \frac{1}{2}\|\Phi(\boldsymbol{p}+\boldsymbol{A}\boldsymbol{e}) - l(\boldsymbol{f})\|_2^2 - \frac{\gamma}{2}\|\boldsymbol{e}\|_2^2, \quad \text{(S.3.6)}$$

where $l(\boldsymbol{f}) = \boldsymbol{f}$ for image-domain post-processing [72-74] and $l(\boldsymbol{f}) = \Phi(\boldsymbol{A}\boldsymbol{f})$ with the end-to-end network (such as AUTOMAP [23] and iRadonMap [75]). Note that (S.3.6) works in the image domain to find perturbations. One generates a reconstructed image using an easy way and then compares the original image with a perturbed one to determine whether the perturbed image is acceptable/unacceptable. Now, we describe the details on how to generate perturbations for a single neural network and the whole ACID system, respectively.

**III.B.2. Stability with Respect to Tiny Perturbation to a Selected Network.** Since the neural network $\Phi: R^m \rightarrow R^N$ is a non-linear function. It is difficult to reach for a global maximum for (S.3.6). Here, we use the same strategy in [6] to search for tiny perturbations. In other words, one usually can reach the local maxima of (S.3.6) using a gradient search method. Especially, one defines the following objective function:

$$D_{\mathbf{p}}^{l}(\boldsymbol{e}) = \frac{1}{2}\|\Phi(\boldsymbol{p}+\boldsymbol{A}\boldsymbol{e}) - l(\boldsymbol{f})\|_2^2 - \frac{\gamma}{2}\|\boldsymbol{e}\|_2^2, \quad \text{(S.3.7)}$$

Regarding the optimization of (S.3.7), the gradient ascent search is a very common method [76]. For an iterative solution to (S.3.7), except the weighting factor $\gamma$, there are two parameters (i.e., $\tau$ and $\vartheta$) to be chosen, which control the perturbation $\boldsymbol{e}$. Algorithm 2 summarized the flowchart to generate tiny perturbations for a single neural network with random initialization $\boldsymbol{e}_0$, where $t_1$ represents a scaling factor to control the range of an input image. The gradient of $D_{\mathbf{p}}^{l}(\boldsymbol{e})$ with respect to $\boldsymbol{e}$ can be given as

$$\nabla_{\boldsymbol{e}} D_{\boldsymbol{p}}^{l}(\boldsymbol{e}) = \boldsymbol{A}^{\mathrm{T}}\nabla_{\boldsymbol{u}_1} z(\boldsymbol{u}_1) - \gamma\boldsymbol{e}, \quad z(\boldsymbol{u}_1) \coloneqq \frac{1}{2}\|\Phi(\boldsymbol{u}_1) - l(\boldsymbol{f})\|_2^2, \quad \text{(S.3.8)}$$

where $\boldsymbol{u}_1 = \boldsymbol{p} + \boldsymbol{A}\boldsymbol{e}$ , $\nabla_{\boldsymbol{u}_1} z(\boldsymbol{u}_1)$ can be implemented as a backpropagation method, and $\boldsymbol{A}^T$ represents the transpose of $\boldsymbol{A}$.

For generating such tiny perturbations via gradient ascent search, the learning rate ($\vartheta > 0$) needs to be optimized. In this study, we search for tiny perturbations guided by the experimental results in [6]. Note that Algorithm 2 needs to be modified slightly for CT reconstruction due to the adjoint operator corresponding to a discretization of the filtered backprojection (FBP) method, and more details for generating tiny perturbations for X-ray CT are given in [6]. The discretization of FBP in our study was done using MATLAB R2017b.

Algorithm 2 Searching tiny perturbation in inverse problems.

```
Input: Image f, neural network Φ, system matrix A, maximum number of iterations K_1
Initialization: p ← Af, v^(0) ← 0, i ← 0, γ,τ > 0 and ϑ > 0, e_0, D_p^l(e) is defined by (S.3.7)
while i ≤ K_1 do
    v^(i+1) ← τ v^(i) + ϑ ∇_e D_p^l(e^(i))
    e^(i+1) = e^(i) + v^(i+1)
    i = i + 1
return e^(K_1)
Output: Perturbation e^(K_1)
```

**III.B.3. Stability with Respect to Tiny Perturbation to Our ACID.** The iterative process of ACID is to find the optimized solution in the intersection of (a) the space of data-driven priors, (b) the space of sparse images, and (c) the space of solutions satisfying the measurement, as shown in Fig. S11. With a tiny perturbation to our proposed ACID workflow, the feedforward propagation of the perturbation is illustrated in Fig. S12. Specifically, the formula of (S.3.7) is converted to

$$D^{l}_{\mathbf{p}^{(0)}}(\boldsymbol{e}) = \frac{1}{2}\left\|\hat{\boldsymbol{f}}\left(\boldsymbol{p}, \boldsymbol{p}^{(\mathbf{0})}\right) - l(\boldsymbol{f})\right\|_2^2 - \frac{\gamma}{2}\|\boldsymbol{e}\|_2^2, \quad \text{(S.3.9)}$$

where $\hat{\boldsymbol{f}}$ is different from $\boldsymbol{f}$ as computed by the neural network and stabilized in the ACID framework. $\hat{\boldsymbol{f}}\left(\boldsymbol{p}, \boldsymbol{p}^{(\mathbf{0})}\right)$ is the solution minimizing the following objective function (for more details, please see Section IV in this **SI**):

$$\underset{\boldsymbol{p},\boldsymbol{f}}{\operatorname{argmin}} \frac{1}{2}\|\Phi(\mathbf{A}\boldsymbol{f} + \boldsymbol{p}) - \boldsymbol{f}\|_2^2 + \frac{\lambda}{2}\left\|\boldsymbol{p}^{(\mathbf{0})} - \boldsymbol{A}\boldsymbol{f} - \boldsymbol{p}\right\|_2^2 + \frac{\mu}{2}\|\boldsymbol{p}\|_2^2 + \xi\|\boldsymbol{H}\boldsymbol{f}\|_1. \quad \text{(S.3.10)}$$

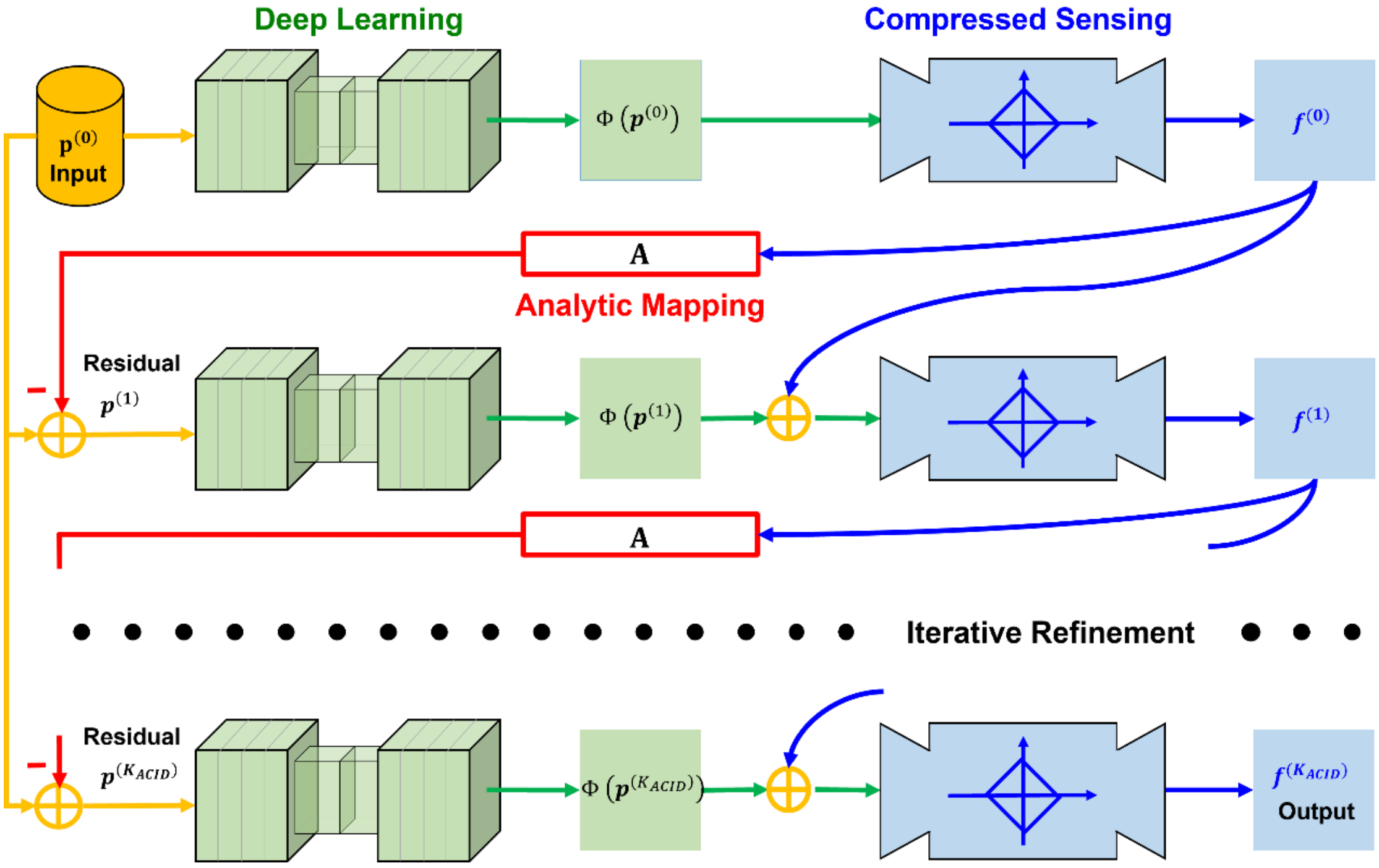

Fig. S11. ACID architecture for stabilizing deep tomographic image reconstruction. ACID consists of the following components: deep reconstruction, compressed sensing-based sparsity promotion, analytic mapping and iterative refinement. $\boldsymbol{p}^{(\mathbf{0})}$ is original tomographic data, and $\boldsymbol{p}^{(k_{ACID})}$, $k_{ACID}$=1, 2, 3, …, $K_{ACID}$, represents an estimated residual dataset in the $k_{ACID}$th iteration between $\boldsymbol{p}^{(\mathbf{0})}$ and the currently reconstructed counterpart. $\Phi\left(\boldsymbol{p}^{(k_{ACID})}\right)$ is an output of the deep reconstruction module, and $\boldsymbol{f}^{(k_{ACID})}$ represents the image after compressed sensing-based regularization.

For the optimization problem (S.3.10), we now compute a tiny perturbation via gradient ascent search. Specifically, we compute

$$D^{l}_{\mathbf{p}^{(0)}+\mathbf{Ae}}(\mathbf{e}) = \frac{1}{2}\left\|\hat{\boldsymbol{f}}\left(\boldsymbol{p}, \boldsymbol{p}^{(\mathbf{0})} + \boldsymbol{A}\boldsymbol{e}\right) - \boldsymbol{f}\right\|_2^2 - \frac{\gamma}{2}\|\boldsymbol{e}\|_2^2. \quad \text{(S.3.11)}$$

The backpropagation process for ACID is shown in Fig. S14. More clearly, we define the cost function of ACID as

$$L_c = \frac{1}{2}\|\boldsymbol{f}_e - \boldsymbol{f}\|_2^2 - \frac{\gamma}{2}\|\boldsymbol{e}\|_2^2 \quad , \quad \text{(S.3.12)}$$

where ***e*** is the perturbation, $\boldsymbol{f}_e = \hat{\boldsymbol{f}}\left(\boldsymbol{p}, \boldsymbol{p}^{(\mathbf{0})} + \boldsymbol{A}\boldsymbol{e}\right)$ is the output of the ACID system with the perturbation **e**, and $\boldsymbol{f}$ is the corresponding output without **e**. To find an effective **e**, we need to compute the gradient $dL_c/d\boldsymbol{e}$, and then refine the perturbation ***e*** using a gradient ascent algorithm. For clarity, the iteration index for ACID is changed to $k_{ACID}$ ($k_{ACID} = 0, \dots, K_{ACID}$) in this subsection. In Fig. S13, there are two branches contributing to $\boldsymbol{f}^{(k_{ACID})}(k_{ACID} = 1, \ldots, K_{ACID} - 1)$; i.e., Branches 1 and 2. To compute the gradient $dL_c/d\boldsymbol{e}$, we take both branches into account.

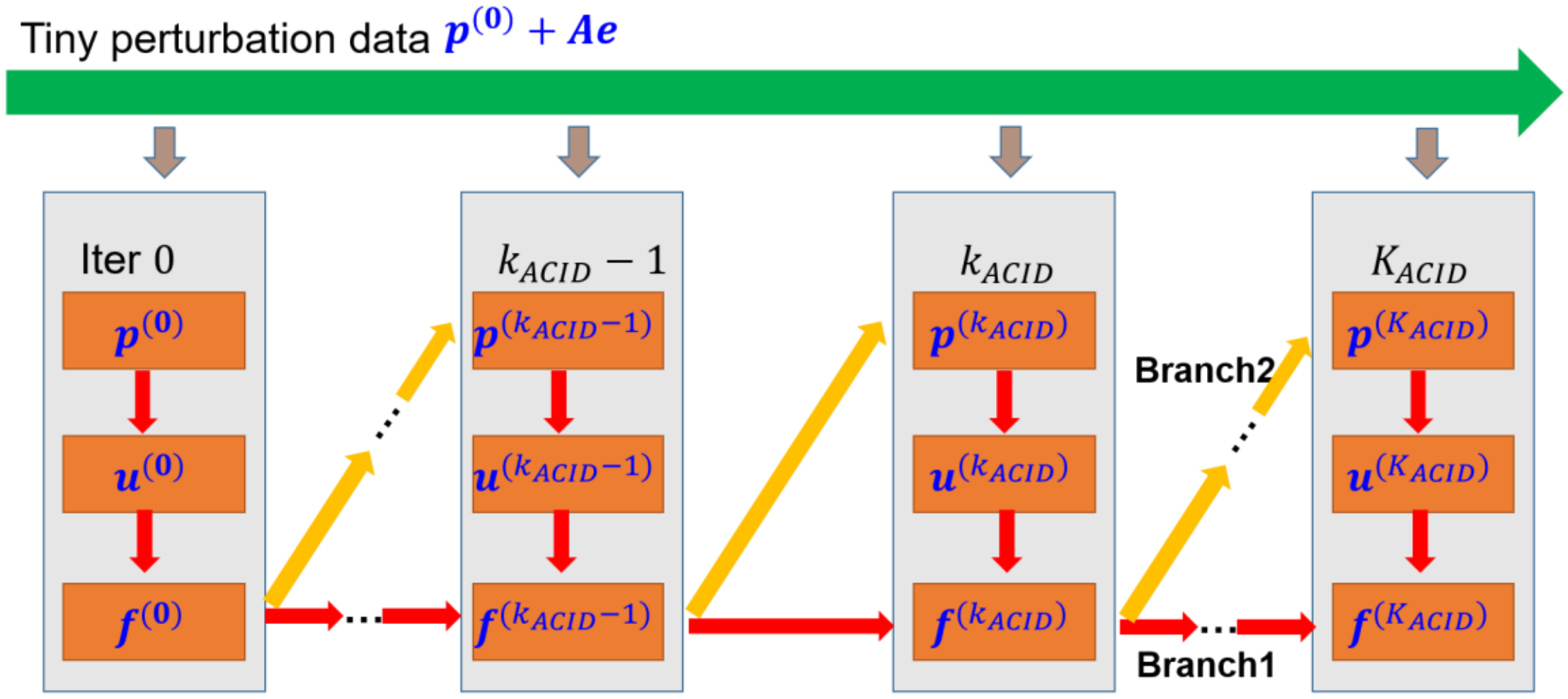

Fig. S12. Feedforward propagation of adversarial data in the ACID framework.

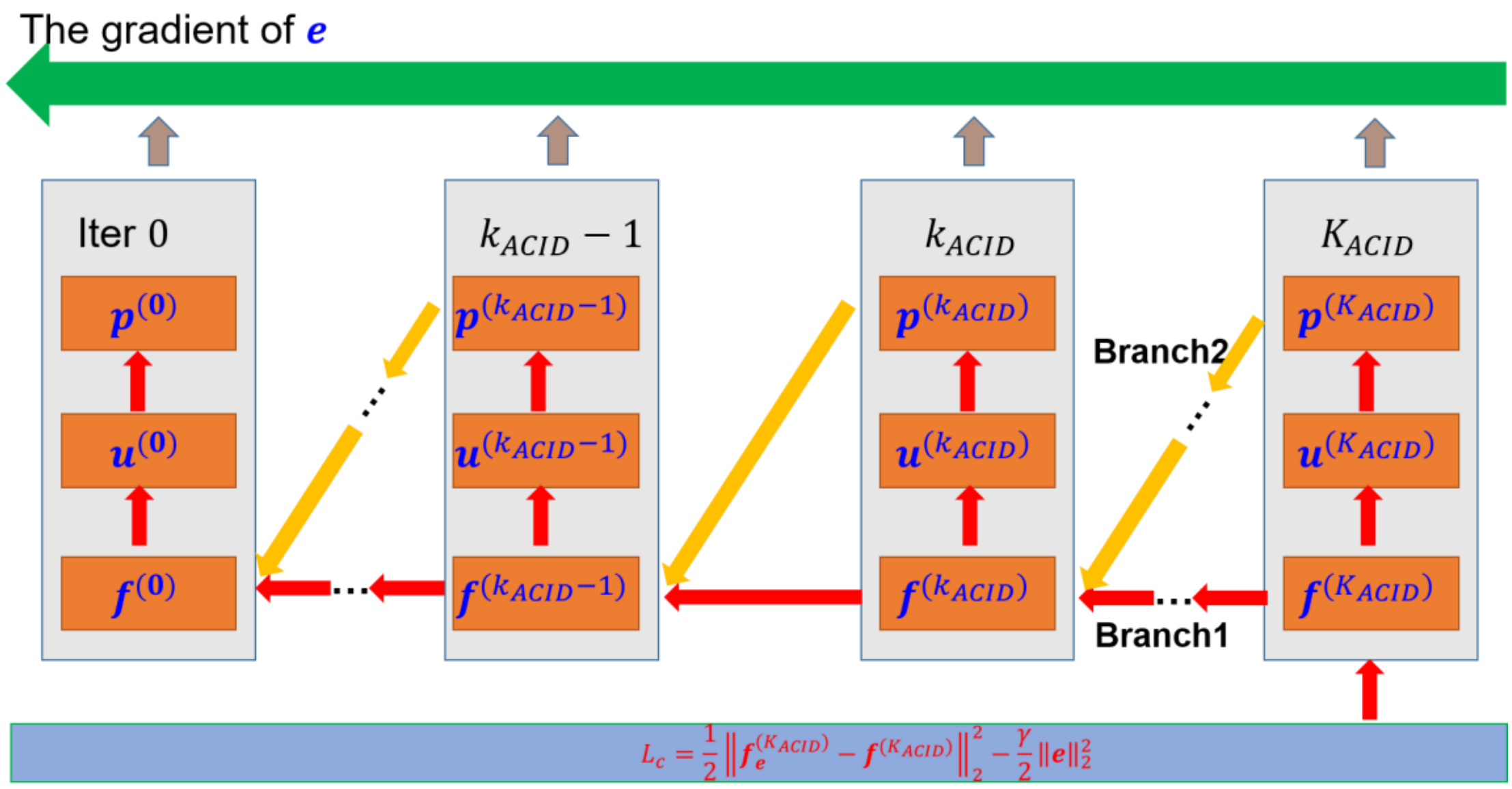

Fig. S13. Backpropagation process of ACID.

Because $dL_{\mathrm{ACID}}/d\boldsymbol{e}$ with the loss function $L_{\mathrm{ACID}} = \frac{1}{2}\|\boldsymbol{f}_e - \boldsymbol{f}\|_2^2$ is complicated, we cannot directly compute the gradient of $L_c$. Fortunately, $dL_c/d\boldsymbol{e}$ can be solved using the backpropagation algorithm [77, 78], which is commonly used in deep learning [79, 80]. Then, $dL_c/d\boldsymbol{e}$ can be split as

$$\frac{dL_C}{d\boldsymbol{e}} = \frac{d(\frac{1}{2}\|\boldsymbol{f}_e - \boldsymbol{f}\|_2^2)}{d\boldsymbol{e}} - \gamma\boldsymbol{e} = \frac{d(L_{ACID})}{d\boldsymbol{e}} - \gamma\boldsymbol{e}, \tag{S.3.13}$$

Now, let us start with the backpropagation process for ACID, as shown in Fig. S13. First, we can decompose the ACID system into the three modules keyed to $\boldsymbol{p}$, $\boldsymbol{u}$ and $\boldsymbol{f}$ respectively, where $\boldsymbol{u} = \Phi(\boldsymbol{p})$, and the whole procedure is shown in Fig. S14. The input and the output of $\boldsymbol{p}$, $\boldsymbol{u}$ and $\boldsymbol{f}$ are denoted as $\boldsymbol{p}_i, \boldsymbol{u}_i, \boldsymbol{f}_i, \boldsymbol{p}_o, \boldsymbol{u}_o$ and $\boldsymbol{f}_o$, respectively. Also, the gradient of $\boldsymbol{p}$, $\boldsymbol{u}$ and $\boldsymbol{f}$ can be denoted as $\frac{d\boldsymbol{p}_o}{d\boldsymbol{p}_i}, \frac{d\boldsymbol{u}_o}{d\boldsymbol{u}_i}$ and $\frac{d\boldsymbol{f}_o}{d\boldsymbol{f}_i}$, respectively.

Following the same steps as that in Algorithm 2, we will use the gradient ascent method to iteratively compute adversarial attacks for the whole ACID system, and the target to be attacked will be changed from a single unstable neural network to our whole ACID workflow. There are two iterative loops: the outer loop is for gradient ascent search, and the inner loop is for ACID feedforward and backpropagation. The stopping criteria of finding an adversarial attack for the whole ACID include (a) the number of iterations reaches the maximum number of iterations for computing an adversarial attack (AA) denoted as $K_{AA}$; or (b) the noise strength of the adversarial attack is greater than that used in attacking the single neural network recorded in our study in terms of the $L_2$-norm. As mentioned above, the maximum number of iterations of the inner loop is $K_{ACID}$ for the ACID feedforward process. Because each whole inner loop can be considered as an intermediate node, we can use the idea of backpropagation to search for a desirable perturbation. To find a tiny perturbation for the whole ACID, the procedure is summarized in Algorithm 3. To generate such tiny perturbations for ACID using the gradient ascent approach, the learning rate ($\vartheta > 0$) needs to be optimized in experiments, and $\boldsymbol{e_0}$ is initialed to random noise. More details can be found at https://zenodo.org/record/5497811.

Algorithm 3: Adversarial attacking approach for ACID.

Input: Image $\boldsymbol{f}$, neural network $\Phi$, system matrix $\boldsymbol{A}$, maximum number of iterations $K_{AA}$;
Initialization: $\boldsymbol{p} \leftarrow \boldsymbol{Af}$, $k_{AA} \leftarrow 0$, $\gamma > 0$ and $\vartheta > 0$, $\boldsymbol{e_0}$, $D^l_{\boldsymbol{p}^{(0)}}(\boldsymbol{e})$ is defined by (S.3.9);
while the stopping criteria are not satisfied

$\boldsymbol{e}^{(k_{AA}+1)} = \boldsymbol{e}^{(k_{AA})} + \vartheta \nabla_{\boldsymbol{e}} \boldsymbol{D}^{\boldsymbol{l}}_{\boldsymbol{p}^{(1)}+\boldsymbol{Ae}}\left(\boldsymbol{e}^{(k_{AA})}\right)$;

$k_{aa} = k_{aa} + 1$;

return $\boldsymbol{e}^{(k_{AA})}$;
Output: Perturbation $\boldsymbol{e}^{(k_{AA})}$.

For Algorithm 3, the key is to compute $\nabla_{\boldsymbol{e}} \boldsymbol{D}^{\boldsymbol{l}}_{\boldsymbol{p}^{(1)}+\boldsymbol{Ae}}\left(\boldsymbol{e}^{(k_{AA})}\right)$ using the backpropagation method. In fact, such a gradient computation could be divided into the following three stages, *i.e.*, initial stage, intermediate stage and final stage.

**Initial Stage (1):** The inner iteration index reaches its maximum value, that is $k_{ACID} = K_{ACID}$. Without causing any confusion, here we omit the index $k_{AA}$ , and simply write $\boldsymbol{e}^{(k_{AA})}$ as $\boldsymbol{e}$. To enable the backpropagation, we perform the following tasks.

**(1.a)** Compute $\frac{d(L_{ACID})}{d\left(\boldsymbol{f}_{\boldsymbol{e}}^{(K_{ACID})}\right)}$. By definition, $\boldsymbol{f}_{\boldsymbol{e}}^{(K_{ACID})} = \boldsymbol{f}_{\boldsymbol{o}}^{(K_{ACID})}$, we have

$$\frac{d(L_{ACID})}{d\left(\boldsymbol{f}_{\boldsymbol{e}}^{(K_{ACID})}\right)} = \frac{d(L_{ACID})}{d\left(\boldsymbol{f}_{o}^{(K_{ACID})}\right)} = \boldsymbol{f}_{\boldsymbol{e}}^{(K_{ACID})} - \boldsymbol{f}^{(K_{ACID})}. \quad \text{(S.3.14)}$$

**(1.b)** Compute $\frac{d(L_{ACID})}{\boldsymbol{f}_{i}^{(K_{ACID})}}$. Since the module $\boldsymbol{f}$ is complicated, we compute the gradient pixel by pixel:

$$\frac{d(L_{ACID})}{d\left(\boldsymbol{f}_{i}^{(K_{ACID})}(j_1,j_2)\right)} = \sum_{s_2=1}^{J_2} \sum_{s_1=1}^{J_1} \frac{d(L_{ACID})}{d\left(\boldsymbol{f}_{o}^{(K_{ACID})}(s_1,s_2)\right)} \times \frac{d\left(\boldsymbol{f}_{o}^{(K_{ACID})}(s_1,s_2)\right)}{d\left(\boldsymbol{f}_{i}^{(K_{ACID})}(j_1,j_2)\right)}, \quad \text{(S.3.15)}$$

where the gradient $\frac{d\left(\boldsymbol{f}_o^{(K_{ACID})}(s_1,s_2)\right)}{d\left(\boldsymbol{f}_i^{(K_{ACID})}(j_1,j_2)\right)}$ is provided by $\frac{d\left(\boldsymbol{f}_o^{(K_{ACID})}\right)}{d\left(\boldsymbol{f}_i^{(K_{ACID})}\right)}$, $s_1, s_2$ and $j_1, j_2$ are the pixel indices, $s_1 = 1, \dots, J_1$, $j_1 = 1, \dots, J_1$, $s_2 = 1, \dots, J_2$, $j_2 = 1, \dots, J_2$, and $J_1$ and $J_2$ are the height and width of the reconstructed image.

**(1.c)** By definition, we have

$$\boldsymbol{f}_i^{(K_{ACID})} = \boldsymbol{u}_o^{(K_{ACID})} + \boldsymbol{f}_o^{(K_{ACID}-1)}, \tag{S.3.16}$$

Therefore,

$$\begin{cases} \frac{d(L_{ACID})}{d\left(\boldsymbol{u}_o^{(K_{ACID})}\right)} = \frac{d(L_{ACID})}{d\left(\boldsymbol{f}_i^{(K_{ACID})}\right)} \\ \frac{d(L_{ACID})}{d\left(\boldsymbol{f}_o^{(K_{ACID}-1)}\right)}|_{Branch=1} = \frac{d(L_{ACID})}{d\left(\boldsymbol{f}_i^{(K_{ACID})}\right)} \end{cases}. \tag{S.3.17}$$

**(1.d)** Compute the change in the loss of ACID with respect to $\boldsymbol{u}_i^{(K_{ACID})}\big((j_1,j_2)\big)$, i.e., $\frac{d(L_{ACID})}{d\left(\boldsymbol{u}_i^{(K_{ACID})}((j_1,j_2))\right)}$. Using the chain rule, we have

$$\frac{d(L_{ACID})}{d\left(\boldsymbol{u}_i^{(K_{ACID})}(j_1,j_2)\right)} = \sum_{s_2=1}^{J_2}\sum_{s_1=1}^{J_1} \frac{d(L_{ACID})}{d\left(\boldsymbol{u}_o^{(K_{ACID})}(s_1,s_2)\right)} \cdot \frac{d\left(\boldsymbol{u}_o^{(K_{ACID})}(s_1,s_2)\right)}{d\left(\boldsymbol{u}_i^{(K_{ACID})}(j_1,j_2)\right)}. \tag{S.3.18}$$

It can be simplified as

$$\frac{d(L_{ACID})}{d\left(\boldsymbol{u}_i^{(K_{ACID})}\right)} = \frac{d(L_{ACID})}{d\left(\boldsymbol{p}_o^{(K_{ACID})}\right)}. \tag{S.3.19}$$

**(1.e)** Because $\boldsymbol{p}_o^{(K_{ACID})} = \boldsymbol{A}^*\left(\frac{\lambda\left(\boldsymbol{A}(\boldsymbol{f}+\boldsymbol{e})-\boldsymbol{A}\boldsymbol{f}_o^{(K_{ACID}-1)}\right)}{1+\lambda}\right)$, we have

$$\frac{d(L_{ACID})}{d\boldsymbol{e}}|_{Iteration=K_{ACID}} = \frac{\lambda}{1+\lambda}\left(\boldsymbol{A}^*\boldsymbol{A}\frac{d(L_{ACID})}{d\boldsymbol{p}_o^{(K_{ACID})}}\right), \tag{S.3.20}$$

and

$$\frac{d(L_{ACID})}{d\left(\boldsymbol{f}_o^{(K_{ACID}-1)}\right)}|_{Branch=2} = \frac{-\lambda}{1+\lambda}\left(\boldsymbol{A}^*\boldsymbol{A}\frac{d(L_{ACID})}{d\left(\boldsymbol{p}_o^{(K_{ACID})}\right)}\right). \tag{S.3.21}$$

**(1.f)** Combining (S.3.7) and (S.3.10), we obtain

$$\frac{dL_{ACID}}{d\left(\boldsymbol{f}_o^{(K_{ACID}-1)}\right)} = \frac{dL_{ACID}}{d\left(\boldsymbol{f}_o^{(K_{ACID}-1)}\right)}|_{Branch=1} + \frac{dL_{ACID}}{d\left(\boldsymbol{f}_o^{(K_{ACID}-1)}\right)}|_{Branch=2}. \tag{S.3.22}$$

**Intermediate Stage (2):** For the iteration index 0< $k_{ACID}$ <$K_{ACID}$, we perform the following tasks:

**(2.a)** $\frac{d(L_{ACID})}{d\left(\boldsymbol{f}_o^{(k_{ACID})}\right)}$ has been obtained in the previous stage. According to $\boldsymbol{f}_i^{(k_{ACID})} = \boldsymbol{u}_o^{(k_{ACID})} + \boldsymbol{f}_o^{(k_{ACID}-1)}$, and we compute

$$\frac{d(L_{ACID})}{d\left(\boldsymbol{f}_i^{(k_{ACID})}(j_1,j_2)\right)} = \sum_{s_2=1}^{J_2}\sum_{s_1=1}^{J_1} \frac{d(L_{ACID})}{d\left(\boldsymbol{f}_o^{(k_{ACID})}(s_1,s_2)\right)} \cdot \frac{d\left(\boldsymbol{f}_o^{(k_{ACID})}(s_1,s_2)\right)}{d\left(\boldsymbol{f}_i^{(k_{ACID})}(j_1,j_2)\right)}. \tag{S.3.23}$$

Then, we have

$$\frac{d(L_{ACID})}{d\left(\boldsymbol{f}_i^{(k_{ACID})}(j_1,j_2)\right)} = \frac{d(L_{ACID})}{d\left(\boldsymbol{f}_o^{(k_{ACID}-1)}(j_1,j_2)\right)}|_{Branch=1} = \frac{d(L_{ACID})}{d\left(\boldsymbol{u}_o^{(K_{ACID})}((j_1,j_2))\right)}. \tag{S.3.24}$$

**(2.b)** Compute $\frac{d(L_{ACID})}{d\left(\boldsymbol{u}_i^{(k_{ACID})}(j_1,j_2)\right)}$ as follows:

$$\frac{d(L_{ACID})}{d\left(\boldsymbol{u}_i^{(k_{ACID})}(j_1,j_2)\right)} = \sum_{s_2=1}^{J_2}\sum_{s_1=1}^{J_1} \frac{d(L_{ACID})}{d\left(\boldsymbol{u}_o^{(k_{ACID})}(s_1,s_2)\right)} \cdot \frac{d\left(\boldsymbol{u}_o^{(k_{ACID})}(s_1,s_2)\right)}{d\left(\boldsymbol{u}_i^{(k_{ACID})}(j_1,j_2)\right)}. \quad \text{(S.3.25)}$$

Since $\boldsymbol{u}_i^{(k_{ACID})} = \boldsymbol{p}_o^{(k_{ACID})}$, we have

$$\frac{d(L_{ACID})}{d\left(\boldsymbol{p}_o^{(k_{ACID})}\right)} = \frac{d(L_{ACID})}{d\left(\boldsymbol{u}_i^{(k_{ACID})}\right)}. \quad \text{(S.3.26)}$$

**(2.c)** With the definition of $\boldsymbol{p}_o^{(k_{ACID})}$, $\frac{d(L_{ACID})}{d\mathbf{e}}$ in the $k_{ACID}$th iteration is computed as

$$\frac{d(L_{ACID})}{d\boldsymbol{e}}|_{Itertion=k_{ACID}} = \frac{\lambda}{1+\lambda+\mu}\left(\boldsymbol{A}^*\boldsymbol{A}\frac{d(L_{ACID})}{d\left(\boldsymbol{p}_o^{(k_{ACID})}\right)}\right). \quad \text{(S.3.27)}$$

That is,

$$\frac{d(L_{ACID})}{d\left(\boldsymbol{f}_o^{(k_{ACID}-1)}\right)}|_{Branch=2} = \frac{-\lambda}{1+\lambda+\mu}\left(\boldsymbol{A}^*\boldsymbol{A}\frac{d(L_{ACID})}{d\left(\boldsymbol{p}_o^{(k_{ACID})}\right)}\right). \quad \text{(S.3.28)}$$

**(2.d)** Finally, we have

$$\frac{d(L_{ACID})}{d\left(\boldsymbol{f}_o^{(k_{ACID}-1)}\right)} = \frac{d(L_{ACID})}{d\left(\boldsymbol{f}_o^{(k_{ACID}-1)}\right)}|_{Branch=1} + \frac{d(L_{ACID})}{d\left(\boldsymbol{f}_o^{(k_{ACID}-1)}\right)}|_{Branch=2}. \quad \text{(S.3.29)}$$

**Final Stage (3):** In this stage (i.e., the inner iteration equal to 0, $k_{ACID} = 0$). We need to finish the following tasks:

**(3.a)** $\frac{d(L_{ACID})}{d\left(\boldsymbol{f}_o^{(k_{ACID})}\right)}$ has been obtained from the previous iteration, and $\frac{d(L_{ACID})}{d\left(\boldsymbol{f}_i^{(k_{ACID})}(j_1,j_2)\right)}$ is computed as

$$\frac{d(L_{ACID})}{d\left(\boldsymbol{f}_i^{(k_{ACID})}(j_1,j_2)\right)} = \sum_{s_2=1}^{J_2}\sum_{s_1=1}^{J_1} \frac{d(L_{ACID})}{d\left(\boldsymbol{f}_o^{(k_{ACID})}(s_1,s_2)\right)} \cdot \frac{d\left(\boldsymbol{f}_o^{(k_{ACID})}(s_1,s_2)\right)}{d\left(\boldsymbol{f}_i^{(k_{ACID})}(j_1,j_2)\right)}. \quad \text{(S.3.30)}$$

Furthermore,

$$\frac{d(L_{ACID})}{d\left(\boldsymbol{u}_o^{(k_{ACID})}((j_1,j_2))\right)} = \frac{d(L_{ACID})}{d\left(\boldsymbol{f}_i^{(k_{ACID})}(j_1,j_2)\right)}. \quad \text{(S.3.31)}$$

**(3.b)** When $k_{ACID} = 0$, since $\boldsymbol{p}_0 = \boldsymbol{A}^*(\boldsymbol{A}(\boldsymbol{f}+\boldsymbol{e}))$, we have

$$\frac{d(L_{ACID})}{d\boldsymbol{e}}|_{Iteration=0} = \boldsymbol{A}^*\boldsymbol{A}\frac{d(L_{ACID})}{d(\boldsymbol{p}_0)}. \quad \text{(S.3.32)}$$

Finally, we compute $\frac{d(L_C)}{d\boldsymbol{e}}$ as follows:

$$\frac{d(L_C)}{d\boldsymbol{e}} = \frac{d(L_{ACID})}{d\boldsymbol{e}} - \gamma\boldsymbol{e} = \sum_{k_{ACID}=0}^{K_{ACID}}\frac{d(L_{ACID})}{d\boldsymbol{e}}|_{Itertion=k_{ACID}} - \gamma\boldsymbol{e}. \quad \text{(S.3.33)}$$

**Gradient with the $\boldsymbol{u}$ module:** The input and the output of the $\boldsymbol{u}^{(k_{ACID})}$ module are $\boldsymbol{u}_i^{(k_{ACID})}$ and $\boldsymbol{u}_o^{(k_{ACID})}$ respectively. $\boldsymbol{u}_o^{(k_{ACID})}$ is computed with $\boldsymbol{u}_o^{(k_{ACID})} = \Phi\left(\boldsymbol{u}_i^{(k_{ACID})}\right)$, where $\Phi$ is a well-trained neural network. Note that, for a deep neural network, such as AUTOMAP, $\boldsymbol{p}_o^{(k_{ACID})} = \boldsymbol{u}_i^{(k_{ACID})} = \boldsymbol{u}_o^{(k_{ACID})}$. $\frac{d\left(\boldsymbol{u}_o^{(k_{ACID})}\right)}{d\left(\boldsymbol{u}_i^{(k_{ACID})}\right)}$ can be obtained directly with the deep-learning-based framework; e.g., TensorFlow (https://www.tensorflow.org/) or PyTorch (https://pytorch.org/). In the CT case, the Ell-50 network was used on the MATLAB platform, and we need to modify the loss function of FBPConvNet so that we can search for the adversarial attack to the whole ACID workflow. More details can be found via the shared link https://zenodo.org/record/5497811.

**Gradient with the $\boldsymbol{f}$ module:** The input and the output of the $\boldsymbol{f}$ module are $\boldsymbol{f}_i^{(k_{ACID})}$ and $\boldsymbol{f}_o^{(k_{ACID})}$. According to the analysis in the main text, we can compute $\boldsymbol{f}_o^{(k_{ACID})}$ by minimizing the following objective function:

$$\boldsymbol{f}_o^{(k_{ACID})} = \underset{\boldsymbol{f}}{argmin}\frac{1}{2}\left\|\boldsymbol{f} - \boldsymbol{f}_o^{(k_{ACID}-1)} - \frac{1}{\lambda}\boldsymbol{u}_o^{(k_{ACID})}\right\|_2^2 + \frac{\xi}{\lambda}\|\boldsymbol{Hf}\|_1,$$
$$\text{s.t.}\,\boldsymbol{f}_i^{(k_{ACID})} = \boldsymbol{f}_o^{(k_{ACID}-1)} + \frac{1}{\lambda}\boldsymbol{\mu}_o^{(k_{ACID})}, \quad \text{(S.3.34)}$$

Let $\bar{\boldsymbol{f}} = \boldsymbol{Hf}$, (S.3.34) can be converted to the following problem:

$$\boldsymbol{f}_o^{(k_{ACID})} = \underset{\boldsymbol{f}}{argmin}\frac{1}{2}\left\|\boldsymbol{H}^*\bar{\boldsymbol{f}} - \boldsymbol{f}_o^{(k_{ACID}-1)} - \frac{1}{\lambda}\boldsymbol{u}_o^{(k_{ACID})}\right\|_2^2 + \frac{\xi}{\lambda}\|\bar{\boldsymbol{f}}\|_1, \quad \text{(S.3.35)}$$

where $\boldsymbol{H}^*$ satisfies $\boldsymbol{H}^*\boldsymbol{H} = \boldsymbol{I}$. Thus, $\boldsymbol{f}_o^{(k_{ACID})} = \boldsymbol{H}^* S_{\frac{\xi}{\lambda}}\left(\boldsymbol{H}\left(\boldsymbol{f}_o^{(k_{ACID}-1)} + \frac{1}{\lambda}\boldsymbol{u}_o^{(k_{ACID})}\right)\right)$, where $S_\xi$ represents the soft thresholding operator [81], which can be written as

$$S_\xi(x) = \begin{cases} 0, & |x| < \xi \\ x - \text{sgn}(x)\xi & \text{otherwise} \end{cases}, \quad \text{(S.3.36)}$$

Then, we can obtain $\frac{d\left(\boldsymbol{f}_o^{(k_{ACID})}\right)}{d\left(\boldsymbol{f}_i^{(k_{ACID})}\right)}$ using the chain rule after combining (S.3.35) and (S.3.36).

## III.C. Lipschitz Convergence with Perturbations

Let the combination of the measurement matrix $\boldsymbol{A}$ and the neural network $\Phi(\cdot)$ be $\Phi_{\boldsymbol{A}}(\cdot)$. According to the definition of the Lipschitz constant, if we employ the $L_2$ norm, the Lipschitz constant is the minimal constant that holds for the following inequality:

$$\|\Phi_{\boldsymbol{A}}(\boldsymbol{f}) - \Phi_{\boldsymbol{A}}(\boldsymbol{f}')\| \le L\|\boldsymbol{f} - \boldsymbol{f}'\|. \quad \text{(S.3.37)}$$

For each fixed $\boldsymbol{f}$, we generated a series of perturbations to obtain $\boldsymbol{f}'$, and computed the value of the ratio $\|\Phi_{\boldsymbol{A}}(\boldsymbol{f}) - \Phi_{\boldsymbol{A}}(\boldsymbol{f}')\|/\|\boldsymbol{f} - \boldsymbol{f}'\|$. Specifically, we computed for many images and found the upper and lower bounds of the Lipschitz constant as our empirically estimated ranges in the CT and MRI cases, respectively. Note that the authors of AUTOMAP did not provide the original data and code for sufficient training and testing, we only performed this experiment on the DAGAN and Ell-50. Here, only 500 pairs of ellipse phantoms were used for Ell-50. Each pair contains $\boldsymbol{f}$ and $\boldsymbol{f}'$, where $\boldsymbol{f}'$ was generated by adding an adversarial attack on the clear image $\boldsymbol{f}$ using the aforementioned adversarial method. Specifically, the lower and upper bounds in the Ell-50 case are 0.4674 and 0.6424, respectively. In contrast, 14,866 pairs of MRI images were used to determine the lower and upper bounds in the DAGAN case, and the corresponding lower and upper bounds are 1.4854 and 12.0737, respectively.

We had shown the convergence of ACID with respect to PSNR in the main text, and here we show the convergence of ACID in terms of the Lipschitz constant with respect to the number of iterations. As representative examples, the convergence curves in the C3 and M4 cases are given in Fig. S14. It can be observed that the Lipschitz constant of ACID for both CT and MRI are monotonically decreasing and finally converge to a constant scale.

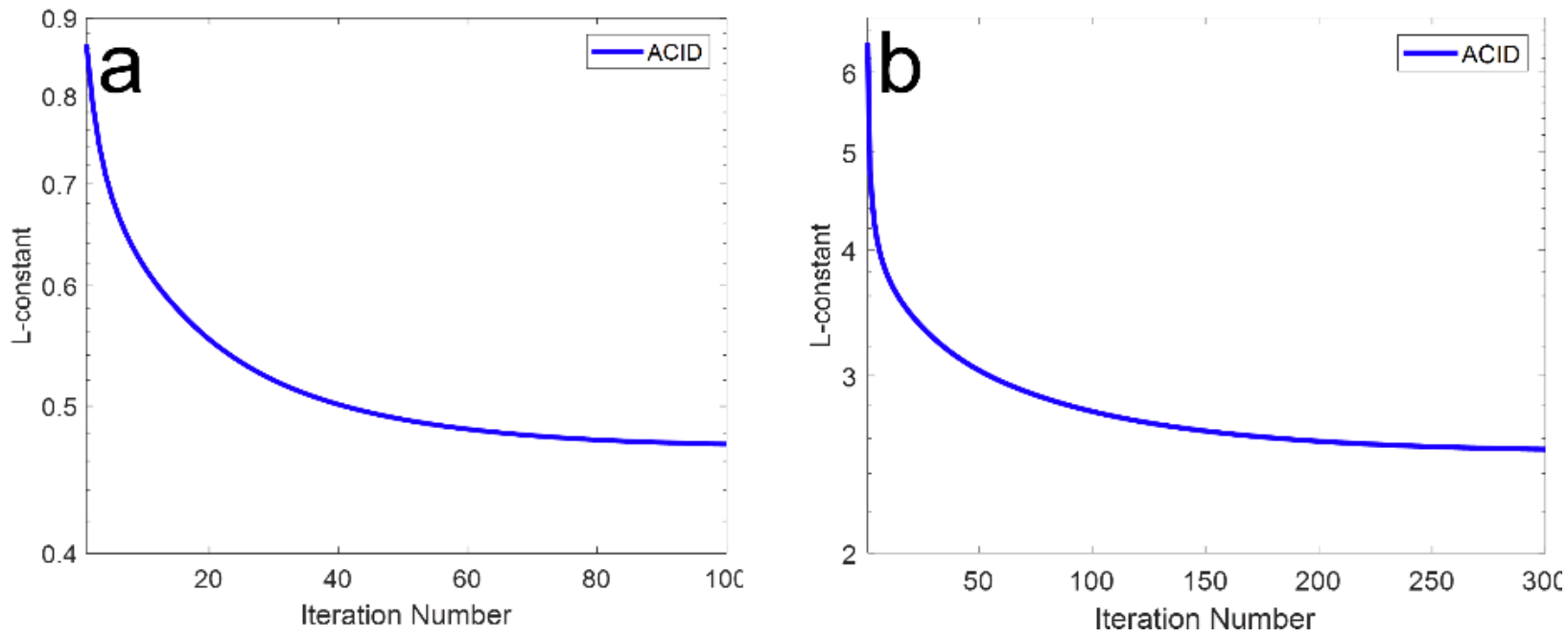

Fig. S14. Convergence curves of ACID in terms of the Lipschitz constant. (a) and (b) show the convergence curves of ACID with respect to the number of iterations in the C3 and M4 cases, respectively.

## III.D. ACID Against Noise

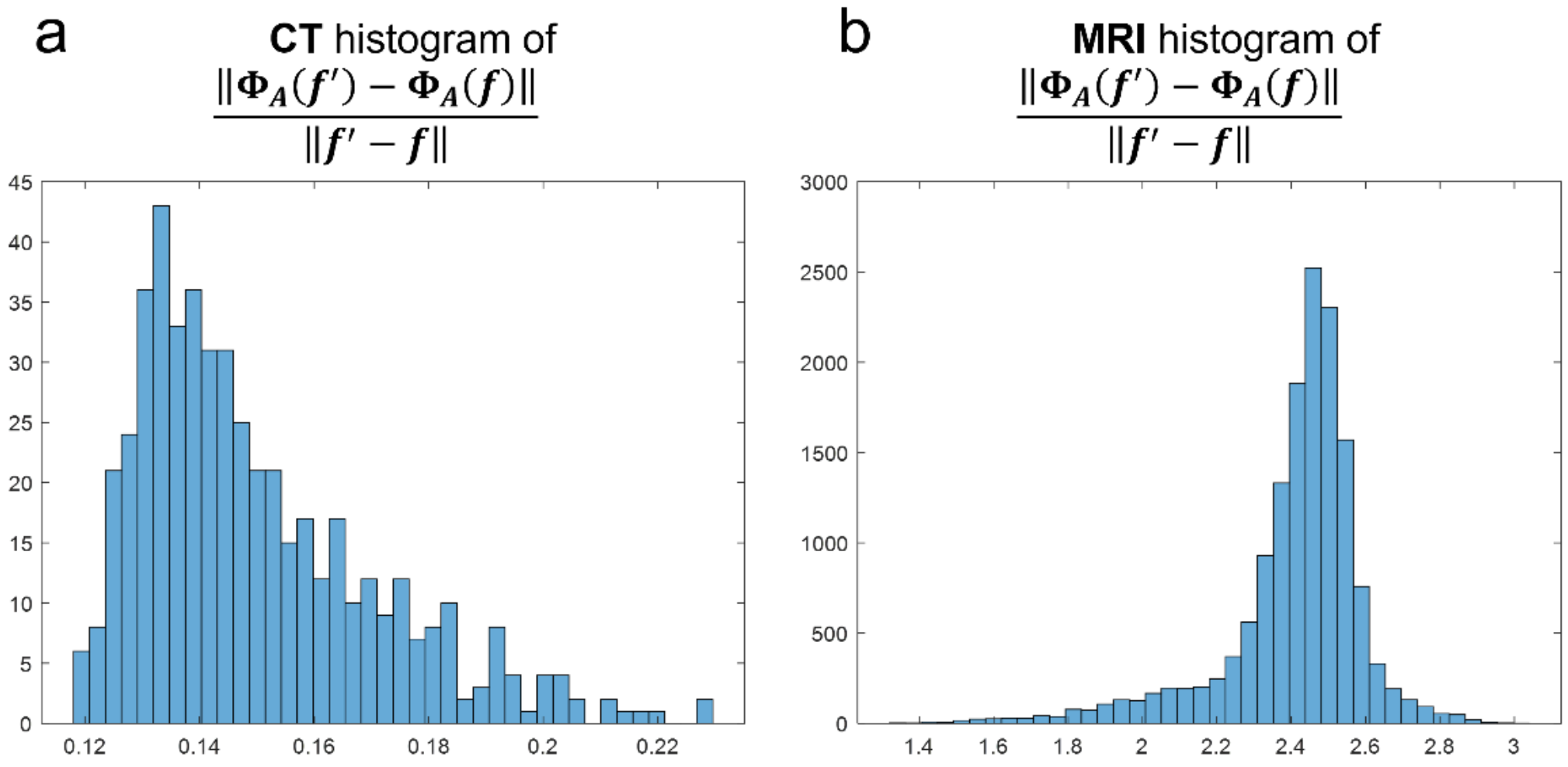

Fig. S15. ACID is locally stable with respect to noise. (a) The histogram of the output-to-input ratio between noise-free and Gaussian input data, where ACID has Ell-50 built-in, giving the maximum value of 0.229. (b) the histogram of the output-to-input ratio between noise-free and Gaussian input data, where ACID has DAGAN embedded, with the maximum ratio of 3.023.

Although some examples about the insensitivity of ACID against noise are reported in the main text, here we followed up the study in [82] and performed a similar local stability test on ACID with DAGAN and Ell-50 respectively built-in. This local robustness was assessed using the maximum ratio between variations in the output space and variations in the imaging object space: $\|\Phi_{\boldsymbol{A}}(\boldsymbol{f}) - \Phi_{\boldsymbol{A}}(\boldsymbol{f}')\| / \|\boldsymbol{f} - \boldsymbol{f}'\|$ for two adjacent images $\boldsymbol{f}$ and $\boldsymbol{f}'$. In this test, 500 pairs of CT phantoms were selected from the Ell-50 test dataset. Then, the additive white Gaussian noise was added, with zero mean and standard derivation 11-30 HU. In this way, we obtained 500 cases. Furthermore, 14,866 image pairs were chosen from the MRI dataset. Similarly, the additive white Gaussian noise was randomly added to each of these images to generate $\boldsymbol{f}'$. A maximum output-input variation ratio of 3.023 was observed for these noisy inputs. The histograms in the CT and MRI cases are given in Fig. S15. The results empirically demonstrate the local stability of the ACID reconstruction against noise.

# IV. Theoretical Analyses on the ACID Workflow

## IV.A. Kernel Awareness & Network Stability

The "kernel awareness" is an important concept. When a reconstruction algorithm lacks the kernel awareness, a "cardinal crime" ("cardinal sin") could be committed [12], which implies that a well-trained network model would potentially produce highly unstable results, defeating the purpose of medical imaging. In that scenario, the trained network would produce significantly different images from essentially identical input datasets, between which there are subtle differences representing invisible perturbations.

Specifically, a deep network is trained on a dataset $D$ using an optimization technique. The learning procedure would normally converge to a network model with optimized parameters, which is usually a continuous transform $T$ such that for $\boldsymbol{f}, \boldsymbol{f}' \in D$

$$\|T(\boldsymbol{Af}) - \boldsymbol{f}\| < \tau, \ \|T(\boldsymbol{Af}') - \boldsymbol{f}'\| < \tau, \tag{S.4.1}$$

where $\|\cdot\|$ is a suitable norm, $\boldsymbol{A}$ is a measurement matrix, and a constant $\tau$ is a bound. To evaluate the stability of the network model, an ε-Lipschitz metric is defined as follows:

$$L^{\varepsilon}(T, \boldsymbol{p}) = \sup_{0<\|\boldsymbol{p}-\boldsymbol{p}'\|<\varepsilon} \frac{\|T(\boldsymbol{p})-T(\boldsymbol{p}')\|}{\|\boldsymbol{p}-\boldsymbol{p}'\|}. \tag{S.4.2}$$

A formula can be derived for a lower bound of the ε-Lipschitz index estimation for $\boldsymbol{p} = \boldsymbol{Af}$ :

$$L^{\varepsilon}(T, \boldsymbol{p}) > \frac{1}{\varepsilon}(\|\boldsymbol{f}' - \boldsymbol{f}\| - 2\tau). \tag{S.4.3}$$

In fact,

$$L^{\varepsilon}(T, \boldsymbol{p}) \geq \frac{\|T(\boldsymbol{Af}') - T(\boldsymbol{Af})\|}{\|\boldsymbol{Af}' - \boldsymbol{Af}\|} \geq \frac{\|\boldsymbol{f}' - \boldsymbol{f}\| - \|T(\boldsymbol{Af}) - \boldsymbol{f}\| - \|T(\boldsymbol{Af}') - \boldsymbol{f}'\|}{\|\boldsymbol{Af}' - \boldsymbol{Af}\|}$$
$$> \frac{1}{\varepsilon}(\|\boldsymbol{f}' - \boldsymbol{f}\| - 2\tau), \tag{S.4.4}$$

for $\varepsilon \geq \tau$. An inverse problem, such as few-view CT and sparse MRI, involves solving: $\boldsymbol{Af} = \boldsymbol{p} + \varepsilon$ , where $\boldsymbol{A}$ is an $m \times N$ matrix, $m < N$ , and $\varepsilon$ is measurement noise. Clearly, the transform $\boldsymbol{A}$ would have a null space (kernel) with $\dim(Null(\boldsymbol{A})) > 1$ . Then, there is a non-zero vector $\boldsymbol{f_0} \in neighbor\ of\ Null(\boldsymbol{A})$ and a scale factor $\sigma$ for a large number $L$ such that

$$\|\boldsymbol{Af}_0\| < \varepsilon, \|\boldsymbol{Af}' - \boldsymbol{p}\| < \varepsilon, \text{ and } \|\boldsymbol{f}' - \boldsymbol{f}\| > L + 2\tau, \tag{S.4.5}$$

where $\boldsymbol{f}' = \boldsymbol{f} + \boldsymbol{\sigma f}_0$ with $\boldsymbol{f} \in (Null(\boldsymbol{A}))^{\perp}$. If the training set has at least two such elements $(\boldsymbol{Af}, \boldsymbol{f})$ and $(\boldsymbol{Af}', \boldsymbol{f}')$ , we have

$$L^{\varepsilon}(T, \boldsymbol{p}) > \frac{L}{\varepsilon}. \tag{S.4.6}$$

From (S.4.6), the instability is intrinsic; that is, when input data are very close to the null space of the associated imaging operator and $\boldsymbol{p}$ is slightly perturbed, a large variation would be induced in the reconstructed image. The instability of the trained network would yield artifacts in reconstructed images, subject to either false positive or false negative diagnosis.

## IV.B. Theoretical Background

The vulnerability of neural networks has been demonstrated with adversarial attacks in all major deep learning tasks, from misclassification examples to deep reconstruction instabilities. In their landmark paper, Antun *et al.* showed that deep reconstruction is unstable due to lack of kernel awareness but sparsity-promoting reconstruction does not have such a problem [6]. To address these instabilities, we design an Analytic, Compressed Iterative Deep (ACID) network. The key idea behind ACID is to combine data-driven priors and sparsity constraints to outperform either simple-minded deep reconstruction networks or established compressed sensing based reconstruction methods. In our study, we have not only experimentally shown the merits of ACID but also theoretically analyzed the rationale of ACID in terms of its converging behavior and solution characteristics. In the following, we put our analysis on ACID in the perspective of others' analyses on general computational optimization in general and existing representative image reconstruction networks in particular.

*There are profound results in non-computability in the field of computer science.* Computational optimization is important not only in the field of computer science but also to our real-world applications, from linear programming to deep learning. The theoretical research on this theme can be traced back to Turing's ground-breaking paper on machine intelligence and Smale's list of problems for the 21st century. Specifically, Smale's 9$^{th}$ problem is that "*Is there a polynomial time algorithm over the real numbers which decides the feasibility of the linear system of inequalities* $\boldsymbol{Ax} \geq \boldsymbol{y}$*?*" [83]. In practice, $\boldsymbol{A}$ and $\boldsymbol{y}$ can only be approximately obtained, and an extended (precision-limited) version of Smale's 9th problem is critically important. Recently, Bastounis, Hansen and Vlačić [84] made remarkable progress in settling this theoretical issue. Their theory bears a major implication for Smale's 18$^{th}$ problem about the boundary of artificial intelligence (AI), *i.e.*, deep learning as the current mainstream of AI. They show that it is in general non-computable to construct a neural network via loss minimization and apply it to testing data, and such a neural network is generally unstable. For example, there are in principle many classification problems for which "*one may have 100% success rate on arbitrarily large training and validation data sets, and yet there are uncountably many points arbitrarily close to the training data for which the trained network will fail*."

*Tomographic reconstruction is an important type of computational optimization problem, and, interestingly enough, deep networks for image reconstruction can and cannot be computed under different conditions.* In the context of these inverse problems, the article by Antun *et al*. reported instabilities of deep reconstruction networks [6] due to the lack of kernel awareness [12]. Then, a comprehensive follow-up analysis by Antun, Colbrook, and Hansen [85] established the boundary of deep learning inspired tomographic reconstruction, which helps address Smale's 18$^{th}$ problem. Among their contributions, the following three points are clearly made on (1) existence, (2) non-existence, and (3) the conditional existence of desirable networks. That

is, while the existence of neural networks is proved in the literature for an excellent functional representation, the non-existence is proved of any algorithm that trains or computes such a neural network in a general setting. However, the conditional existence is also proved of such an algorithm to compute an accurate and stable network that solves meaningful inverse problems such as Fourier imaging from sparse data. Specifically, the existence of a network for a universal representation is well known (Theorem 2.1 in [85]), but how to train a network to achieve an accurate and stable approximation is a difficult issue. It has been shown that a counterexample can always be found in a general setting so that the accuracy and robustness of a network cannot be simultaneously obtained (Theorem 2.2 in [85]). On the other hand, under certain conditions, such as sparsity in levels, an accurate and stable network can be indeed obtained (Theorems 5.5 and 5.10 in [85]), with the FIRENET network as a good example [85]. At the core of the construction of FIRENET is the kernel awareness. Clearly, training the network defined in Subsection 5.1 in [85] cannot obtain kernel awareness and is subject to the phase transition of solutions to the inverse problems. In other words, if the difference between the two images lies close to the null space of the measurement matrix and is bounded from below, the Lipschitz constant of the inverse mapping can be very large, yielding a poor imaging performance. Fortunately, an algorithm can be used to utilize sparsity in levels and find a stable and accurate neural network (Theorems 5.5 and 5.10 in [85], with uniform recovery guarantees, geometric convergence, bounds on the number of samples and the number of layers of a network for a pre-specified accuracy).

*In addition to the excellent work by Antun et al., active research efforts have been going on to develop deep networks for accurate and stable deep tomographic reconstruction.* Representative results include the LEARN ("*Learned Experts' Assessment-based Reconstruction Network*") network [86], ItNet network [87], Momentum-Net [88], null space network [89], as well as deep equilibrium networks [90].

In [86], an iterative reconstruction algorithm in the CS framework was unrolled and trained in an end-to-end fashion. The experimental results from the resultant LEARN network on the Mayo Clinic low-dose CT dataset are competitive with representative methods in terms of artifact reduction, feature preservation, and computational speed. In [87], an iterative deep-learning-based reconstruction network was designed to solve underdetermined inverse problems accurately and stably (ItNet shown in Fig. 1 in [87]). In comparison with total-variation minimization, their results reveal that standard end-to-end network architectures are not only resilient against statistical noise but also adversarial perturbations. In [88], another iterative neural network, referred to as Momentum-Net, was prototyped by combining data-driven regression and model-based image reconstruction (MBIR). Momentum-Net is convergent under reasonable conditions (quadratic majorization via M-Lipschitz continuous gradients). Their results show that Momentum-Net outperformed MBIR and several other networks, but the effect of adversarial attacks on Momentum-Net was not evaluated. In [89], a null space network was studied to offer a theoretical justification to deep learning-based tomographic reconstruction via so-called Φ-regularization. The convergence of the overall reconstruction workflow is proved, assuming a Lipschitz continuity and preserving the data consistency (illustrated in Fig. 1 in [89]). In [90], the deep equilibrium models were adapted to find the fixed point with guaranteed convergence under the ε-Lipschitz continuity. Subsequently, the trade-off can be made between reconstruction quality and computational cost.

*In connection with the above results, our ACID network has significant merits and unique features.* First, ACID is dedicated to overcoming the instabilities of neural networks in the same settings using the same datasets as that described in [6]. As a result, we have made a solid step forward along the direction of stabilizing deep reconstruction networks, showing that accurate and stable deep reconstruction is feasible, and remains an exciting research opportunity. Second, the ACID network is the first prototype that combines an established sparsity-oriented algorithm, a data-driven direct reconstruction network, and an iterative data fidelity enforcement (for example, LEARN [86] ignores data consistency; ItNet network [87] lacks kernel awareness, Momentum-Net [88] misses a learned mapping from data to images, null space network [89] uses no sparsity, and deep equilibrium networks [90] focuses only on the fixed point that does not imply image sparsity nor data fidelity). Third, the converging behavior and solution characteristics of ACID have been analyzed under a reasonable assumption. The assumption is called the bounded relative error norm (BREN), which is a special case of a Lipschitz continuity. The Lipschitz continuity we used in our convergence analysis, which is practically interpreted as the BREN property and experimentally verified in our study, is consistent with the previous studies on non-convex optimization such as in the aforementioned network convergence analyses in [88-90]. Furthermore, note that we do not request the measurement matrix must satisfy a compressed sensing condition such as the restricted isometry property. This means that a standard sparsity-promotion algorithm may not give a unique solution. In this case, ACID promises to outperform the sparsity-minimization reconstruction alone, because data-prior plays a significant role to fill in the gap in deep reconstruction. Last but not the least, in addition to an accurate reconstruction performance, ACID has stability in the two related aspects: (a) ACID can stabilize an unstable deep reconstruction network (by putting it in the ACID framework), and (b) ACID as a whole is resilient against adversarial attacks. Both aspects of the ACID stability have been systematically shown in our studies.

## IV.C. Bounded Relative Error Norm (BREN) Property

An $N$-dimensional signal $\boldsymbol{f}^* \in \mathbb{R}^N$ is called sparse if the number of its nonzero components is much smaller than the signal length $N$. Suppose that $\|\boldsymbol{f}^*\|_0$ denotes the number of nonzero components of the signal $\boldsymbol{f}^*$; if $\|\boldsymbol{f}^*\|_0 \leq s \ll N$, then we say the signal $\boldsymbol{f}^*$ is $s$-sparse. If a measurement matrix $\boldsymbol{A} \in \mathbb{R}^{m\times N}$ satisfies the Restricted Isometry Property (RIP) of order $1 \leq s \ll N$, an $s$-sparse vector $\boldsymbol{f}^* \in \mathbb{R}^N$ can be successfully and efficiently recovered from the measurement $\boldsymbol{A}\boldsymbol{f}^*$ [91], and the $s$-sparse vector $\boldsymbol{f}^*$ is observable. Our analysis requires the following bounded relative error norm (BREN) property of a reconstruction neural network to reconstruct $\boldsymbol{f}$ from measurement $\boldsymbol{p} = \boldsymbol{A}\boldsymbol{f}^*$. If a reconstruction network satisfies the BREN property, we call it a well-designed and well-trained reconstruction network, or a proper network.

**Definition**: A reconstruction network has the BREN property if the ratio between the $L_2$ norm of the reconstruction error and the $L_2$ norm of the corresponding ground truth is less than $(1-\sigma)$ with $0<\sigma<1$. For an s-sparse observable image $\boldsymbol{f}^*$, there are different ways to formulate the Lipschitz continuity such as our BREN property. Let us assume that the function $\Phi(\cdot)$ models a well-trained neural network. Denote the output of the neural network $\Phi(\boldsymbol{A}\boldsymbol{f}^*)=\boldsymbol{f}^*+\boldsymbol{f}^{ob}+\boldsymbol{f}^{nl}$, where the second and third terms are observable and null-space components of the error image associated with the ground-truth image $\boldsymbol{f}^*$ and the measurement matrix $\boldsymbol{A}$, the BREN property is defined as

$$\frac{\|\Phi(\mathbf{A}\boldsymbol{f}^*)-\boldsymbol{f}^*\|}{\|\boldsymbol{f}^*\|}=\frac{\|\boldsymbol{f}^{ob}+\boldsymbol{f}^{nl}\|}{\|\boldsymbol{f}^*\|}\leq(1-\sigma). \quad \text{(S.4.7)}$$

(S.4.7) implies that $\|\boldsymbol{f}^{ob}+\boldsymbol{f}^{nl}\|\leq(1-\sigma)\|\boldsymbol{f}^*\|$.

**Remark 1**: In the literature of deep imaging, including the paper on instabilities of deep reconstruction [6], a reconstruction network, even if it is unstable, will still produce an output not too far from the ground truth in the sense of the BREN property. The involved errors of types I and II have significant clinical impacts but the norm of these errors in combination is assumed to be small relative to that of the underlying image. This is how a proper reconstruction network is defined and commonly expected in practice. For example, the most popular loss function of a reconstruction network is in the $L_2$ norm so that a reconstructed image should be close to the ground truth in the sense of the $L_2$ norm without an adversarial attack. Furthermore, in the adversarial attack cases of our interest, the BREN property is assumed to be valid as the condition for our convergence analysis below.

**Remark 2:** For deep reconstruction in the supervised mode, a training dataset is typically in the format of $(\boldsymbol{p}(i),\boldsymbol{f}(i))$, $i=1,2,\cdots,I_{trn}$. We assume that the imaging model is linear, and we can augment the training dataset to $(\alpha\boldsymbol{p}(i),\alpha\boldsymbol{f}(i))$, $i=1,2,\cdots,I_{trn}$, where $\alpha$ is any constant within a reasonable range. With the augmented data, the network will map the input of a small norm to an output of a proportionally small norm. Alternatively, we can include the normalization layer(s) in the reconstruction network so that the network performance is insensitive to the magnitude of data and images.

**Remark 3**: Our assumption of the BREN property is needed for our convergence analysis below, just like the case for CS theory where RIP/rNSP is required for unique image recovery. If the requirement is not met, the theoretical arguments below will not be valid. We have shown that our BREN ratio is substantially less than 1 for the datasets in this PNAS study [6].

Specifically, all the experimental results with perturbations were repeated in the CT and MRI cases reported in [6]. Then, the BREN ratios were computed using different reconstruction networks with various perturbations. It is found that all these ratios in CT and MRI experiments are substantially less than 1. As shown in Table S1 and Figs. S16-19, the AUTOMAP seems more sensitive to the perturbations; *i.e.*, small perturbations cause large changes in the sense of the $L_2$-norm. Clearly, the BREN property is satisfied in this context.

Table S1. BREN ratios (%) associated with different reconstruction networks.

| Methods | $r_1$ | $r_2$ | $r_3$ | $r_4$ |
|---|---|---|---|---|
| Med-50 | 2.90 | x | x | x |
| AUTOMAP | 10.39 | 23.09 | 47.85 | 85.86 |
| Deep MRI | 2.73 | 8.03 | 13.28 | x |
| MRI-VN | 3.53 | x | x | x |

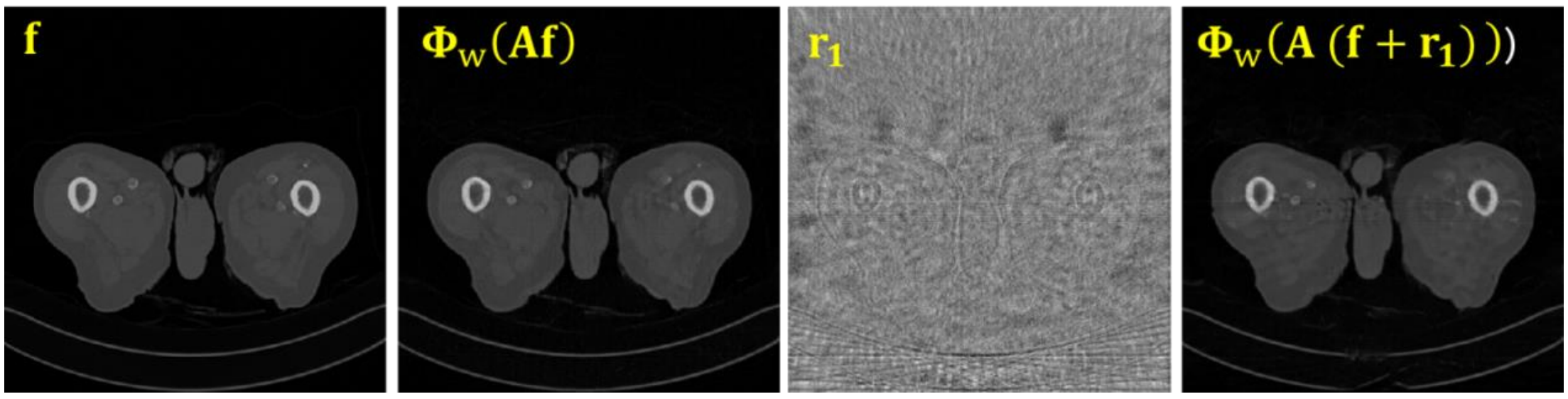

Fig. S16. Reconstruction results using MED-50 [1] from [6]. The 1st-4th images represent the original, MED-50 without perturbation, perturbation, and perturbed MED-50 results, respectively.

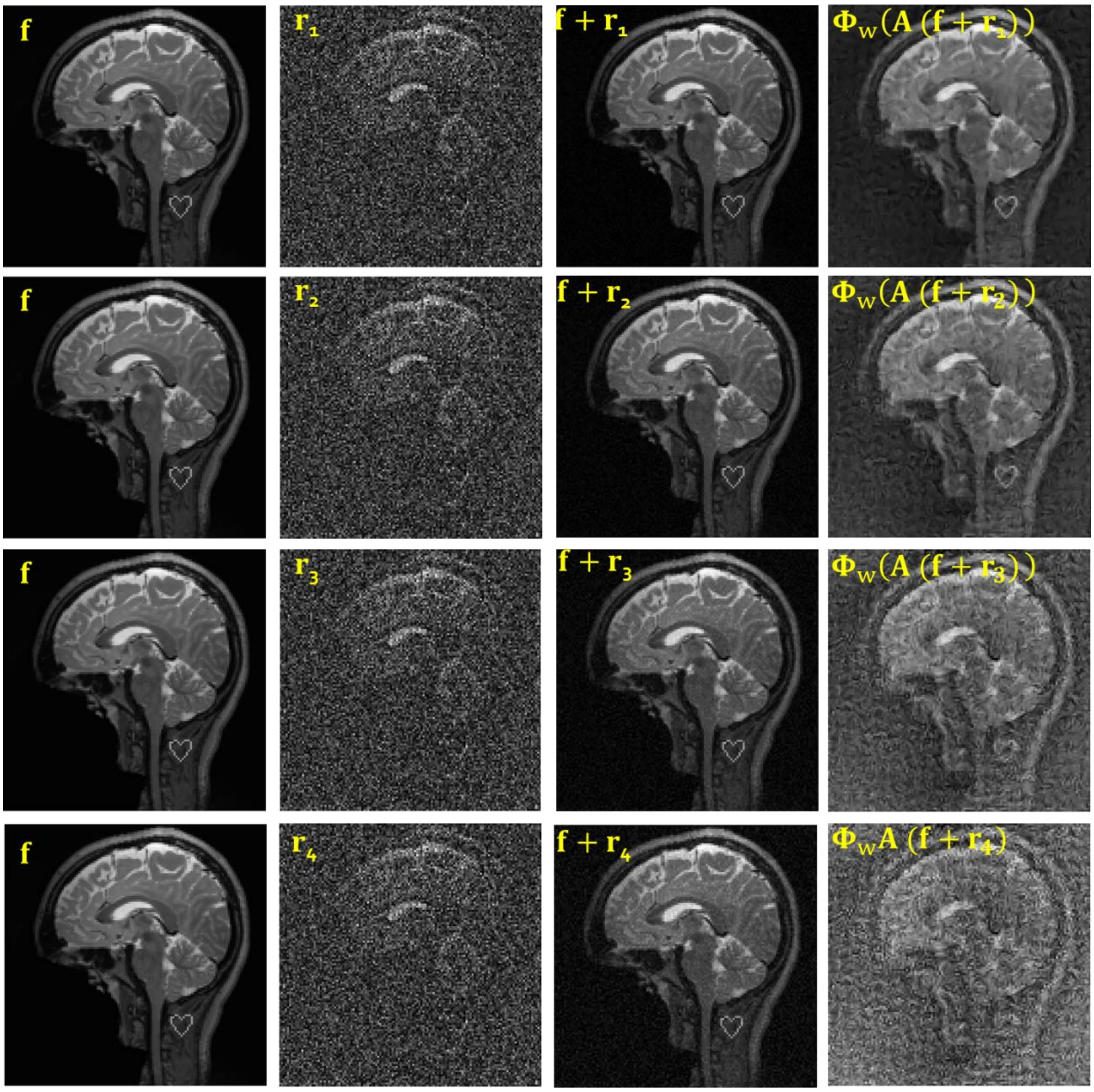

Fig. S17. Reconstruction results using AUTOMAP[23] from [6]. The 1st-4th columns represent the original, perturbation, original plus perturbation, and perturbed AUTOMAP images, respectively. The 1st-4th rows represent different strengths of perturbation, where $\|r_1\|_F^2 < \|r_2\|_F^2 < \|r_3\|_F^2 < \|r_4\|_F^2$.

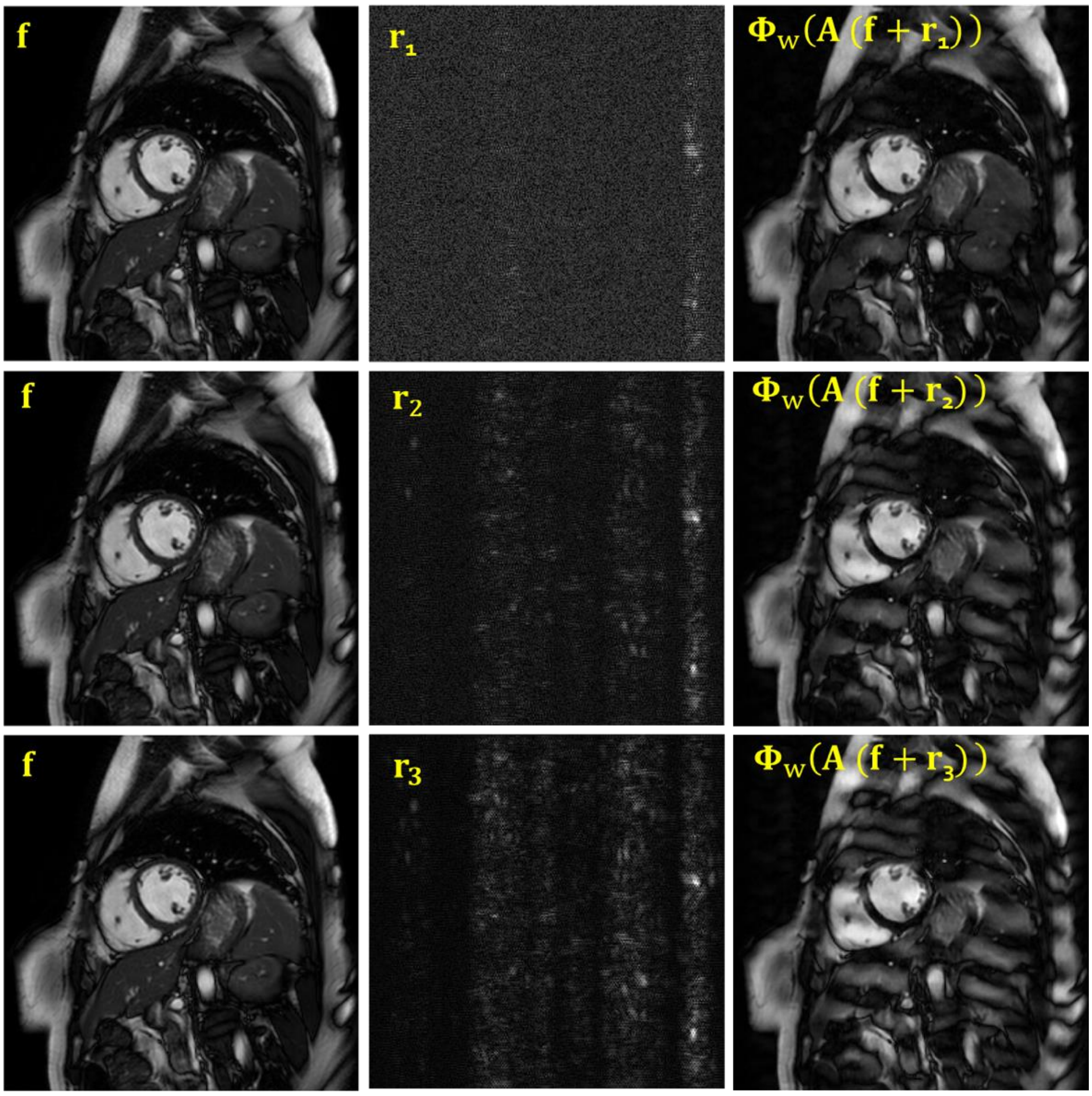

Fig. S18. Reconstruction results using Deep MRI (DM) [92] . These results were adapted from [6]. The 1st-3rd columns represent the original, perturbation, and perturbed DM results, respectively. The 1st-3rd rows present different strengths of perturbation, where $\|r_1\|_F^2 < \|r_2\|_F^2 < \|r_3\|_F^2$.

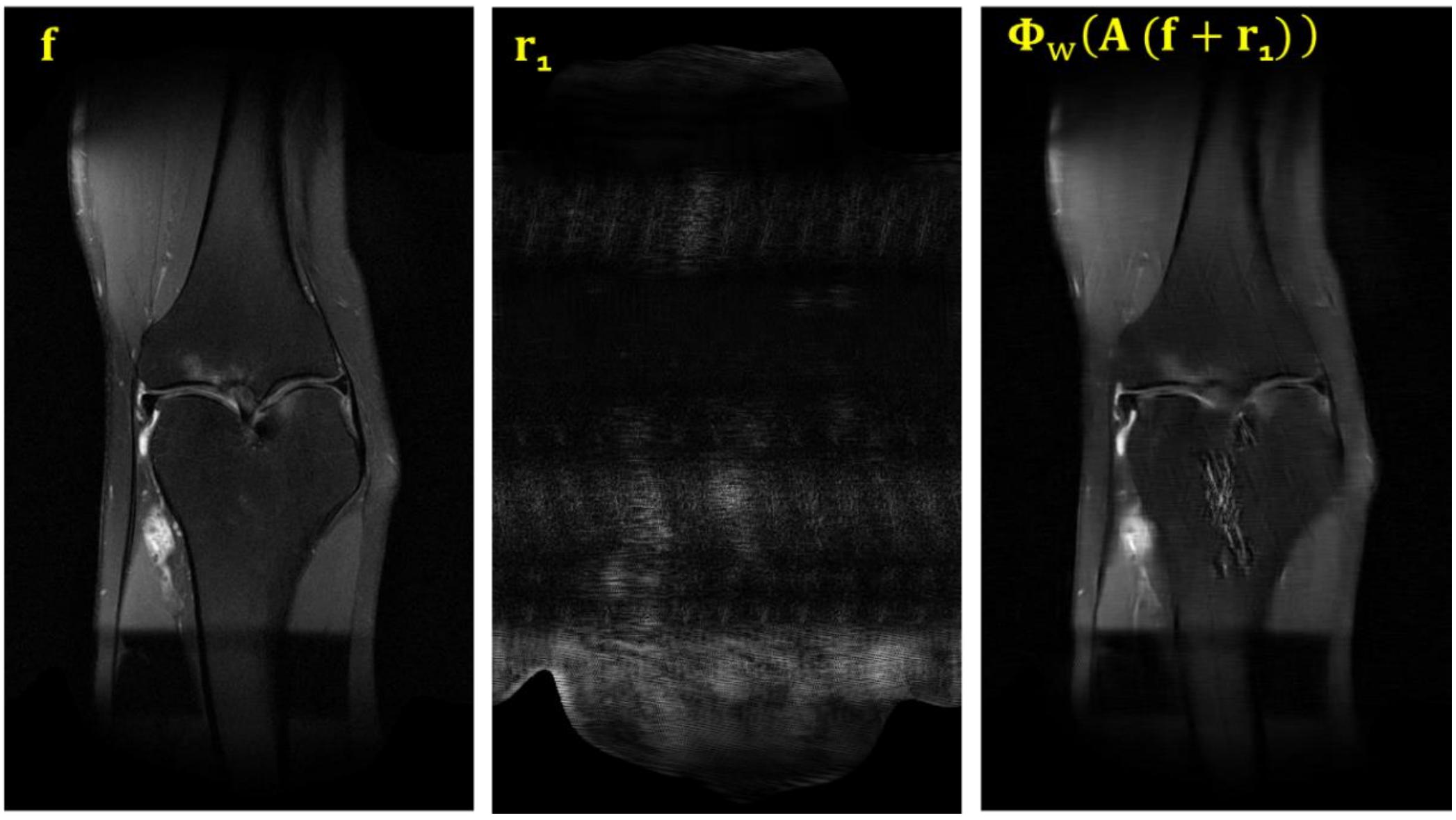

Fig. S19. Reconstruction results using MRI-VN [93] from [6]. The 1st-4th images represent the original, perturbation, and perturbed MRI-VN results, respectively.

It is easy to observe that the perturbed images contain artifacts; for example, the MED-50, AUTOMAP and MRI-VN results. In these cases, the sparsity of reconstructed images was corrupted, and the feedforward data estimation based on these reconstruction results are usually not consistent with the original measurement. Searching for a feasible solution within the space of sparse solution is central to the traditional iterative reconstruction. Furthermore, the ACID searches for a reconstruction good in the three aspects: image sparsity, big data-driven prior, and iterative calibration to eliminate unexplained residual data. When the final image satisfies all these three constraints, it will be our best possible solution.

To analyze the convergence of ACID under the BREN condition, we will first present our heuristic analysis in the next subsection, and then mathematical analysis under several approximations in the subsections after the next one. Hopefully, our insights through these two layers will help better understand the mechanism of the ACID scheme, despite the fact that our analyses are not, at this stage, thorough and rigorous.

## IV.D. Heuristic Analysis on the ACID Convergence

Let us first introduce notations. Let $\boldsymbol{p}^{(\mathbf{0})}$ denotes measured data, which is generally incomplete, inconsistent, and noisy. Specifically, the data can be sinogram or k-space data. Then, we need the three key functions in the ACID scheme. First, an imaging model $\boldsymbol{A}$ is the forward model from an underlying image to tomographic data, which is assumed to be linear without loss of generality. Second, the recon-net $\Phi(\cdot)$ consists of a data-enhancement sub-net and a direct-reconstruction sub-net. This network may be unstable. Note that even if the recon-net is unstable, we assume that it respects the BREN property for our convergence analysis. Third, the CS module $\Theta$ can be a standard CS algorithm or a network-version of the CS algorithm. This module is an image post-processor that maps an image reconstructed by the recon-net to a refined image within the space of sparse solutions. The loss function of the CS module can be a weighted sum of the fidelity term and the sparsity term. The fidelity term can be in the $L_2$ norm of the difference of the input and output images. Let $k$ be the index for iteration, $k = 0,1,2,\cdots$, and we define the following variables.

$\boldsymbol{d}^{(k)}$: Tomographic image produced by the recon-net $\Phi(\cdot)$. $\boldsymbol{d}^{(0)} = \Phi(\boldsymbol{p}^{(\mathbf{0})})$, which is assumed to be a good initial image based on the BREN assumption. $\boldsymbol{d}^{(k)}$, $k = 1,2,\cdots$ represents successive refinements to $\boldsymbol{d}^{(0)}$.

$\boldsymbol{f}^{(k)}$: Tomographic image refined by the CS module $\Theta$, $k = 0,1,2,\cdots$, which should be in the space of sparse solutions in the CS framework, and may or may not be the ground truth, depending on if RIP/rNSP is satisfied or not.

$\boldsymbol{m}^{(k)}$: Estimated data based on the output of the CS module, $k = 0,1,2,\cdots$, which should eventually become as close as possible to the measured data $\boldsymbol{p}^{(\mathbf{0})}$.

$\boldsymbol{p}^{(k)}$: Unexplained residual errors based on $\boldsymbol{f}^{(k-1)}$ in reference to the measured data $\boldsymbol{p}^{(\mathbf{0})}$, $k = 1,2,\cdots$. Specifically, this residual is defined as $\boldsymbol{p}^{(k)} = \boldsymbol{p}^{(\mathbf{0})} - \boldsymbol{A}\boldsymbol{f}^{(k-1)}$. This data residual will be small when $k$ is sufficiently large to obtain a good image quality.

Now, let us analyze the first cycle of the ACID workflow in the following steps.

The first step is to generate $\boldsymbol{d}^{(0)}$ from the original data $\boldsymbol{p}^{(\mathbf{0})}$, which is done by the recon-net $\Phi(\cdot)$: $\boldsymbol{d}^{(0)} = \Phi(\boldsymbol{p}^{(\mathbf{0})})$. Since the recon-net $\Phi(\cdot)$ may be unstable, $\boldsymbol{d}^{(0)}$ can be generally decomposed into the following three components: (a) $\boldsymbol{f}^{*}$: the ground truth image, which is assumed to be s-sparse, (b) $\boldsymbol{f}^{(sps,0)}$: artifacts in the space of sparse solutions of the CS module $\Theta$, which cannot be eliminated based on the sparsity consideration, and (c) $\boldsymbol{f}^{(nsps,0)}$: artifacts not in the space of sparse solution that can be suppressed by the CS module $\Theta$ . That is, $\boldsymbol{d}^{(0)} = \boldsymbol{f}^{*} + \boldsymbol{f}^{(sps,0)} + \boldsymbol{f}^{(nsps,0)}$.

Then, $\boldsymbol{d}^{(0)}$ is processed by the CS module $\Theta$ to obtain $\boldsymbol{f}^{(0)}$. That is, $\boldsymbol{f}^{(0)} = \Theta(\boldsymbol{d}^{(0)})$ so that the difference between $\boldsymbol{d}^{(0)}$ and $\boldsymbol{f}^{(0)}$ is minimized subject to that $\boldsymbol{f}^{(0)}$ is in the space of sparse solutions of $\Theta$ under the constraint of the measurement. As a result, $\boldsymbol{f}^{(0)} = \boldsymbol{f}^{*} + \boldsymbol{f}^{(sps,0)}$ (without loss of generality, here we assume that the sparsity can be perfectly achieved). Without loss of generality, let us take CT as an example.

Based on $\boldsymbol{f}^{(0)}$, $\boldsymbol{m}^{(0)}$ can be estimated with the forward imaging model as $\boldsymbol{m}^{(0)} = \boldsymbol{A}\boldsymbol{f}^{(0)}$ . Generally, we consider $\boldsymbol{f}^{(sps,0)} = \boldsymbol{f}^{(ob,0)} + \boldsymbol{f}^{(nl,0)}$ where two components $\boldsymbol{f}^{(ob,0)}$and $\boldsymbol{f}^{(nl,0)}$ are observable and unobservable, respectively (an unobservable image $\boldsymbol{f}^{(nl,0)}$ is in the null space of $\boldsymbol{A}$). When $\boldsymbol{f}^{(ob,0)}$ is non-zero, the estimated data and the measured data $\boldsymbol{p}^{(\mathbf{0})}$ must be inconsistent. This discrepancy is quantified as the data residual $\boldsymbol{p}^{(1)} = \boldsymbol{p}^{(\mathbf{0})} - \boldsymbol{A}\boldsymbol{f}^{(0)}$. When $\boldsymbol{A}$ does not satisfy RIP/rNSP, the intersection of the data constrained space and the data prior space may contain many solutions, and thus it could be possible that $\boldsymbol{p}^{(1)} = \mathbf{0}$ but $\boldsymbol{f}^{(0)} = \boldsymbol{f}^{*} + \boldsymbol{f}^{(nl,0)}$ (that is, $\boldsymbol{f}^{*} + \boldsymbol{f}^{(nl,0)}$ and $\boldsymbol{f}^{*}$ explain the data $\boldsymbol{p}^{(0)}$ equally well). Nevertheless, it is highly unlikely in practice that the residual data becomes zero, and the ACID iterative process will not converge immediately.

The nonzero data residual can be further reconstructed into the image increment $\Delta\boldsymbol{f}^{(1)}$ using the recon-net $\Phi(\cdot)$; that is, $\Delta\boldsymbol{f}^{(1)} = \Phi(\boldsymbol{p}^{(1)})$. Then, the current tomographic image is updated to $\boldsymbol{d}^{(1)} = \boldsymbol{f}^{(0)} + \Delta\boldsymbol{f}^{(1)}$ (the sum of two prior-consistent images are assumed to be still consistent with data-driven prior, which can be alternatively achieved by applying the recon-net to the augmented data $\boldsymbol{p}^{(\mathbf{0})} + \boldsymbol{p}^{(k)}$). Generally speaking, $\boldsymbol{d}^{(1)}$ will be closer to the ground truth $\boldsymbol{f}^{*}$ than the previous image $\boldsymbol{d}^{(0)}$, since our reconstructed image should explain as much as possible data $\boldsymbol{p}^{(\mathbf{0})}$. With $\boldsymbol{d}^{(1)}$, the residual error in the data domain will be reduced.

Now, we can describe the converging mechanism of ACID, assuming that the recon-net $\Phi(\cdot)$ satisfies the BREN property. Our key arguments include the following three steps:

1) After we have $\boldsymbol{f}^{(0)} = \Theta(\Phi(\boldsymbol{p}^{(\mathbf{0})}))$, by BREN we have $\frac{\|\boldsymbol{f}^{(sps,0)}+\boldsymbol{f}^{(nsps,0)}\|}{\|\boldsymbol{f}^{*}\|} < 1$. That is, $\frac{\|\boldsymbol{f}^{(ob,0)}\|}{\|\boldsymbol{f}^{*}\|} = \alpha^{(0)} < 1$ and $\frac{\|\boldsymbol{f}^{(nl,0)}\|}{\|\boldsymbol{f}^{*}\|} = \beta^{(0)} < 1$ , because $\boldsymbol{f}^{(sps,0)} = \boldsymbol{f}^{(ob,0)} + \boldsymbol{f}^{(nl,0)}$ is orthogonal to $\boldsymbol{f}^{(\boldsymbol{nsps},0)}$, and $\boldsymbol{f}^{(ob,0)}$ is orthogonal to $\boldsymbol{f}^{(nl,0)}$. That is, both $\|\boldsymbol{f}^{(ob,0)}\|$ and $\|\boldsymbol{f}^{(nl,0)}\|$ are fractions of $\|\boldsymbol{f}^{*}\|$.

2) We use the forward model $\boldsymbol{A}$ to synthesize the unexplained residual data $\boldsymbol{p}^{(\mathbf{1})}$. Then, $\boldsymbol{p}^{(\mathbf{1})}$ is fed to the recon-net to reconstruct $\Delta\boldsymbol{f}^{(1)} = \Phi(\boldsymbol{p}^{(1)})$. Since $\boldsymbol{p}^{(\mathbf{1})}$ is due to $\boldsymbol{f}^{(ob,0)}$. Then, $-\boldsymbol{f}^{(ob,0)}$can be reconstructed up to a new artifact image $\boldsymbol{f}^{(ob,1)} + \boldsymbol{f}^{(nl,1)} + \boldsymbol{f}^{(nsps,1)}$. By BREN again, $\frac{\|\boldsymbol{f}^{(ob,1)}\|}{\|\boldsymbol{f}^{(ob,0)}\|} = \alpha^{(1)} < 1$ and $\frac{\|\boldsymbol{f}^{(nl,1)}\|}{\|\boldsymbol{f}^{(ob,0)}\|} = \beta^{(1)} < 1$. That is, both $\|\boldsymbol{f}^{(ob,1)}\|$ and $\|\boldsymbol{f}^{(nl,1)}\|$ are fractions of $\|\boldsymbol{f}^{(ob,0)}\|$.
3) We can repeat this process for $k \to \infty$, we have that $\boldsymbol{d}^{(\infty)} = \boldsymbol{f}^{(\infty)} = \boldsymbol{f}^* + \boldsymbol{f}^{(ob,\infty)} + \sum_{k=0}^{\infty} \boldsymbol{f}^{(nl,k)}$ . Because the norm of $\boldsymbol{f}^{(ob,\infty)}$ is less than $\|\boldsymbol{f}^*\| \prod_{k=0}^{\infty} \alpha^{(k)} \to 0$, $\boldsymbol{f}^{(ob,\infty)} \to 0$. Meanwhile $\boldsymbol{f}^{(nl,\infty)} \to 0$ since
$$\|\boldsymbol{f}^{(nl,k)}\| = \|\boldsymbol{f}^{(ob,k-1)}\|\beta^{(k)} = \|\boldsymbol{f}^*\| \prod_{k'=0}^{k-1} \alpha^{(k')} \beta^{(k)}. \quad \text{(S.4.8)}$$
Noting that both $\alpha^{(k)}$ and $\beta^{(k)}$ are smaller than 1, we have
$$\left\|\sum_{k=0}^{\infty} \boldsymbol{f}^{(nl,k)}\right\| < \sum_{k=0}^{\infty}\|\boldsymbol{f}^{(nl,k)}\| = \sum_{k=0}^{\infty}\|\boldsymbol{f}^*\| \prod_{k'=0}^{k-1} \alpha^{(k')} \beta^{(k)} < \infty. \quad \text{(S.4.9)}$$

That is, the ACID will converge to $\boldsymbol{f}^* + \sum_{k=0}^{\infty} \boldsymbol{f}^{(nl,k)}$ which will be in the intersection of the space of solutions satisfying measured data, the space of sparse solutions, and the space of data-driven solutions. While this ACID scheme may converge to an image still containing a nonzero null space component when RIP/rNSP is not satisfied, the key point is that under the same condition (i.e., RIP/rNSP is not satisfied) a sparsity-promoting algorithm cannot eliminate such a nonzero null space component either, and more importantly ACID has enforced the powerful deep prior so that the space of feasible solutions is greatly reduced relative to that permitted with a sparsity-promoting algorithm alone. In our experiments, we have shown that ACID with the kernel awareness embedded consistently outperforms the selected sparsity-promoting algorithms that do not utilize big data-driven prior. In other words, the data prior is instrumental in recovering the nonzero null space component that cannot be measured by the system matrix.

Although the above analysis is not mathematically rigorous, it indeed sheds light on the inner working of ACID. This analysis adds value, especially in the current situation that a general non-convex optimization theory is yet to be developed. In the above analysis, we have assumed the BREN property of the recon-net. As a result, even if the network is not ideal (which means producing a substantial non-zero artifact image), the convergence is still guaranteed, as long as the relative error is under control (less than 100%) in the $L_2$ norm, which is a practically motivated condition. On the other hand, it is underlined that if the network is indeed optimized or nearly optimized so that artifact image is small in the first place, the iterative process will converge rapidly, and in that case the whole ACID workflow can be unrolled into a compact feedforward network.

## IV.E. Theoretical Interpretation of the ACID Convergence

While our ACID architecture is heuristically obtained, here we interpret the ACID algorithm by casting it as the iterative solution that minimizes an overall objective function. Although the following theoretical analysis is under several approximations, our findings do improve our understanding of the initially heuristically-derived ACID scheme. It is underlined that given the nonconvex nature of this complicated optimization problem, a closed-form solution is impossible, and in the iterative process such approximations are appropriate to derive computationally efficient iterative formulas, and the errors due to the approximations will be suppressed via ACID iterations so that the ACID algorithm will converge to a desirable solution in the intersection of the space constrained by measured data, the space of sparse solutions, and the space of data-driven deep priors. This mechanism is similar to the conventional ART/SART algorithm whose convergence was rigorously proved for convex optimization [94, 95].

In the imaging field, we often assume that the measurement $\boldsymbol{p}^{(\mathbf{0})} = \boldsymbol{A}\boldsymbol{f}^* + \boldsymbol{e}$, where $\boldsymbol{A} \in \mathbb{R}^{m\times N}$ is a measurement matrix (for example, $\boldsymbol{A}$ is the Radon transform for CT [96] and the Fourier transform for MRI [97]). $\boldsymbol{p}^{(\mathbf{0})} \in \mathbb{R}^m$ is an original dataset, $\boldsymbol{f}^* \in \mathbb{R}^m$ is the ground truth image, $\boldsymbol{e} \in \mathbb{R}^m$ is data noise, $\|\boldsymbol{e}\|_2 \leq \eta$ for some $\eta \geq 0$, and most relevantly $m < N$, meaning that the inverse problem is underdetermined. In the under-deterministic case, additional prior knowledge must be introduced to uniquely and stably recover the original image. Without loss of generality, we assume that $\boldsymbol{H} \in \mathbb{R}^{N\times N}$ is unitary, $\boldsymbol{H}^*$ is the adjoint of $\boldsymbol{H}$, and $\boldsymbol{H}^*$ is equal to the inverse of $\boldsymbol{H}$. $\boldsymbol{A}$ satisfies the RIP of order $s$, and $\boldsymbol{H}\boldsymbol{f}^*$ is s-sparse. We further assume that the function $\Phi(\cdot)$ models a well-trained neural network, and it continuously maps measurement data to an image. Although $\Phi(\cdot)$ is to solve the problem $\boldsymbol{f}$ from the measurement $\boldsymbol{p} = \mathbf{A}\boldsymbol{f}$, we can only approximately have $\boldsymbol{A}\Phi(\boldsymbol{p}) \cong \boldsymbol{p}$ in some sense such as satisfying the aforementioned BREN, there is not only any guarantee that $\Phi(\mathbf{A}\boldsymbol{f}) = \boldsymbol{f}$ but also a real risk that $\Phi(\cdot)$ is unstable. This is because the system matrix $\boldsymbol{A}$ is underdetermined and the neural network is venerable and may generate an artifact image so that $\Phi(\mathbf{A}\boldsymbol{f}) = \boldsymbol{f} + \boldsymbol{f}^{ob} + \boldsymbol{f}^{nl}$, where $\boldsymbol{f}^{ob}$ is observable and $\boldsymbol{f}^{nl}$ is in the null space of $\boldsymbol{A}$. $\boldsymbol{f}^{nl}$ satisfies $\boldsymbol{A}\boldsymbol{f}^{nl} = 0$, and $\|\boldsymbol{A}\boldsymbol{f}^{ob}\| \neq 0$ if $\|\boldsymbol{f}^{ob}\| \neq 0$.

In this work, our goal is to design an iterative framework to stabilize an unstable neural network aided by a CS-based sparsity-promoting module. As an idealized setting to show the essential idea, we assume that the input to the neural network is a noise-free dataset $\boldsymbol{p}^{(0)} - \boldsymbol{e}$, and the output of the CS-module is $\boldsymbol{f}$ .

Let us introduce a residual error correction mechanism in the projection domain and assume the residual error is $\boldsymbol{p}$. There will be $\boldsymbol{e} = \boldsymbol{p}^{(0)} - \mathbf{A}\boldsymbol{f} - \boldsymbol{p}$. Then, we want to minimize the following objective function:

$$\underset{\boldsymbol{p},\boldsymbol{f}}{\operatorname{argmin}} \frac{1}{2}\|\Phi(\mathbf{A}\boldsymbol{f} + \boldsymbol{p}) - \boldsymbol{f}\|_2^2 + \frac{\lambda}{2}\left\| \boldsymbol{p}^{(0)} - \mathbf{A}\boldsymbol{f} - \boldsymbol{p}\right\|_2^2 + \frac{\mu}{2}\|\boldsymbol{p}\|_2^2 + \xi\|\boldsymbol{H}\boldsymbol{f}\|_1, \quad \text{(S.4.10)}$$

where $\lambda > 0$, $\mu \geq 0$ and $\xi > 0$ are hyperparameters, the first term is the difference between the outputs of the neural network and the CS-based sparsifying module, the second term is the measured noise energy, the third term is the residual error energy also in the projection domain, and the last term is to enforce the sparsity of the output image of the CS module, which is subject to the data-fidelity constraint in the projection domain. Let us define

$$L(\boldsymbol{p},\boldsymbol{f}) \coloneqq \frac{1}{2}\|\Phi(\mathbf{A}\boldsymbol{f} + \boldsymbol{p}) - \boldsymbol{f}\|_2^2 + \frac{\lambda}{2}\left\| \boldsymbol{p}^{(0)} - \mathbf{A}\boldsymbol{f} - \boldsymbol{p}\right\|_2^2 + \frac{\mu}{2}\|\boldsymbol{p}\|_2^2 + \xi\|\boldsymbol{H}\boldsymbol{f}\|_1. \quad \text{(S.4.11)}$$

Then, we can use the block coordinate descent method [98] to optimize (S.4.11) as follows:

$$\begin{cases} \boldsymbol{p}^{(\boldsymbol{k}+\mathbf{1})} = \underset{\boldsymbol{p}}{\operatorname{argmin}}\, L(\boldsymbol{p}, \boldsymbol{f}^{(\boldsymbol{k})}) \\ \boldsymbol{f}^{(\boldsymbol{k}+\mathbf{1})} = \underset{\boldsymbol{f}}{\operatorname{argmin}}\, L\left(\boldsymbol{p}^{(\boldsymbol{k}+\mathbf{1})}, \boldsymbol{f}\right) \end{cases}. \tag{S.4.12}$$

To update $\boldsymbol{p}$, we need to solve the following problem:

$$\boldsymbol{p}^{(\boldsymbol{k}+\mathbf{1})} = \underset{\boldsymbol{p}}{\operatorname{argmin}} \left(\frac{1}{2}\left\|\Phi\left(\mathbf{A}\boldsymbol{f}^{(\boldsymbol{k})} + \boldsymbol{p}\right) - \boldsymbol{f}^{(\boldsymbol{k})}\right\|_2^2 + \frac{\lambda}{2}\left\|\boldsymbol{p}^{(0)} - \mathbf{A}\boldsymbol{f}^{(\boldsymbol{k})} - \boldsymbol{p}\right\|_2^2\right) + \frac{\mu}{2}\|\boldsymbol{p}\|_2^2. \tag{S.4.13}$$

Computing the partial derivative of the right side of (S.4.13), we have

$$\left(\frac{\partial\Phi\left(\mathbf{A}\boldsymbol{f}^{(\boldsymbol{k})}+\boldsymbol{p}\right)}{\partial\left(\mathbf{A}\boldsymbol{f}^{(\boldsymbol{k})}+\boldsymbol{p}\right)}\right)^T \left(\Phi\left(\mathbf{A}\boldsymbol{f}^{(\boldsymbol{k})} + \boldsymbol{p}\right) - \boldsymbol{f}^{(\boldsymbol{k})}\right) + \lambda\left(\mathbf{A}\boldsymbol{f}^{(\boldsymbol{k})} + \boldsymbol{p} - \boldsymbol{p}^{(0)}\right) + \mu\boldsymbol{p} = 0. \tag{S.4.14}$$

Because the neural network is well-trained to solve the problem $\boldsymbol{Af} = \boldsymbol{p}$, we assume $\boldsymbol{A}\Phi(\boldsymbol{p}) \cong \boldsymbol{p}$ (at least on a training dataset). By performing derivative on both sides of $\boldsymbol{A}\Phi(\boldsymbol{p}) \cong \boldsymbol{p}$, we have $\boldsymbol{A}\left(\frac{\partial\Phi(\boldsymbol{p})}{\partial\boldsymbol{p}}\right) \cong \boldsymbol{I}$, where $\boldsymbol{I}$ is the identity matrix. This means $\frac{\partial\Phi(\boldsymbol{p})}{\partial\boldsymbol{p}} \cong \boldsymbol{A}^{-\mathbf{1}}$ (in the sense of a pseudo inverse for an underdetermined matrix $\boldsymbol{A}$ which can be obtained by classical methods such as truncated SVD). In the classic and modern iterative CT reconstruction methods (e.g., SART), while $\boldsymbol{A}^T\boldsymbol{A} \neq \boldsymbol{I}$, the residual error correction mechanism and the resultant cumulative effect of the whole iterative process will make the final solution converge to an optimal solution for projections that are sufficiently sampled [94, 95]. In this sense, treating a backprojection operator $\boldsymbol{A}^T$ as an approximate inverse to the projection operator $\boldsymbol{A}$ in each iteration is reasonable. Furthermore, in the ACID iterative framework, we also make the approximation $\boldsymbol{A}^T \cong \boldsymbol{A}^{-1}$. Hence, we have the approximation that $\frac{\partial\Phi(\boldsymbol{p})}{\partial\boldsymbol{p}} \cong \boldsymbol{A}^{-\mathbf{1}} \cong \boldsymbol{A}^T$ and $\left(\frac{\partial\Phi(\boldsymbol{p})}{\partial(\boldsymbol{p})}\right)^T \cong \boldsymbol{A}$.

In (S.4.14), $\left(\frac{\partial\Phi\left(\mathbf{A}\boldsymbol{f}^{(\boldsymbol{k})}+\boldsymbol{p}\right)}{\partial\left(\mathbf{A}\boldsymbol{f}^{(\boldsymbol{k})}+\boldsymbol{p}\right)}\right)^T \cong \boldsymbol{A}$ is the operator transforming a reconstructed image into a measurement dataset, and it is approximated as $\boldsymbol{A}$. By ignoring the observable artifact image from the neural network (since in the iterative correction, the artifact image will be gradually reduced; see the next subsection for justification), we have $\boldsymbol{A}\Phi\left(\mathbf{A}\boldsymbol{f}^{(\boldsymbol{k})} + \boldsymbol{p}\right) \cong \mathbf{A}\boldsymbol{f}^{(\boldsymbol{k})} + \boldsymbol{p}$. Therefore, (S.4.14) can be simplified as

$$\boldsymbol{p}^{(k+1)} \cong \frac{\lambda\left(\boldsymbol{p}^{(0)} - \mathbf{A}\boldsymbol{f}^{(\boldsymbol{k})}\right)}{1+\lambda+\mu}. \tag{S.4.15}$$

To update $\boldsymbol{f}$, we solve the following problem:

$$\boldsymbol{f}^{(\boldsymbol{k}+\mathbf{1})} = \underset{\boldsymbol{f}}{\operatorname{argmin}} \left(\frac{1}{2}\left\|\Phi\left(\mathbf{A}\boldsymbol{f} + \boldsymbol{p}^{(\boldsymbol{k}+\mathbf{1})}\right) - \boldsymbol{f}\right\|_2^2 + \frac{\lambda}{2}\left\|\boldsymbol{p}^{(0)} - \mathbf{A}\boldsymbol{f} - \boldsymbol{p}^{(\boldsymbol{k}+\mathbf{1})}\right\|_2^2 + \xi\|\boldsymbol{Hf}\|_1\right). \tag{S.4.16}$$

With $\bar{\boldsymbol{f}} = \boldsymbol{Hf}$ and $\boldsymbol{f} = \boldsymbol{H}^*\bar{\boldsymbol{f}}$, (S.4.16) is rewritten as follows:

$$\bar{\boldsymbol{f}}^{(\boldsymbol{k}+\mathbf{1})} = \underset{\bar{\boldsymbol{f}}}{\operatorname{argmin}} \left(\frac{1}{2}\left\|\Phi\left(\boldsymbol{A}\boldsymbol{H}^*\bar{\boldsymbol{f}} + \boldsymbol{p}^{(\boldsymbol{k}+\mathbf{1})}\right) - \boldsymbol{H}^*\bar{\boldsymbol{f}}\right\|_2^2 + \frac{\lambda}{2}\left\|\boldsymbol{p}^{(0)} - \boldsymbol{A}\boldsymbol{H}^*\bar{\boldsymbol{f}} - \boldsymbol{p}^{(\boldsymbol{k}+\mathbf{1})}\right\|_2^2 + \xi\left\|\bar{\boldsymbol{f}}\right\|_1\right). \tag{S.4.17}$$

Computing the partial derivative of the right side of (S.4.17), we have

$$\left(\boldsymbol{H}\boldsymbol{A}^{T}\left(\frac{\partial\Phi\left(\mathbf{A}\boldsymbol{H}^{*}\bar{\boldsymbol{f}}+\boldsymbol{p}^{(\boldsymbol{k}+\mathbf{1})}\right)}{\partial\left(\mathbf{A}\boldsymbol{H}^{*}\bar{\boldsymbol{f}}+\boldsymbol{p}^{(\boldsymbol{k}+\mathbf{1})}\right)}\right)^{T}-\boldsymbol{H}\right)\left(\Phi\left(\mathbf{A}\boldsymbol{H}^{*}\bar{\boldsymbol{f}}+\boldsymbol{p}^{(\boldsymbol{k}+\mathbf{1})}\right)-\boldsymbol{H}^{*}\bar{\boldsymbol{f}}\right)+\lambda\boldsymbol{H}\boldsymbol{A}^{\boldsymbol{T}}\left(\boldsymbol{A}\boldsymbol{H}^{*}\bar{\boldsymbol{f}}-\boldsymbol{p}^{(0)}+\boldsymbol{p}^{(\boldsymbol{k}+\mathbf{1})}\right)+\xi sgn\left(\bar{\boldsymbol{f}}\right)=0. \quad \text{(S.4.18)}$$

Similarly treating $\left(\frac{\partial\Phi\left(\mathbf{A}\boldsymbol{H}^{*}\bar{\boldsymbol{f}}+\boldsymbol{p}^{(\boldsymbol{k}+\mathbf{1})}\right)}{\partial\left(\mathbf{A}\boldsymbol{H}^{*}\bar{\boldsymbol{f}}+\boldsymbol{p}^{(\boldsymbol{k}+\mathbf{1})}\right)}\right)^{T}$ as $\boldsymbol{A}$ and $\boldsymbol{A}^{T}\cong\boldsymbol{A}^{-1}$ , (S.4.18) can be simplified as

$$\lambda\bar{\boldsymbol{f}}+\lambda\boldsymbol{H}\boldsymbol{A}^{\boldsymbol{T}}\left(\boldsymbol{p}^{(\boldsymbol{k}+\mathbf{1})}-\mathbf{p}^{(0)}\right)+\xi sgn\left(\bar{\boldsymbol{f}}\right)\cong 0. \quad \text{(S.4.19)}$$

From (S.4.15), we have

$$\boldsymbol{p}^{(0)}\cong\frac{(1+\lambda+\mu)\boldsymbol{p}^{(k+1)}}{\lambda}+\boldsymbol{A}\boldsymbol{f}^{(\boldsymbol{k})}. \quad \text{(S.4.20)}$$

By substituting (S.4.20) into (S.4.19), we have

$$\lambda\bar{\boldsymbol{f}}-(1+\mu)\boldsymbol{H}\boldsymbol{A}^{\boldsymbol{T}}\boldsymbol{p}^{(k+1)}-\lambda\boldsymbol{H}\boldsymbol{f}^{(\boldsymbol{k})}+\xi sgn\left(\bar{\boldsymbol{f}}\right)\cong 0. \quad \text{(S.4.21)}$$

Noting that

$$\boldsymbol{A}^{T}\boldsymbol{p}^{(k+1)}\cong\boldsymbol{A}^{-1}\boldsymbol{p}^{(k+1)}\cong\Phi\left(\boldsymbol{p}^{(\boldsymbol{k}+\mathbf{1})}\right), \quad \text{(S.4.22)}$$

(S.4.21) can be simplified as

$$\lambda\bar{\boldsymbol{f}}-(1+\mu)\boldsymbol{H}\Phi\left(\boldsymbol{p}^{(\boldsymbol{k}+\mathbf{1})}\right)-\lambda\boldsymbol{H}\boldsymbol{f}^{(\boldsymbol{k})}+\xi sgn\left(\bar{\boldsymbol{f}}\right)\cong 0. \quad \text{(S.4.23)}$$

By rewriting (S.4.23) as

$$\bar{\boldsymbol{f}}\cong\boldsymbol{H}\boldsymbol{f}^{(\boldsymbol{k})}+\frac{(1+\mu)}{\lambda}\boldsymbol{H}\Phi\left(\boldsymbol{p}^{(\boldsymbol{k}+\mathbf{1})}\right)-\frac{\xi}{\lambda}sgn\left(\bar{\boldsymbol{f}}\right) \quad \text{(S.4.24)}$$

we have $\boldsymbol{f}^{(\boldsymbol{k}+\mathbf{1})}$ via soft-threshold filtering:

$$\boldsymbol{f}^{(\boldsymbol{k}+\mathbf{1})}\cong\boldsymbol{H}^{*}\boldsymbol{S}_{\frac{\xi}{\lambda}}\left(\boldsymbol{H}\left(\boldsymbol{f}^{(\boldsymbol{k})}+\frac{1+\mu}{\lambda}\Phi\left(\boldsymbol{p}^{(\boldsymbol{k}+\mathbf{1})}\right)\right)\right), \quad \text{(S.4.25)}$$

where the soft-thresholding kernel is defined as

$$S_{\varepsilon}(x)=\begin{cases}0, & |x|<\varepsilon\\ x-sgn(x)\varepsilon & otherwise\end{cases}. \quad \text{(S.4.26)}$$

Combining (S.4.15) and (S.4.25), we obtain a set of formulas:

$$\begin{cases}\boldsymbol{p}^{(k+1)}\cong\frac{\lambda\left(\mathbf{p}^{(0)}-\mathbf{A}\boldsymbol{f}^{(\boldsymbol{k})}\right)}{1+\lambda+\mu}\\ \boldsymbol{f}^{(\boldsymbol{k}+\mathbf{1})}\cong\boldsymbol{H}^{*}\boldsymbol{S}_{\frac{\xi}{\lambda}}\left(\boldsymbol{H}\left(\boldsymbol{f}^{(\boldsymbol{k})}+\frac{1+\mu}{\lambda}\Phi\left(\boldsymbol{p}^{(\boldsymbol{k}+\mathbf{1})}\right)\right)\right)\end{cases}. \quad \text{(S.4.27)}$$

Let us denote $\frac{\xi}{\lambda}=\varepsilon$ and simplify (S.4.27) as

$$\begin{cases}\boldsymbol{p}^{(k+1)}\cong\frac{\lambda\left(\mathbf{p}^{(0)}-\mathbf{A}\boldsymbol{f}^{(\boldsymbol{k})}\right)}{1+\lambda+\mu}\\ \boldsymbol{f}^{(\boldsymbol{k}+\mathbf{1})}\cong\boldsymbol{H}^{*}\boldsymbol{S}_{\varepsilon}\left(\boldsymbol{H}\left(\boldsymbol{f}^{(\boldsymbol{k})}+\frac{1+\mu}{\lambda}\Phi\left(\boldsymbol{p}^{(\boldsymbol{k}+\mathbf{1})}\right)\right)\right)\end{cases}. \quad \text{(S.4.28)}$$

Clearly, (S.4.28) agrees with our heuristically-derived ACID network by setting $\mu=0$. In other words, ACID is a special case of (S.4.28) after the weighting parameters are properly selected.

Although a unitary property is assumed for the sparse transform $\boldsymbol{H}$ to obtain (S.4.28), as is the case of the orthogonal wavelet decomposition, similar results can be also obtained in non-unitary cases. In [99], Poon studied the problem of recovering a 1D or 2D discrete signal that is approximately sparse in its gradient transform from an incomplete subset of its Fourier coefficients. To obtain a high quality reconstruction with high probability, robust to noise and stable to inexact gradient sparsity of order $s$, Poon proved that it is sufficient to draw $O(s\log N)$ of the available Fourier coefficients uniformly at random [99]. With Poon's results, we can extend (S.4.28) to a non-unitary discrete gradient transform for TV minimization.

Specifically, the term $\|\boldsymbol{Hf}\|_1$ in (S.4.10) is specialized as a total variation function $g(\boldsymbol{f})$ based on discrete gradient transform

$$g(\boldsymbol{f}) = \sum_{i_w=2}^{I_w}\sum_{i_h=2}^{I_h}(|\boldsymbol{f}(i_w,i_h) - \boldsymbol{f}(i_w-1,i_h)| + |\boldsymbol{f}(i_w,i_h) - \boldsymbol{f}(i_w,i_h-1)|), \quad \text{(S.4.29)}$$

where $I_w$ and $I_h$ represent the width and height of a reconstructed image, and the gradients on the image border are assumed to be zero. An FFT-based algorithm, FTVd [100], can be employed to find the sparse solution for $\boldsymbol{f}$. Note that the generic TV favors piecewise constant regions, while high-order TV encourages piecewise polynomials [100]. Here, the input to the CS-based sparsifying module is normalized to [0, 1] to facilitate the selection of the regularized parameters, which requires de-normalization of the output of the CS module. In the CS framework, the robust null space property ensures the stability of sparsity regularized recovery [12, 101].

Let us denote

$$\boldsymbol{f}^{\left(k+\frac{1}{2}\right)} = \boldsymbol{f}^{(k)} + \frac{1+\mu}{\lambda}\Phi\left(\boldsymbol{p}^{(k+1)}\right), \quad \text{(S.4.30)}$$

(S.4.28) can be rewritten as

$$\begin{cases} \boldsymbol{p}^{(k+1)} \cong \frac{\lambda\left(\mathbf{p}^{(0)} - \mathbf{A}\boldsymbol{f}^{(\boldsymbol{k})}\right)}{1+\lambda+\mu} \\ \boldsymbol{f}^{\left(k+\frac{1}{2}\right)} = \boldsymbol{f}^{(\boldsymbol{k})} + \frac{1+\mu}{\lambda}\Phi\left(\boldsymbol{p}^{(\boldsymbol{k}+\mathbf{1})}\right). \\ \boldsymbol{f}^{(\boldsymbol{k}+\mathbf{1})} \cong \boldsymbol{H}^*\boldsymbol{S}_\varepsilon\left(\boldsymbol{Hf}^{\left(k+\frac{1}{2}\right)}\right) \end{cases} \quad \text{(S.4.31)}$$

The discrete gradient function $g(\boldsymbol{f})$ defined in (S.4.29) can be interpreted as $\|\boldsymbol{Hf}\|_1$ with a non-unitary transform matrix $\boldsymbol{H}$. Because $\boldsymbol{H}$ is not invertible, the adjoint matrix $\boldsymbol{H}^*$ in (S.4.31) is not the inverse of $\boldsymbol{H}$. However, due to the fact that both $\boldsymbol{H}^*$ and $\boldsymbol{H}$ work as a pair before and after the soft-thresholding filtering in (S.4.31), $\boldsymbol{H}^*$ can be interpreted as a pseudo-inverse of the discrete gradient transform $\boldsymbol{H}$ [102]. Hence, each pixel of $\boldsymbol{f}^{(k+1)}$ at the position $(i_w,i_h)$ in (S.4.31) can be expressed as follows [102]:

$$\boldsymbol{f}^{(k+1)}(i_w,i_h) = \frac{\mathbf{1}}{\mathbf{4}}\left(S_{\boldsymbol{\varepsilon}}^{-\mathbf{1}}\left(\boldsymbol{f}^{\left(k+\frac{1}{2}\right)}(i_w,i_h), \boldsymbol{f}^{\left(k+\frac{1}{2}\right)}(i_w+1,i_h)\right) + S_{\boldsymbol{\varepsilon}}^{-\mathbf{1}}\left(\boldsymbol{f}^{\left(k+\frac{1}{2}\right)}(i_w,i_h), \boldsymbol{f}^{\left(k+\frac{1}{2}\right)}(i_w,i_h+1)\right) + S_{\boldsymbol{\varepsilon}}^{-\mathbf{1}}\left(\boldsymbol{f}^{\left(k+\frac{1}{2}\right)}(i_w,i_h), \boldsymbol{f}^{\left(k+\frac{1}{2}\right)}(i_w-1,i_h)\right) + S_{\boldsymbol{\varepsilon}}^{-\mathbf{1}}\left(\boldsymbol{f}^{\left(k+\frac{1}{2}\right)}(i_w,i_h), \boldsymbol{f}^{\left(k+\frac{1}{2}\right)}(i_w,i_h-1)\right)\right), \quad \text{(S.4.32)}$$

where $S_{\boldsymbol{\varepsilon}}^{-\mathbf{1}}(\cdot,\cdot)$ is the pseudo-inverse of the soft-thresholding kernel $S_\varepsilon(x)$ for a given threshold $\varepsilon$. The pseudo-inverse $S_{\boldsymbol{\varepsilon}}^{-\mathbf{1}}(\cdot,\cdot)$ is defined as [102]:

$$S_{\boldsymbol{\varepsilon}}^{-\mathbf{1}}(v_a,v_b) = \begin{cases} \frac{v_a+v_b}{2}, & if\ |v_a - v_b| \le \varepsilon \\ v_a - \frac{\varepsilon}{2}, & if\ v_a - v_b > \varepsilon \\ v_a + \frac{\varepsilon}{2}, & if\ v_a - v_b < -\varepsilon \end{cases}. \quad \text{(S.4.33)}$$

With the pseudo-inverse (S.4.33), although the discrete gradient transform is neither unitary nor invertible, the iterative framework (S.4.28) can still be applied for TV minimization using a compressed sensing technique [99].

Under practically reasonable conditions such as noisy and insufficient data, the ACID iteration will converge to a feasible solution subject to an uncertain range proportional to the noise level (see the convergence analysis below for more details).

## IV.F. Mathematical Analysis on the ACID Convergence

In the theoretical iterative framework (S.4.28) and with the BREN property of the neural network, we will show that the final solution $\boldsymbol{f}^{(k+1)}$ will converge to an optimal image, and particularly the ground truth $\boldsymbol{f}^{*}$ assuming that RIP/rNSP is satisfied, subject to a noise-induced uncertainty distance in terms of the L2 norm and the null space component. While this convergence analysis is not rigorous, it helps rationalize the ACID workflow, and in this context the convergence to the optimal solution implies stability.

Now, let us analyze the convergence of our ACID scheme. Denoting $M_1 = \frac{\lambda}{1+\lambda+\mu}$ and $M_2 = \frac{1+\mu}{\lambda}$, we have $M = M_1 M_2 = \frac{\lambda}{1+\lambda+\mu}\frac{1+\mu}{\lambda} = \frac{1+\mu}{1+\lambda+\mu} < 1$. (S.4.28) can be simplified to our heuristically-designed ACID iteration:

$$\begin{cases} \boldsymbol{p}^{(k+1)} = M_{\mathbf{1}}\left(\boldsymbol{p}^{(0)} - \mathbf{A}\boldsymbol{f}^{(k)}\right) \\ \boldsymbol{f}^{(k+1)} = \boldsymbol{H}^{*}\boldsymbol{S}_{\varepsilon}\left(\boldsymbol{H}\left(\boldsymbol{f}^{(\boldsymbol{k})} + M_{\mathbf{2}}\Phi\left(\boldsymbol{p}^{(\boldsymbol{k+1})}\right)\right)\right) \end{cases}. \tag{S.4.34}$$

In this subsection, we replace "≅" in (S.4.28) with "=" in (S.4.34) abusing the notation a bit. Let us analyze the convergence of our ACID network as follows. Because the input to the neural network is $\boldsymbol{p}^{(0)} - \boldsymbol{e}$ in our objective function (S.4.10), we initially assume that the noise is $\boldsymbol{e} = 0$ and $\boldsymbol{p}^{(0)} = \boldsymbol{A}\boldsymbol{f}^{*}$, where $\boldsymbol{f}^{*}$ is the ground truth. Assuming an initial image $\Phi\left(\boldsymbol{p}^{(\mathbf{0})}\right) = \boldsymbol{f}^{*} + \boldsymbol{f}^{(ob,0)} + \boldsymbol{f}^{(nl,0)}$. By the BREN property, we have

$$\left\|\boldsymbol{f}^{(ob,0)} + \boldsymbol{f}^{(nl,0)}\right\| < (1-\sigma)\|\boldsymbol{f}^{*}\|. \tag{S.4.35}$$

Since $\boldsymbol{H}\boldsymbol{f}^{(ob,0)}$ and $\boldsymbol{H}\boldsymbol{f}^{(nl,0)}$ are orthogonal, we have

$$\begin{aligned} \left\|\boldsymbol{f}^{(ob,0)} + \boldsymbol{f}^{(nl,\mathbf{0})}\right\| &= \left\|\boldsymbol{H}\boldsymbol{f}^{(ob,0)} + \boldsymbol{H}\boldsymbol{f}^{(nl,0)}\right\| \\ &= \left\|\boldsymbol{H}\boldsymbol{f}^{(ob,0)}\right\| + \left\|\boldsymbol{H}\boldsymbol{f}^{(nl,0)}\right\| \\ &= \left\|\boldsymbol{f}^{(ob,0)}\right\| + \left\|\boldsymbol{f}^{(nl,0)}\right\|. \end{aligned} \tag{S.4.36}$$

This implies that

$$\left\|\boldsymbol{f}^{(ob,0)}\right\| < (1-\sigma)\|f^{*}\|. \tag{S.4.37}$$

Since $\boldsymbol{f}^{(0)}$ is the output of the soft-thresholding filtering, it can be expressed as

$$\boldsymbol{f}^{(0)} = \Phi\left(\boldsymbol{p}^{(0)}\right) - \boldsymbol{H}^{*}\bar{\boldsymbol{f}}^{(e,0)}, \tag{S.4.38}$$

where $\bar{\boldsymbol{f}}^{(\boldsymbol{e},\mathbf{0})}$ is a noise background in the transform domain. If we denote $\bar{f}_n^{(\boldsymbol{e},0)}$ as the $n^{th}$ component of $\bar{\boldsymbol{f}}^{(e,0)}$, there will be $\left|\bar{f}_n^{(\boldsymbol{e},0)}\right| \leq \varepsilon$, which is a noise floor. Without loss of generality, in the transform domain we assume the first $s$ components span the s-sparse space of $\boldsymbol{H}\boldsymbol{f}^{*}$. Because only the first $s$ components of $\bar{\boldsymbol{f}}^{(e,0)}$ is observable, let us decompose $\bar{\boldsymbol{f}}^{(e,0)}$ into two parts $\left(\bar{\boldsymbol{f}}^{(e,0)}\right)_{\boldsymbol{n}\leq s}$ and $\left(\bar{\boldsymbol{f}}^{(e,0)}\right)_{\boldsymbol{n}>s}$, where $\left(\bar{\boldsymbol{f}}^{(e,0)}\right)_{\boldsymbol{n}\leq s}$ is observable and $\left(\bar{\boldsymbol{f}}^{(e,0)}\right)_{\boldsymbol{n}>s}$ is in the null space of $\boldsymbol{A}$. Then, (S.4.38) can be rewritten as

$$\boldsymbol{f}^{(0)} = \boldsymbol{f}^{*} + \boldsymbol{f}^{(ob,0)} - \boldsymbol{H}^{*}\left(\bar{\boldsymbol{f}}^{(e,0)}\right)_{\boldsymbol{n}\leq s} + \boldsymbol{g}^{(nl,0)}, \tag{S.4.39}$$

where $\boldsymbol{g}^{(nl,0)} = \boldsymbol{f}^{(nl,0)} - \boldsymbol{H}^*\left(\bar{\boldsymbol{f}}^{(e,0)}\right)_{\boldsymbol{n}>s}$ is in the null space of $\boldsymbol{A}$.

**For the case $k = 0$:**

From (S.4.34) and (S.4.39), we have

$$\boldsymbol{p}^{(1)} = M_1\left(\boldsymbol{p}^{(0)} - \mathbf{A}\boldsymbol{f}^{(0)}\right) = -M_1\boldsymbol{A}\boldsymbol{f}^{(ob,0)} + M_1\boldsymbol{A}\boldsymbol{H}^*\left(\bar{\boldsymbol{f}}^{(e,0)}\right)_{\boldsymbol{n}\le s}, \quad \text{(S.4.40)}$$

$$\Phi\left(\boldsymbol{p}^{(1)}\right) = -M_1\boldsymbol{f}^{(ob,0)} + M_1\boldsymbol{H}^*\left(\bar{\boldsymbol{f}}^{(e,0)}\right)_{\boldsymbol{n}\le s} + \boldsymbol{f}^{(ob,1)} + \boldsymbol{f}^{(nl,1)}, \quad \text{(S.4.41)}$$

$$\left\|\boldsymbol{f}^{(ob,1)}\right\| < (1-\sigma)\left\|M_1\boldsymbol{f}^{(ob,0)} - M_1\boldsymbol{H}^*\left(\bar{\boldsymbol{f}}^{(e,0)}\right)_{\boldsymbol{n}\le s}\right\|, \quad \text{(S.4.42)}$$

$$\begin{aligned}\boldsymbol{f}^{(1)} &= \boldsymbol{f}^{(0)} + M_2\Phi\left(\boldsymbol{p}^{(1)}\right) - \boldsymbol{H}^*\bar{\boldsymbol{f}}^{(e,1)} \\ &= \boldsymbol{f}^* + (\mathbf{1}-M)\left(\boldsymbol{f}^{(ob,0)} - \boldsymbol{H}^*\left(\bar{\boldsymbol{f}}^{(e,0)}\right)_{\boldsymbol{n}\le s}\right) \\ &\quad + M_2\boldsymbol{f}^{(ob,1)} - \boldsymbol{H}^*\left(\bar{\boldsymbol{f}}^{(e,1)}\right)_{\boldsymbol{n}\le s} + \boldsymbol{g}^{(nl,1)}, \end{aligned} \quad \text{(S.4.43)}$$

where $\boldsymbol{g}^{(nl,1)} = \boldsymbol{g}^{(nl,0)} + M_2\left(\boldsymbol{f}^{(nl,1)}\right) - \boldsymbol{H}^*\left(\bar{\boldsymbol{f}}^{(e,1)}\right)_{\boldsymbol{n}>s}$ is the null space of $\boldsymbol{A}$.

**For the case $k = 1$:**

From (S.4.34) and (S.4.43), we have

$$\begin{aligned}\boldsymbol{p}^{(2)} = M_1\left(\boldsymbol{p}^{(0)} - \mathbf{A}\boldsymbol{f}^{(1)}\right) &= -(1-M)M_1\boldsymbol{A}\left(\boldsymbol{f}^{(ob,0)} - \boldsymbol{H}^*\left(\bar{\boldsymbol{f}}^{(e,0)}\right)_{\boldsymbol{n}\le s}\right) \\ &\quad - M\boldsymbol{A}\boldsymbol{f}^{(ob,1)} + M_1\boldsymbol{A}\boldsymbol{H}^*\left(\bar{\boldsymbol{f}}^{(e,\mathbf{1})}\right)_{\boldsymbol{n}\le s}, \end{aligned} \quad \text{(S.4.44)}$$

$$\begin{aligned}\Phi\left(\boldsymbol{p}^{(2)}\right) &= -(1-M)M_1\left(\boldsymbol{f}^{(ob,0)} - \boldsymbol{H}^*\left(\bar{\boldsymbol{f}}^{(e,0)}\right)_{\boldsymbol{n}\le s}\right) \\ &\quad - M\boldsymbol{f}^{(ob,1)} + M_1\boldsymbol{H}^*\left(\bar{\boldsymbol{f}}^{(e,1)}\right)_{\boldsymbol{n}\le s} + \boldsymbol{f}^{(ob,2)} + \boldsymbol{f}^{(nl,2)}, \end{aligned} \quad \text{(S.4.45)}$$

$$\begin{aligned}\left\|\boldsymbol{f}^{(ob,2)}\right\| < (1-\sigma)\Big\|(1-M)M_1\left(\boldsymbol{f}^{(ob,0)} - \boldsymbol{H}^*\left(\bar{\boldsymbol{f}}^{(e,\mathbf{0})}\right)_{\boldsymbol{n}\le s}\right) + M\boldsymbol{f}^{(ob,1)} - \\ M_1\boldsymbol{H}^*\left(\bar{\boldsymbol{f}}^{(e,\mathbf{1})}\right)_{\boldsymbol{n}\le s}\Big\|, \end{aligned} \quad \text{(S.4.46)}$$

$$\begin{aligned}\boldsymbol{f}^{(2)} &= \boldsymbol{f}^{(\mathbf{1})} + M_2\Phi\left(\boldsymbol{p}^{(2)}\right) - \boldsymbol{H}^*\bar{\boldsymbol{f}}^{(e,2)} \\ &= \boldsymbol{f}^* + (1-M)^{\mathbf{2}}\left(\boldsymbol{f}^{(ob,0)} - \boldsymbol{H}^*\left(\bar{\boldsymbol{f}}^{(e,\mathbf{0})}\right)_{\boldsymbol{n}\le s}\right) + \\ &\quad (1-M)\left(M_2\boldsymbol{f}^{(ob,1)} - \boldsymbol{H}^*\left(\bar{\boldsymbol{f}}^{(e,1)}\right)_{\boldsymbol{n}\le s}\right) + \\ &\quad M_2\boldsymbol{f}^{(ob,2)} - \boldsymbol{H}^*\left(\bar{\boldsymbol{f}}^{(e,2)}\right)_{\boldsymbol{n}\le s} + \boldsymbol{g}^{(nl,\mathbf{2})}, \end{aligned} \quad \text{(S.4.47)}$$

with $\boldsymbol{g}^{(nl,\mathbf{2})} = \boldsymbol{g}^{(nl,\mathbf{1})} + M_2\boldsymbol{f}^{(nl,\mathbf{2})} - \boldsymbol{H}^*\left(\bar{\boldsymbol{f}}^{(e,2)}\right)_{\boldsymbol{n}>s}$ .

If we continue the above procedure, for $k > 1$, it is easy to obtain that

$$\begin{aligned}\boldsymbol{p}^{(\boldsymbol{k}+\mathbf{1})} &= -(1-M)^{\boldsymbol{k}}M_1\boldsymbol{A}\left(\boldsymbol{f}^{(ob,0)} - \boldsymbol{H}^*\left(\bar{\boldsymbol{f}}^{(e,\mathbf{0})}\right)_{\boldsymbol{n}\le s}\right) \\ &\quad - \sum_{\boldsymbol{k}'=\mathbf{1}}^{\boldsymbol{k}}(1-M)^{\boldsymbol{k}-\boldsymbol{k}'}\boldsymbol{A}\left(M\boldsymbol{f}^{(ob,\boldsymbol{k}')} - M_1\boldsymbol{H}^*\left(\bar{\boldsymbol{f}}^{(e,\boldsymbol{k}')}\right)_{\boldsymbol{n}\le s}\right) \\ &= (1-M)\boldsymbol{p}^{(\boldsymbol{k})} - \boldsymbol{A}\left(M\boldsymbol{f}^{(ob,k)} - M_1\boldsymbol{H}^*\left(\bar{\boldsymbol{f}}^{(e,\boldsymbol{k})}\right)_{\boldsymbol{n}\le s}\right). \end{aligned} \quad \text{(S.4.48)}$$

Denoting the ground-truth image of $\boldsymbol{p}^{(k)}$ as $\boldsymbol{f}^{(*,k)}$, that is $\boldsymbol{p}^{(k)} = \boldsymbol{A}\boldsymbol{f}^{(*,k)}$, we have

$$\boldsymbol{f}^{(*,k+1)} = (1-M)\boldsymbol{f}^{(*,k)} - M\boldsymbol{f}^{(ob,k)} + M_1\boldsymbol{H}^*\left(\bar{\boldsymbol{f}}^{(e,k)}\right)_{\boldsymbol{n}\le s}. \quad \text{(S.4.49)}$$

$$\left\|\boldsymbol{f}^{(ob,k+1)}\right\| < (1-\sigma)\left\|\boldsymbol{f}^{(*,k+1)}\right\|. \quad \text{(S.4.50)}$$

Because each component of $\left(\bar{\boldsymbol{f}}^{(e,k)}\right)_{\boldsymbol{n}\le s}$ is bounded by $\varepsilon$, we have

$$\left\|\boldsymbol{H}^*\left(\bar{\boldsymbol{f}}^{(e,k)}\right)_{\boldsymbol{n}\le s}\right\| = \left\|\left(\bar{\boldsymbol{f}}^{(e,k)}\right)_{\boldsymbol{n}\le s}\right\| \le \varepsilon\sqrt{s}. \quad \text{(S.4.51)}$$

(S.4.49), (S.4.50) and (S.4.51) imply:

$$\left\|\boldsymbol{f}^{(*,1)}\right\| = \left\|M_1\boldsymbol{f}^{(ob,0)} + M_1\boldsymbol{H}^*\left(\bar{\boldsymbol{f}}^{(e,0)}\right)_{\boldsymbol{n}\le s}\right\| \le M_1\left\|\boldsymbol{f}^{(ob,0)}\right\| + M_1\varepsilon\sqrt{s}, \quad \text{(S.4.52)}$$

$$\left\|\boldsymbol{f}^{(*,2)}\right\| = \left\|(1-M)\boldsymbol{f}^{(*,1)} - M\boldsymbol{f}^{(ob,1)} + M_1\boldsymbol{H}^*\left(\bar{\boldsymbol{f}}^{(e,1)}\right)_{\boldsymbol{n}\le s}\right\|$$

$$\leq (1-M)\|\boldsymbol{f}^{(*,1)}\| + M\|\boldsymbol{f}^{(ob,1)}\| + M_1\varepsilon\sqrt{s}$$
$$\leq (1-M)\|\boldsymbol{f}^{(*,1)}\| + M(1-\sigma)\|\boldsymbol{f}^{(*,1)}\| + M_1\varepsilon\sqrt{s}$$
$$\leq (1-M\sigma)\left(\boldsymbol{M_1}\|\boldsymbol{f}^{(ob,0)}\| + \boldsymbol{M_1}\varepsilon\sqrt{s}\right) + M_1\varepsilon\sqrt{s}$$
$$\leq (1-M\sigma)\boldsymbol{M_1}\|\boldsymbol{f}^{(ob,0)}\| + ((1-M\sigma)+1)M_1\varepsilon\sqrt{s}. \quad \text{(S.4.53)}$$

If we continue this process, we can reach

$$\|\boldsymbol{f}^{(*,k+1)}\| \leq (1-M\sigma)^k\boldsymbol{M_1}\|\boldsymbol{f}^{(ob,0)}\| + \sum_{\boldsymbol{k'=0}}^{\boldsymbol{k}}(1-M\sigma)^{k'}\boldsymbol{M_1}\varepsilon\sqrt{s}$$
$$= (1-M\sigma)^k\boldsymbol{M_1}\|\boldsymbol{f}^{(ob,0)}\| + \frac{1-(1-M\sigma)^{k+1}}{M\sigma}\boldsymbol{M_1}\varepsilon\sqrt{s}, \quad \text{(S.4.54)}$$

$$\|\boldsymbol{f}^{(ob,k+1)}\| < (1-\sigma)\left((1-M\sigma)^k\boldsymbol{M_1}\|\boldsymbol{f}^{(ob,0)}\| + \frac{1-(1-M\sigma)^{k+1}}{M_2\sigma}\varepsilon\sqrt{s}\right). \quad \text{(S.4.55)}$$

When $k \to \infty$, (S.4.55) shows

$$\|\boldsymbol{f}^{(ob,\infty)}\| < \frac{(1-\sigma)\sqrt{s}}{M_2\sigma}\varepsilon. \quad \text{(S.4.56)}$$

Because the parameter $\varepsilon$ for the soft-thresholding kernel should match the system tolerance level, it is a noise floor. (S.4.56) implies that $\boldsymbol{f}^{(ob,k)}$ will converge to a noise-induced uncertainty range of the imaging system. For ideal noise-free case, the matching $\varepsilon \to 0$ and $\|\boldsymbol{f}^{(ob,\infty)}\| \to 0$. The bound (S.4.54) will monotonously decrease if it satisfies $\frac{\varepsilon\sqrt{s}}{\|\boldsymbol{f}^*\|} < \frac{(1-\sigma)M\sigma}{(1-M\sigma)}$. In other words, if the image is not too noisy, the ACID algorithm will converge to a solution in the intersection of the space constrained by measured data, the space of sparse solutions, and the space of deep priors.

In the above analysis, we have assumed that the input to the neural network is noise-free, that is, $\boldsymbol{p}^{(0)} = \boldsymbol{A}\boldsymbol{f}^*$. When there is a noise component in the projection data, this noise $\boldsymbol{e}$ can be decomposed into two parts: $\boldsymbol{e}_1$ and $\boldsymbol{e}_2$, where $\boldsymbol{e}_1$ satisfying $\boldsymbol{e}_1 = \boldsymbol{A}\boldsymbol{n}^*$ with $\boldsymbol{n}^*$ being the observable image corresponding to the noise so that $\boldsymbol{f}^* + \boldsymbol{n}^*$ is still consistent to both the data-driven prior and the sparsity condition, and $\boldsymbol{e}_2 = \boldsymbol{e} - \boldsymbol{e}_1$ as a complement of $\boldsymbol{e}_1$. Because the image $\boldsymbol{n}^*$ can be absorbed by $\boldsymbol{f}^*$, we can ignore $\boldsymbol{e}_1$ and only consider $\boldsymbol{e}_2$. Because $\boldsymbol{e}_2$ is outside the intersection of the three spaces constrained by (1) data-driven prior, (2) sparsity condition, and (3) measurement data, and thus makes no contribution to the final image, we can just modify the system tolerance level $\varepsilon$ accordingly to accommodate the effect of the noise $\boldsymbol{e}$ without affecting the above convergence analysis.

## IV.G. BREN and Lipschitz Continuity

Assuming the BREN property, our analysis shows that ACID is stable against adversarial attacks. In fact, BREN can be viewed as a special case of the Lipschitz continuity, *i.e.*, they are consistent.

Let us first define measurement and reconstruction operators $\boldsymbol{\mathcal{M}}$ and $\boldsymbol{\mathcal{R}}$ on two metric spaces $(\boldsymbol{F}, d_F)$ and $(\boldsymbol{P}, d_{\boldsymbol{P}})$, respectively. Let us measure an image $\boldsymbol{f} \in \boldsymbol{F}$ tomographically to obtain a measurement $\boldsymbol{p} \in \boldsymbol{P}$. We assume that each image in $\boldsymbol{F}$ is non-trivial in that sense that $\|\boldsymbol{f}\| > 0$. Let us denote the measurement operator $\boldsymbol{\mathcal{M}}: \boldsymbol{F} \to \boldsymbol{P}$; that is $\boldsymbol{p} = \boldsymbol{\mathcal{M}}(\boldsymbol{f})$. Suppose that the measurement process is totally transparent to us. Thus, we know a 1-to-1 correspondence $\boldsymbol{P} \leftrightarrow \boldsymbol{F}$ perfectly. For example, in our case, the measurement matrix $\boldsymbol{A}$ satisfies the RIP of order $s$, $\boldsymbol{f}$ is s-sparse, $\boldsymbol{p} = \boldsymbol{Af}$, and there exists a 1-to-1 map. Then, we can define the ideal reconstruction operator $\boldsymbol{f} = \boldsymbol{\mathcal{R}}(\boldsymbol{p})$. Reasonably, we assume that the operator $\boldsymbol{\mathcal{R}}$ is a Lipschitz continuous (LC) function $\boldsymbol{\mathcal{R}}: \boldsymbol{P} \to \boldsymbol{F}$ which satisfies $d_F(\boldsymbol{\mathcal{R}}(\boldsymbol{p_1}), \boldsymbol{\mathcal{R}}(\boldsymbol{p_2})) \leq L_1 d_F(\boldsymbol{f_1}, \boldsymbol{f_2})$ for a constant $L_1 > 0$.

With a big dataset, we can train a deep network $\Phi$ to approximate the ideal reconstruction operator $\boldsymbol{\mathcal{R}}$, $\Phi: \boldsymbol{P} \to \boldsymbol{F}$ is a LC function which satisfies $d_F(\Phi(\boldsymbol{p_1}), \Phi(\boldsymbol{p_2})) \leq L_2 d_F(\boldsymbol{f_1}, \boldsymbol{f_2})$ for a constant $L_2 > 0$. Furthermore, we assume that network $\Phi$ is well-designed and well-trained so that for a training tomographic dataset, we have $\|\Phi(\boldsymbol{p}(i)) - \boldsymbol{\mathcal{R}}(\boldsymbol{p}(i))\| < \delta_n$, $i = 1,2,\cdots, I_{trn}$. For a new dataset $\boldsymbol{p}'$ from an underlying image $\boldsymbol{f}'$, the BREN property requires that $\frac{\|\Phi(\boldsymbol{p}') - \boldsymbol{\mathcal{R}}(\boldsymbol{p}')\|}{\|\boldsymbol{\mathcal{R}}(\boldsymbol{p}')\|} = \frac{\|\Phi(\boldsymbol{Af}') - \boldsymbol{f}'\|}{\|\boldsymbol{f}'\|} \leq 1 - \sigma$. Suppose that the image $\boldsymbol{f}'$ is close to an image $\boldsymbol{f}(i_0)$ in the training dataset. In this setting, we have the following relations:

$$\|\Phi(\boldsymbol{p}(i_0)) - \boldsymbol{\mathcal{R}}(\boldsymbol{p}(i_0))\| < \delta_n, \tag{S.4.57}$$

$$d_F\left(\boldsymbol{\mathcal{R}}(\boldsymbol{p}(i_0)), \boldsymbol{\mathcal{R}}(\boldsymbol{p}')\right) \leq L_1 d_F(\boldsymbol{f}(i_0), \boldsymbol{f}'), \tag{S.4.58}$$

$$d_F\left(\Phi(\boldsymbol{p}(i_0)), \Phi(\boldsymbol{p}')\right) \leq L_2 d_F(\boldsymbol{f}(i_0), \boldsymbol{f}'), \tag{S.4.59}$$

where (S.4.57) is due to the fact the network is well-designed and well-trained, and (S.4.58) and (S.4.59) are due to the Lipschitz continuity of $\boldsymbol{\mathcal{R}}$ and $\Phi$. Therefore, we have

$$\begin{aligned} &\|\Phi(\boldsymbol{p}') - \boldsymbol{\mathcal{R}}(\boldsymbol{p}')\| \\ &= \|\Phi(\boldsymbol{p}') - \Phi(\boldsymbol{p}(i_0)) + \Phi(\boldsymbol{p}(i_0)) - \boldsymbol{\mathcal{R}}(\boldsymbol{p}(i_0)) + \boldsymbol{\mathcal{R}}(\boldsymbol{p}(i_0)) - \boldsymbol{\mathcal{R}}(\boldsymbol{p}')\| \\ &\leq \|\Phi(\boldsymbol{p}') - \Phi(\boldsymbol{p}(i_0))\| + \|\Phi(\boldsymbol{p}(i_0)) - \boldsymbol{\mathcal{R}}(\boldsymbol{p}(i_0))\| + \|\boldsymbol{\mathcal{R}}(\boldsymbol{p}(i_0)) - \boldsymbol{\mathcal{R}}(\boldsymbol{p}')\| \\ &< L_1 d_F(\boldsymbol{f}(i_0), \boldsymbol{f}') + L_2 d_F(\boldsymbol{f}(i_0), \boldsymbol{f}') + \delta_n. \end{aligned} \tag{S.4.60}$$

Therefore, under the condition that

$$\frac{L_1 d_F(\boldsymbol{f}(i_0), \boldsymbol{f}') + L_2 d_F(\boldsymbol{f}(i_0), \boldsymbol{f}') + \delta_n}{\|\boldsymbol{f}'\|} < 1, \tag{S.4.61}$$

we have the BREN. The condition can be simplified to $\frac{(L_{1+} + L_2) d_F(\boldsymbol{f}(i_0), \boldsymbol{f}') + \delta_n}{\|\boldsymbol{f}'\|} < 1$, which is roughly $\frac{(L_{1+} + L_2) d_F(\boldsymbol{f}(i_0), \boldsymbol{f}')}{\|\boldsymbol{f}'\|} < 1$. That is, as long as an image is fairly close to the training dataset, the BREN property is satisfied. Heuristically, if the image norm is greater than the product of the LC constant $(L_{1+} + L_2)$ and the distance between an image to be reconstructed and its closest reference point, we have the BREN property. For a big dataset, $d_F(\boldsymbol{f}(i_0), \boldsymbol{f}')$ is small, so that $L$ can be large, which is especially true if we interpret $\boldsymbol{F}$ and $\boldsymbol{P}$ as appropriate low-dimensional manifolds.

Because there is a 1-to-1 correspondence $\boldsymbol{P} \leftrightarrow \boldsymbol{F}$ perfectly, we can treat the combination of the measurement matrix $\boldsymbol{A}$ and the neural network $\Phi$ as a new LC function $\Phi_{\boldsymbol{A}}$ which satisfies

$$d_F(\Phi_{\boldsymbol{A}}(\boldsymbol{f_1}), \Phi_{\boldsymbol{A}}(\boldsymbol{f_2})) \leq L d_F(\boldsymbol{f_1}, \boldsymbol{f_2}). \tag{S.4.62}$$

The Lipschitz continuity assumption is useful to assess the convergence of a deep reconstruction algorithm. In our previous section IV.C, we have verified the BREN property for the data used in [6]. Those results support the practical relevance of the BREN property. More importantly, one can calculate the Lipschitz constant directly for both the MRI and CT data using (S.4.62).